\documentclass[a4paper,twoside,twocolumn,9pt]{article}
\usepackage{extsizes}
\usepackage[super,sort&compress,comma]{natbib} 
\usepackage[version=3]{mhchem}
\usepackage[left=1.5cm, right=1.5cm, top=1.785cm, bottom=2.0cm]{geometry}
\usepackage{balance}
\usepackage{mathptmx}
\usepackage{sectsty}
\usepackage{graphicx} 
\usepackage{lastpage}
\usepackage[format=plain,justification=justified,singlelinecheck=false,font={stretch=1.125,small,sf},labelfont=bf,labelsep=space]{caption}
\usepackage{float}
\usepackage{fancyhdr}
\usepackage{fnpos}
\usepackage[english]{babel}
\addto{\captionsenglish}{%
  
}
\usepackage{array}
\usepackage{droidsans}
\usepackage{charter}
\usepackage[T1]{fontenc}
\usepackage[usenames,dvipsnames]{xcolor}
\usepackage{setspace}
\usepackage[compact]{titlesec}
\usepackage{hyperref}

\newcommand{\rev}[1]{\textcolor{black}{#1}}

\usepackage{multirow,booktabs,subcaption,amsmath,amsfonts,amssymb,bm,tikz,tikz-3dplot,xcolor, wrapfig}
\definecolor{cream}{RGB}{222,217,201}

\graphicspath{{./}}

\begin{document}

\pagestyle{fancy}
\thispagestyle{plain}
\fancypagestyle{plain}{
\renewcommand{\headrulewidth}{0pt}
}

\makeFNbottom
\makeatletter
\renewcommand\LARGE{\@setfontsize\LARGE{15pt}{17}}
\renewcommand\Large{\@setfontsize\Large{12pt}{14}}
\renewcommand\large{\@setfontsize\large{10pt}{12}}
\renewcommand\footnotesize{\@setfontsize\footnotesize{7pt}{10}}
\makeatother

\renewcommand{\thefootnote}{\fnsymbol{footnote}}
\renewcommand\footnoterule{\vspace*{1pt}%
\color{cream}\hrule width 3.5in height 0.4pt \color{black}\vspace*{5pt}} 
\setcounter{secnumdepth}{5}

\makeatletter 
\renewcommand\@biblabel[1]{#1}            
\renewcommand\@makefntext[1]%
{\noindent\makebox[0pt][r]{\@thefnmark\,}#1}
\makeatother 
\renewcommand{\figurename}{\small{Fig.}~}
\sectionfont{\sffamily\Large}
\subsectionfont{\normalsize}
\subsubsectionfont{\bf}
\setstretch{1.125} 
\setlength{\skip\footins}{0.8cm}
\setlength{\footnotesep}{0.25cm}
\setlength{\jot}{10pt}
\titlespacing*{\section}{0pt}{4pt}{4pt}
\titlespacing*{\subsection}{0pt}{15pt}{1pt}

\fancyfoot{}
\fancyfoot[RO]{\footnotesize{\sffamily{\hspace{1cm}\thepage}}}
\fancyfoot[LE]{\footnotesize{\sffamily{\thepage}}}
\fancyhead{}
\renewcommand{\headrulewidth}{0pt} 
\renewcommand{\footrulewidth}{0pt}
\setlength{\arrayrulewidth}{1pt}
\setlength{\columnsep}{6.5mm}
\setlength\bibsep{1pt}

\makeatletter 
\newlength{\figrulesep} 
\setlength{\figrulesep}{0.5\textfloatsep} 

\newcommand{\topfigrule}{\vspace*{-1pt}%
\noindent{\color{cream}\rule[-\figrulesep]{\columnwidth}{1.5pt}} }

\newcommand{\botfigrule}{\vspace*{-2pt}%
\noindent{\color{cream}\rule[\figrulesep]{\columnwidth}{1.5pt}} }

\newcommand{\dblfigrule}{\vspace*{-1pt}%
\noindent{\color{cream}\rule[-\figrulesep]{\textwidth}{1.5pt}} }

\makeatother

\twocolumn[
  \begin{@twocolumnfalse}
\vspace{1em}
\sffamily

\noindent\LARGE{\textbf{Defect-influenced particle advection in highly confined liquid crystal flows$^\dag$}} \\

\noindent\large{Magdalena Lesniewska\textit{$^{a}$}, Nigel Mottram\textit{$^{b}$}, and Oliver Henrich\textit{$^{a}$}} \\

\noindent\normalsize{
We study the morphology of the Saturn ring defect and director structure around a colloidal particle with normal anchoring conditions and within the flow of the nematic host phase through a rectangular duct of comparable size to the particle. The changes in the defect structures and director profile influence the advection behaviour of the particle, which we compare to that in a simple Newtonian host phase. These effects lead to a non-monotonous dependence of the differential velocity of particle and fluid, also known as retardation ratio, on the Ericksen number.}\\


 \end{@twocolumnfalse} \vspace{0.6cm}

  ]

\renewcommand*\rmdefault{bch}\normalfont\upshape
\rmfamily
\section*{}
\vspace{-1cm}


\footnotetext{\textit{$^{a}$~Department of Physics, University of Strathclyde, Glasgow G4 0NG, UK.}}
\footnotetext{\textit{$^{b}$~School of Mathematics \& Statistics, University of Glasgow, Glasgow G12 8QQ, UK. }}
\footnotetext{\ddag~Email: oliver.henrich@strath.ac.uk}





\definecolor{bla1}{HTML}{1b9e77}
\definecolor{bla2}{HTML}{d95f02}
\definecolor{bla3}{HTML}{7570b3}
\definecolor{bla4}{HTML}{e7298a}

\definecolor{bla1_1}{HTML}{666666}
\definecolor{bla1_2}{HTML}{cf9901}

\section{Introduction}

Microfluidics is concerned with the manipulation and control of fluid flow at the microscale and sits as a  multidisciplinary field at the intersection of physics, engineering, and biology. During the last two decades it has seen a tremendous rise in importance as it entered the main stream with many practical applications ranging from medical diagnostics \cite{Chin2011,Hu2014} and drug delivery \cite{Zhang2013,Golombek2018} to chemical synthesis \cite{Seemann2012} and lab-on-a-chip technologies \cite{Mark2010,Qin2010,Sackmann2014}.

The essential physics of a microfluidic system is dictated by a competition between various phenomena, which is captured by a series of dimensionless numbers expressing their relative importance \cite{Squires2005}. The Reynolds number, for instance, is often considered to be small in microfluidic applications. However, relatively recently a focal point has been on inertial microfluidics \cite{DiCarlo2007,Amini2014,Zhang2016}, which gives rise to some interesting and counter-intuitive phenomena.

When particle-laden flows are considered, aspects of confinement become central to microfluidics. Characteristic dimensions of channels and chambers are in the range of tens to hundreds of micrometres, and can be comparable to the size of the particles being transported, and  interactions between the fluid and solid boundaries become increasingly important. In the simplest case the fluid is Newtonian, without internal structure, and the geometry is a uniform duct. The first theoretical results for migrating, rigid spheres in unidirectional, two-dimensional flow were provided by Ho and Leal \cite{Ho1974}. These were later extended to three dimensions and refined by Ganatos \cite{Ganatos1980} and Staben \cite{Staben2003}, who also verified their theoretical results with an experimental study \cite{Staben2005}. Owing to the rise of computer power and the advent of sophisticated simulation methodologies, increasingly complex geometries can now be investigated \cite{Nikoubashman2013}.

For similar reasons, the study of particle \rev{advection (i.e. movement in flow direction)} in non-Newtonian fluids with internal order structure has evolved only slowly.
Suspensions of colloidal particles in a nematic liquid crystalline host phase may serve here as prototype of systems that cannot be described with standard continuum theories, but require additional order parameters to capture the microstructure and its change under flow conditions.
After first theoretical studies on the drag of colloidal particles in nematic hosts by Stark \cite{Stark2002}, \rev{a main research focus in colloid-liquid crystal suspension has been on confinement effects \cite{Grollau2003,Garlea2015,Garlea2016} and topology properties \cite{Tkalec2013,Copar2021}. Confinement effects, often in combination with the behaviour in external electric fields, were also investigated in liquid crystalline emulsions \cite{Sulaiman2006}} as well as in droplets and shells \cite{Urbanski2017}. \rev{The study of flowing liquid crystals covered primarily pure and confined phases} \cite{Wiese2016,Mondal2018a,Kos2020,Copar2021,Fedorowicz2023} \rev{as they appear frequently in microfluidic setups}.
  
On the theoretical side, studies have been extended to multiple, explicitly resolved particles and the full nemato-hydrodynamic problem that solves for the tensor order parameter and velocity field \cite{Mondal2018b}. These approaches are now complemented by various simulation methods \cite{Stieger2014, Stieger2015, Lesniewska2022}. 
\rev{Recently, the dynamics of anisotropic particles in nematic liquid crystals under shear flow was investigated \cite{Hijar2020}.} However, \rev{the advection of colloidal particles in pressure driven flow and} extreme confinement, which forms the focus of this work, \rev{has to our knowledge so far not been addressed.}

Our paper is organised as follows:~Section \ref{sec2} introduces our theoretical modelling framework, the Landau-de Gennes free energy and Beris-Edwards model, while Section \ref{sec3} gives details of our lattice Boltzmann simulation method. Section \ref{sec4} shows simulation results for scalar order parameter and director field at various confinement ratios and flow velocities and them to those obtained for simple Newtonian fluids. Section \ref{sec5} summarises our results and conclusions.

\section{Theory}\label{sec2}

\subsection{Landau-de Gennes free energy}

The local order of the liquid crystal is described by a traceless and symmetric second-order tensor $\bm{Q}(\bm{r}, t)$ \cite{deGennes, WrightMermin:1989}. Its largest eigenvalue $q<2/3$ is referred to as the scalar order parameter and provides a measure of the liquid crystalline order at a certain position and time. The eigenvector, $\bm{d}$, associated with $q$ is called the director and describes the corresponding average orientation of the liquid crystal molecules.

In equilibrium, the liquid crystal order is determined through minimisation of its free energy, commonly described by the Landau-de Gennes free energy functional 
\begin{equation}
{\cal F}[\bm{Q}] = \int_V f(\bm{Q})\, dV + \int_S f_s(\bm{Q})\, dS,
\end{equation}
which includes the volume contribution $f=f_b + f_g$, that itself consists of a bulk contribution $f_b$ and a gradient contribution $f_g$, and a surface contribution $f_s$. The bulk free energy density is given by
\begin{eqnarray}
f_b(\bm{Q})=
{\frac{A_0}{2}}\left(1 - \dfrac{\gamma}{3}\right) Q_{\alpha\beta}^2 -{\frac{A_0}{3}}\gamma \,Q_{\alpha\beta} Q_{\beta\pi} Q_{\pi\alpha}+ {\frac{A_0}{4}}  \gamma \,(Q_{\alpha\beta}^2)^2,
\label{eq-lc-fed-bulk}
\end{eqnarray}
where we use the Einstein summation convention, so that Greek indices that appear twice are summed over. In Eq.~\ref{eq-lc-fed-bulk}, $A_0$ is a constant that sets the overall energy scale and the parameter $\gamma$ controls the temperature difference from the isotropic-nematic transition, and is related to a reduced temperature $\tau$ by
\begin{equation}
\tau = \frac{27}{\gamma}\left(1-\frac{\gamma}{3}\right).
\end{equation}
For $\gamma>3$ or $\tau<0$ the ordered, nematic state is the equilibrium phase, whereas for $2.7 \le \gamma \le 3$ or $0\le \tau\le 1$ the nematic state is metastable. For $\gamma<2.7$ or $\tau >1$ the isotropic state is the equilibrium phase.
  
The gradient free energy density $f_g$ contains the contributions of splay, bend and twist deformations of the director field as well as order-elastic effects due to gradients of the scalar order parameter,
\begin{eqnarray}
f_g(\bm{Q}) =
{\frac{1}{2}} \kappa_0 (\partial_\alpha Q_{\alpha\beta})^2+ {\frac{1}{2}} \kappa_1 (\epsilon_{\alpha\sigma\nu} \partial_\sigma Q_{\nu\beta})^2,
\label{equation-lc-gradient-fe}
\end{eqnarray}
where $\partial_\alpha=\partial/\partial x_\alpha$ and $\epsilon_{\alpha\sigma\nu}$ is the Levi-Civita symbol in three dimensions. In principle, the elastic constants $\kappa_0$ for splay and bend deformations, and $\kappa_1$ for twist deformations can be different. However, in our simulations we use the one-elastic-constant approximation $\kappa_0=\kappa_1$.

The director is assumed to have a preferred normal orientation to the wall surfaces and to the surface of the colloidal particle, known as a homeotropic anchoring, and is described using a surface free energy term 
\begin{equation}
f_s(\bm{Q}) = {\textstyle\frac{1}{2}} w (Q_{\alpha\beta} - Q_{\alpha\beta}^0)^2,
\label{surface_fe}
\end{equation}
where $w$ is the surface anchoring strength with values $w_{wall}$ and $w_{part}$ at the wall and particle surfaces, respectively. The preferred orientation $Q^0_{\alpha\beta}$ is assumed to be uniaxial and is given by
\begin{equation}
Q^0_{\alpha\beta} = {\textstyle \frac{1}{2}} S_0 (3\,n_\alpha n_\beta - \delta_{\alpha\beta}),
\end{equation}
where $\bm{n}$ is the surface unit normal, $\delta_{\alpha\beta}$ is the Kronecker delta and $S_0$ is the preferred surface scalar order parameter given by
\begin{equation}
\label{equation-lc-amplitude}
S_0=\frac{2}{3}\left(\frac{1}{4}+\frac{3}{4}\sqrt{1 - \frac{8}{3\gamma}}\right).
\end{equation}

The anchoring strength at the surface of a colloidal particle is often compared to the bulk fluid elastic constant by means of the dimensionless parameter
\begin{equation}
\omega = \frac{w\,R}{\kappa},
\end{equation}
where $R$ and $\kappa$ are the radius of the particle and the elastic constant, respectively.
For small values of this parameter, the presence of a particle surface should have little impact on the local bulk liquid crystalline ordering.

\subsection{Beris-Edwards model}

The time evolution of $Q_{\alpha\beta}$ is governed by the Beris-Edwards equation \cite{BerisEdwards} 
\begin{eqnarray}
\partial_t Q_{\alpha\beta} + \partial_{\rev{\pi}} (u_{\rev{\pi}} Q_{\alpha\beta})
- S_{\alpha\beta}(\bm{W},\bm{Q}) = \Gamma\,  H_{\alpha\beta},
\label{equation-lc-beris-edwards}
\end{eqnarray}
where $\partial_t=\partial/\partial t$, $\bm{u}$ is the flow velocity, $\bm{S}(\bm{W},\bm{Q})$ denotes the response to shear, $\bm{W}$ is the velocity gradient tensor, $\bm{H}$ is the molecular field and $\Gamma$ is a mobility parameter. The shear term is given by 
\begin{align}
S_{\alpha\beta}& (\bm{W},\,\bm{Q}) =  (\xi D_{\alpha\pi} + \Omega_{\alpha\pi})(Q_{\pi\beta} + {\textstyle \frac{1}{3}} \delta_{\pi\beta}) \nonumber\\
&+ (Q_{\alpha\pi} + {\textstyle \frac{1}{3}} \delta_{\alpha\pi})(\xi D_{\pi\beta} - \Omega_{\pi\beta}) - 2\xi(Q_{\alpha\beta} + {\textstyle\frac{1}{3}}\delta_{\alpha\beta})Q_{\pi\sigma}W_{\sigma\pi}
\end{align}
where $D_{\alpha\beta} = \frac{1}{2}(W_{\alpha\beta} + W_{\beta\alpha})$ and $\Omega_{\alpha\beta} = \frac{1}{2}(W_{\alpha\beta} - W_{\beta\alpha})$ are the symmetric and antisymmetric contributions to the velocity gradient tensor  $W_{\alpha \beta} = \partial_\alpha u_\beta$, respectively, and $\xi$ is the so-called flow alignment parameter, a material constant representing an effective molecular aspect ratio which determines whether the liquid crystal exhibits a flow-aligned state at the Leslie angle or tumbling state. 
The molecular field $\bm{H}$ is the functional derivative of the free energy functional with respect to the order parameter,
\begin{equation}
H_{\alpha\beta} = -
  \frac{\delta \cal F} {\delta Q_{\alpha\beta}}
+ \frac{\delta_{\alpha\beta}}{3} {\rm Tr} \frac{\delta \cal F} {\delta Q_{\alpha\beta}}.
\label{molecular_field}
\end{equation}
The second term in Eq. \ref{molecular_field} involving the trace ensures tracelessness of the tensor order parameter as it evolves through Eq.\ref{equation-lc-beris-edwards}. This leads to the following molecular field:
\begin{eqnarray}
\label{eq-lc-h-full}
H_{\alpha\beta} = \hspace*{-0.6cm}&&-A_0 (1 - \gamma/3) Q_{\alpha\beta} + A_0 \gamma\, (Q_{\alpha\sigma} Q_{\sigma\beta} - {\textstyle\frac{1}{3}} Q_{\sigma\nu}^2\delta_{\alpha\beta})\nonumber\\ 
\hspace*{-0.6cm}&&-A_0 \gamma\, Q_{\sigma\nu}^2 Q_{\alpha\beta} +\kappa_0 \partial_\alpha \partial_\sigma Q_{\sigma\beta}
+ \kappa_1 \partial_\sigma(\partial_\sigma Q_{\alpha\beta} - \partial_\alpha Q_{\sigma\beta})\nonumber\\
\end{eqnarray}
The governing  equations of hydrodynamic motion are the equation of mass conservation, also known as the continuity equation, and the Navier-Stokes equation that describes the \rev{balance between the rate of change of linear momentum density and the gradients of the pressure and viscous stresses}. In tensor notation they read
\begin{equation}
\partial_t \rho + \partial_\alpha (\rho u_\alpha) = 0
\label{eq_mass1}
\end{equation}
and
\begin{align}
\hspace*{-0.2cm}\partial_t (\rho & u_\alpha) = \partial_\beta \Pi^{(LC)}_{\alpha\beta}+ \partial_\beta \Pi^{(HD)}_{\alpha\beta},
\label{eq_momentum1}
\end{align}
respectively.
Eq. \ref{eq_mass1} relates the local rate of change of the density $\rho$ to the advection of mass by the fluid velocity $\bm{u}$. Eq. \ref{eq_momentum1} is Newton's second law of momentum change for the fluid and involves the thermotropic stress tensor $\Pi^{(LC)}_{\alpha\beta}$ and the hydrodynamic stress tensor $\Pi_{\alpha\beta}^{(HD)}$.
The thermotropic stress arises due to the liquid crystal and is given by

\begin{equation}
\Pi^{(LC)}_{\alpha\beta} = \sigma_{\alpha\beta} + \tau_{\alpha\beta} - \partial_\alpha Q_{\sigma\nu} \frac{\delta {\cal F}}{ \delta \partial_\beta Q_{\sigma\nu}}.
\label{equation-lc-stress}
\end{equation} 

In Eq.~\ref{equation-lc-stress}, $\sigma_{\alpha\beta}$ and $\tau_{\alpha\beta}$ are the symmetric and antisymmetric stress contributions, respectively, defined as

\begin{eqnarray}
\hspace*{-0.6cm}\sigma_{\alpha\beta} &=& -p_0\, \delta_{\alpha\beta} - \xi H_{\alpha\sigma}(Q_{\sigma\beta} + {\textstyle\frac{1}{3}}\delta_{\sigma\beta})- \xi (Q_{\alpha\sigma} + {\textstyle\frac{1}{3}}\delta_{\alpha\sigma}) H_{\sigma\beta}\nonumber\\
&& + 2\xi (Q_{\alpha\beta} + {\textstyle \frac{1}{3}}\delta_{\alpha\beta}) Q_{\sigma\nu} H_{\sigma\nu},
\label{sym_stress}
\end{eqnarray}

where $p_0=-\left(\partial {\cal F} / \partial V\right)_T = -f$ is the isotropic contribution from the nematic liquid crystal to the total pressure, and

\begin{equation}
\tau_{\alpha\beta} = Q_{\alpha\sigma} H_{\sigma\beta} - H_{\alpha\sigma} Q_{\sigma\beta}.
\label{antisym_stress}
\end{equation}
The final term in   Eq. \ref{equation-lc-stress} may be expanded as
\begin{equation}
\begin{split}
\partial_\alpha Q_{\sigma\nu} \frac{\delta {\cal F}}{ \delta \partial_\beta Q_{\sigma\nu}} =& -\kappa_0 \partial_\alpha Q_{\sigma\beta} \partial_\nu Q_{\sigma\nu}\\
&-\kappa_1 \partial_\alpha Q_{\sigma\nu} \left( \partial_\beta Q_{\sigma\nu} - \partial_\sigma Q_{\nu\beta}\right).
\label{delgradq_stress}
\end{split}
\end{equation}
The hydrodynamic stress tensor is defined as
\begin{equation}
\Pi_{\alpha\beta}^{(HD)}=-p\,\delta_{\alpha\beta}-\rho u_\alpha u_\beta +
\mu(\partial_\beta u_\alpha + \partial_\alpha u_\beta) + \zeta \partial_\sigma u_\sigma \delta_{\alpha\beta},
\label{equation-hydro-stress}
\end{equation}
where $\mu$ and $\zeta$ are the dynamic and bulk viscosity, respectively. The hydrostatic pressure $p$ is related to the density via an ideal gas equation of state as $p=c_s^2\rho$ with $c_s$ as lattice speed of sound as is standard in the lattice Boltzmann method. The last term vanishes in incompressible fluids as Eq. \ref{eq_mass1} becomes $\partial_\alpha u_\alpha=0$. 

No-slip and no-penetration boundary conditions are applied on the walls and particle surfaces, and 
the boundary conditions for $\bm{Q}$ are found from the minimisation of the free energy \cite{Skarabot2007}

\begin{equation}
n_\gamma \frac{\partial f}{\partial Q_{\alpha\beta,\gamma}}
+ \frac{\partial f_s}{\partial Q_{\alpha\beta}} = 0,
\label{equation-lc-general-bc}
\end{equation}

where $Q_{\alpha\beta,\gamma}=\partial Q_{\alpha\beta}/\partial x_\gamma$.

\section{Simulation Method}\label{sec3}

\subsection{Simulation Setup}

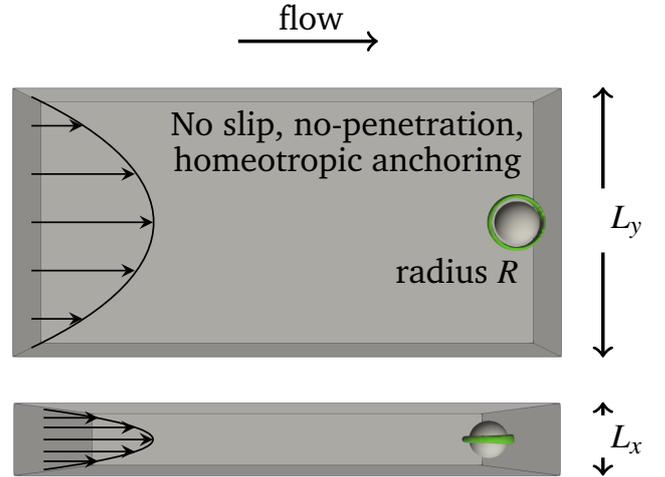
\begin{figure}[htbp]
\centering
\scalebox{1.6}{
\begin{tikzpicture}
\node[inner sep=0pt] at (0,0) {\includegraphics[trim={1cm 2.5cm 1cm 2.5cm},clip, scale=0.1]{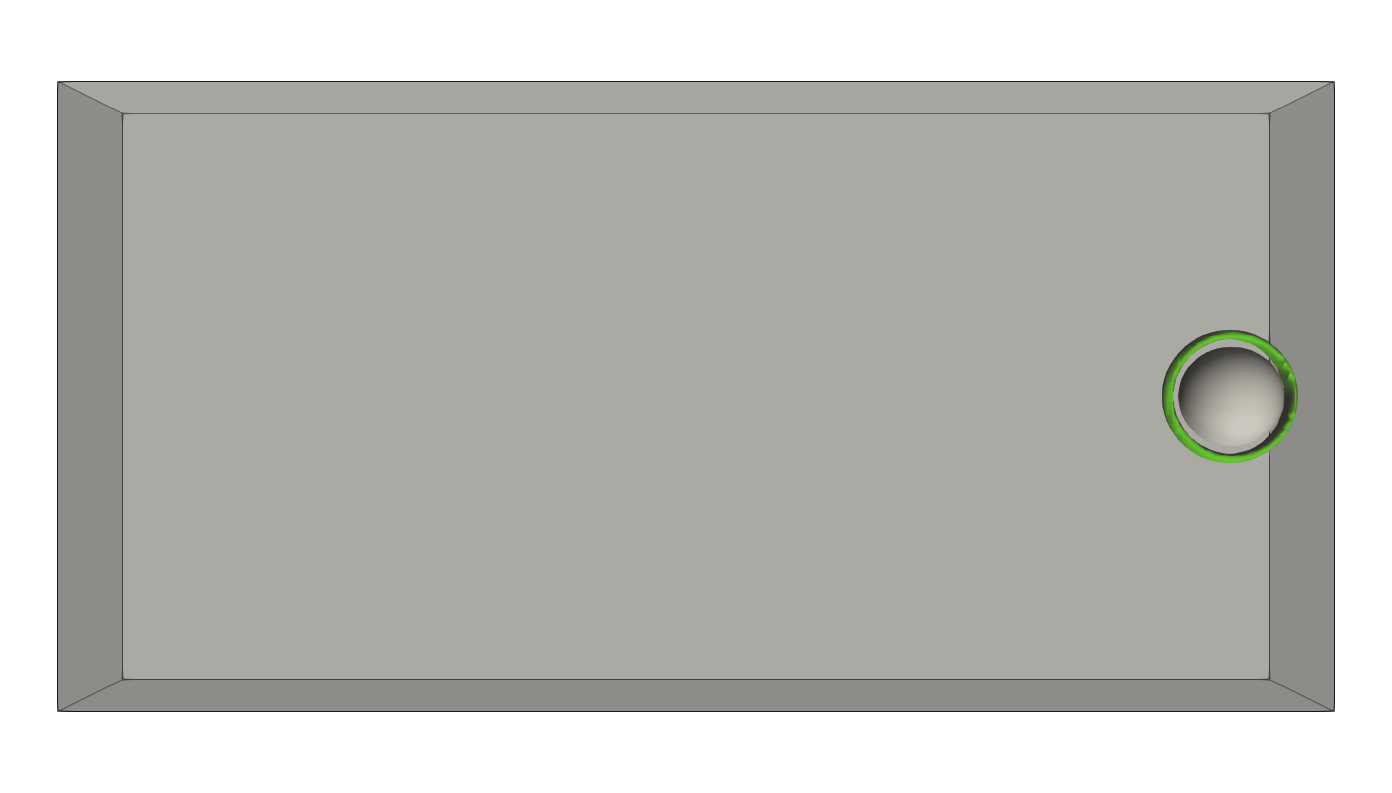}} ;
\node[inner sep=0pt] at (0,-1.8) {\includegraphics[trim={3cm 11cm 3cm 11cm},clip,scale=0.106]{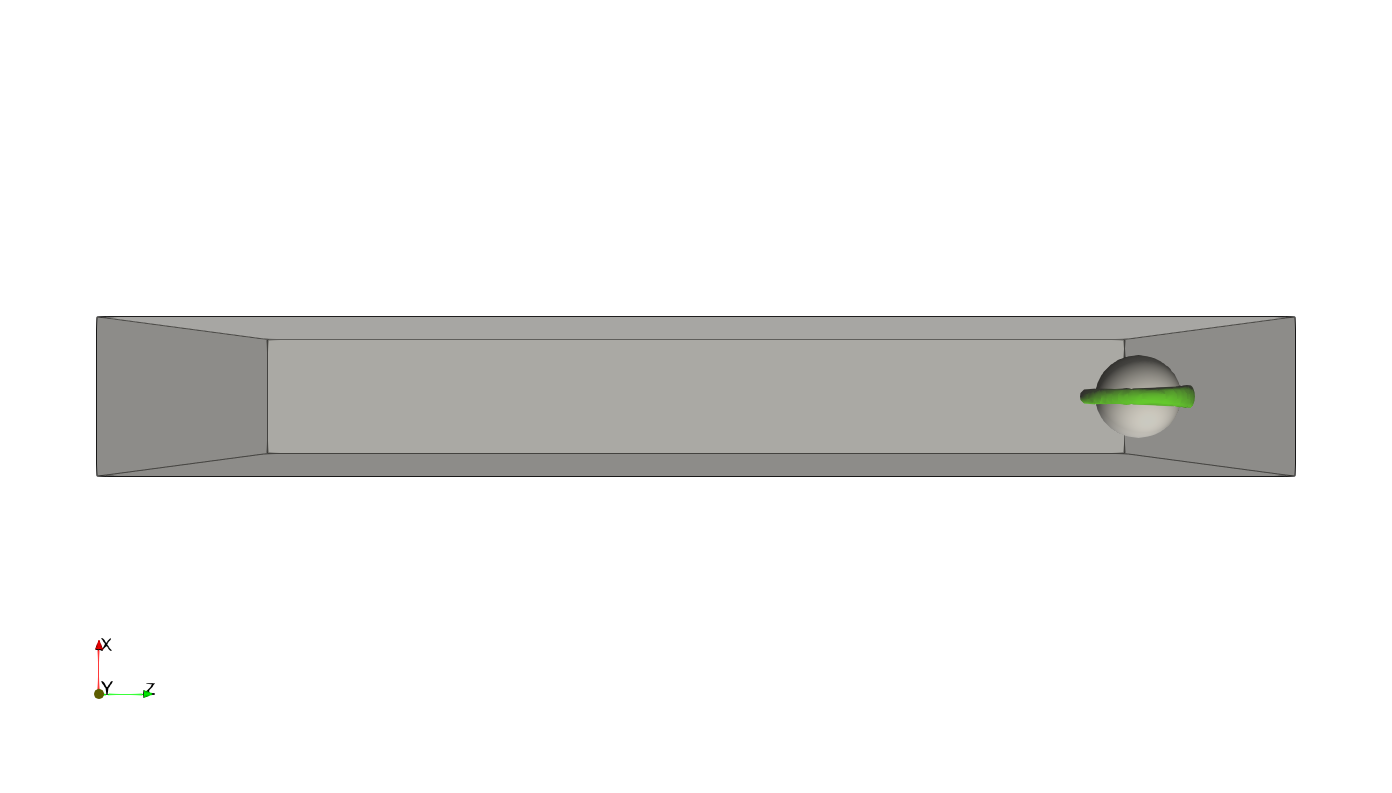}} ;
\node[] at (2.8,0) {\small{$L_y$} };
\draw[->, thick] (2.6,0.28)  -- (2.6,1.12) ;
\draw[->, thick] (2.6,-0.25)  -- (2.6,-1.12); 
\node[] at (2.8,-1.8) {\small{$L_x$} };
\draw[->, thick] (2.6,-1.95)  -- (2.6,-2.1) ;
\draw[->, thick] (2.6,-1.6)  -- (2.6,-1.49); 
\draw[thick,->] (-0.40,1.5) -- (0.75,1.5);
\node[] at (1.4,-0.4) {\small radius $R$ };
\node[] at (0.5,0.8) {\small No slip, no-penetration,};
\node[] at (0.5,0.5) {\small homeotropic anchoring};
\node[] at (0.2,1.7) {\small flow};  
\path[draw] (-2,-2.05) .. controls (-0.8,-1.883) and (-0.8,-1.716) .. (-2,-1.55);
\draw [-stealth](-2,-1.8) -- (-1.1,-1.8);
\draw [-stealth] (-2,-1.7) -- (-1.25,-1.7);
\draw [-stealth] (-2,-1.9) -- (-1.25,-1.9);
\draw [-stealth] (-2,-1.62) -- (-1.55,-1.62);
\draw [-stealth] (-2,-1.98) -- (-1.55,-1.98);
\path[draw] (-2.1,-1.04) .. controls (-0.76,-0.4 ) and (-0.76,0.4) .. (-2.1,1.04);
\draw [-stealth](-2.1,0) -- (-1.1,0);
\draw [-stealth](-2.1,0.4) -- (-1.25,0.4);
\draw [-stealth](-2.1,-0.4) -- (-1.25,-0.4);
\draw [-stealth](-2.1,0.8) -- (-1.68,0.8);
\draw [-stealth](-2.1,-0.8) -- (-1.68,-0.8);
\end{tikzpicture}}
\caption{ Overview of the computational geometry: We apply no-slip and no-penetration boundary conditions and homeotropic anchoring conditions at the walls which bound the region in $x$-direction and $y$-direction and at the particle surface, with periodic boundary conditions at the $z$-boundaries. The top part shows the top view, looking along the $x$-direction, and the bottom part shows the side view, looking along the $y$-direction.}
\label{fig1}
\end{figure}

Fig~\ref{fig1} shows a diagram of the three-dimensional computational geometry, which consists of a duct of $L_x=24$, $32$ or $48$ and $L_y \times L_z =  256 \times 384$ lattice sites.
Solid walls are positioned at $x=0$ and $x=L_x$, $y=0$ and $y=L_y$. We define the measure of confinement as the ratio of the particle diameter to the height of the duct, which leads in our case to confinement ratios $2R/L_x=0.8$, $0.6$, and $0.4$. The value of $L_y$ means that, with the particle at the centre of the duct, the system is effectively unconfined in $y$-direction since  $2R/L_y=0.075$. Periodic boundary conditions are applied in the $z$-direction with the $z$-boundaries acting as inlet and outlet of the duct.  A pressure gradient $\Psi=\Delta p/L_z$ is applied in $z$-direction, leading to a body force density acting on all sites.

We use a hybrid lattice Boltzmann scheme \cite{Marenduzzo2007} that applies a finite-difference method for the dynamics of the $\bm{Q}$-tensor order parameter and solves the hydrodynamic part of the problem by means of the lattice Boltzmann method.
The colloidal particle is discretised as a solid, mobile particle with a radius $R=9.6$. \rev{The longitudinal and angular momenta of the colloidal particle are evolved according to Newton's second law of motion. We use a mixed explicit-implicit velocity update, which minimises the number of linear equations that must be solved, while maintaining absolute stability \cite{Nguyen2002}}.
On both the walls and the particle surface no-slip and no-penetration boundary conditions are applied by using a bounce-back on links scheme \cite{Ladd:1994a, Ladd:1994b, LBBook2017}. \rev{Lubrication corrections are applied normal to the walls within a distance of 0.1 lattice sites \cite{Nguyen2002}.} The surface free energy in Eq.~\ref{surface_fe} invokes a homeotropic anchoring condition with a preferred orientation of the director normal to the surfaces.
 
There are technical limitations to our model that should be borne in mind. While the centre of mass of the particle is integrated off-grid according to Newton's equation, the particle itself is discretised using a stair-case geometry. This can result in some inaccuracies, especially for highly confined regimes. For instance in our case $2R/L_x=0.8$ only about two lattice sites are between the particle and the walls surfaces at its narrowest point.
\rev{The ideal gas equation of state that the pressure obeys, as well as the modelling of the constant pressure gradient through an additional body force density on all sites are both common treatments in the lattice Boltzmann methodology and allow for an accurate modelling of a weakly compressible fluid.} But the assumption of a constant pressure gradient represents a simplification over the real situation. Thermal fluctuations have not been included since our simulations were carried out using a temperature well away from the isotropic-nematic transition line, and so elastic forces from the anchoring of the liquid crystal dominate over thermal forces by orders of magnitude. \rev{The one-elastic-constant approximation} is commonly used as first approach and does not compromise our results qualitatively. However, relaxing this approximation could lead to quantitative differences, and potentially also richer phenomenology.  
The Beris-Edwards model uses a simplified approach to viscosities compared to the Ericksen-Leslie theory, which has six viscosity coefficients $\alpha_1,\dots,\alpha_6$ \cite{deGennes} (only five are independent as the Parodi relation applies \cite{Parodi1970}). The viscosities in the Beris-Edwards model are implicitly given through the isotropic dynamic shear viscosity $\mu$, the rotational diffusion constant $\Gamma$, the flow alignment parameter $\xi$, and the scalar order parameter $S_0$. They can be directly related to the Ericksen-Leslie viscosities $\alpha_1,\dots,\alpha_6$ \cite{Marenduzzo2007}, but parameterise only a subset of possible values.

The simulations were run with our lattice Boltzmann code for complex fluids {\it Ludwig} version 0.15.0 \cite{Ludwig}. A typical simulation is first initialised with no applied pressure gradient for $5 \times 10^{4}$ iteration steps for each anchoring strength. After this initial equilibration phase, the simulations are restarted with various pressure gradients that are kept constant for $4 \times 10^{5}$ iteration steps. Typical runtimes are approximately 26 hours using a hybrid MPI/OpenMP parallelisation with 2 MPI tasks each running on 20 OpenMP threads. The overview of the simulation parameters is included in Table \ref{tab1}. For further information about the exact implementation used in this work, we guide the reader to the Ludwig code repository and related literature \cite{Desplat2001, Adhikari2005}. 

\begin{table}[htbp]
\normalsize
\centering
\begin{tabular}{|| l | c | c ||}
\hline
\hline
Bulk energy scale & $A_0$ & $0.01$\\
\hline
Effective temperature & $\tau$ & $-0.29$\\
\hline
Elastic constants & $\kappa_0, \kappa_1$ & $0.01$\\
\hline
Wall anchoring strength & $w_{wall}$ & $0.02$\\
\hline
Particle anchoring strength & $w_{part}$ & 0, 0.001, 0.01, 0.05\\
\hline
Anchoring parameter & $\omega$ & 0, 0.96, 9.6, 48\\ 
\hline
Flow alignment parameter & $\xi$ & $0.7$\\
\hline
Mobility parameter& $\Gamma$ & $0.5$\\
\hline
Density & $\rho$ & 1.0\\
\hline
Shear viscosity & $\mu$ & $5/6$\\
\hline
Bulk viscosity & $\zeta$ & $5/6$\\
\hline
Particle radius & $R$ &  $9.6$\\
\hline
\hline
\end{tabular}
\caption{Overview of simulation parameters}
\label{tab1}
\end{table}

\subsection{Parameter Mapping}

Our simulation units can be mapped to physical units by calibrating the units of pressure, time, and length. To achieve this, we assign the lattice spacing $\Delta x$, the algorithmic time step $\Delta t$ and the reference pressure $p^*$ from unity in lattice Boltzmann units (LBU) to their corresponding values in SI units. The principle of this parameter mapping was also shown in our previous work \cite{Lesniewska2022} using a different characteristic length scale.

The calibration of the length scale is straightforward as it is simply set by considering the dimensions of the microfluidic duct. If we associate the narrowest gap size $L_x=24$ in LBU corresponds to $L_x \,\widehat{=}\, 1.2\times 10^{-6}$ m in SI units, we obtain an LBU length of $\Delta x \,\widehat{=}\, 5\times 10^{-8}$ m $= 50$ nm in SI units.

To obtain the pressure scale, we use the measurements of the Landau-de Gennes parameters \cite{WrightMermin:1989} (Appendix D therein) which give 
\begin{eqnarray*}
\frac{27}{2\,A_0\,\gamma}&\simeq& 5\times 10^{-6} \mbox{ J}^{-1} \mbox{m}^3=5\times 10^{-6} \mbox{ Pa}^{-1}.
\end{eqnarray*}
Using $A_0=0.01$ and $\gamma=3.1$ in our simulations results in a reference pressure of $p^* = 1$ LBU $\widehat{=}\, 10^8$ Pa in SI units.

For the calibration of the timescale we use the following formula, which relates the rotational viscosity $\gamma_1$ of the director to the equilibrium scalar order parameter $q$ and the order parameter mobility $\Gamma$:
\begin{equation*}
   \gamma_1 = \frac{2 q^2}{\Gamma}
\end{equation*}
We use $\Gamma= 0.5$ in LBU and bulk energy density parameters that lead to $q \approx 0.5$ since it is assumed that the system is well within the nematic phase. Therefore, the rotational viscosity $\gamma_1=1$ in LBU. Typical values for liquid crystals in SI units are $\gamma_1=0.1$ Pa \nolinebreak s \cite{deGennes}. Together with $1$ Pa equating to a pressure of $10^{-8}$ in LBU, we obtain for the algorithmic time step $\Delta t\,\widehat{=}\,10^{-9}$ s $= 1$ ns.

The Ericksen number characterises the ratio of viscous to elastic forces and is defined as
\begin{equation*}
\rev{\mbox{Er}} = \frac{\eta\,u\,\Lambda}{\kappa},
\end{equation*} 
where $u$ is a characteristic flow velocity, in our case the velocity at the centre of the duct $U_c$, $\eta$ is the dynamic viscosity, $\Lambda$ is a characteristic length scale which is set by the narrowest gap size $L_x$ (see Table \ref{tab2} for $\Lambda=2R$, which allows direct comparison with our previous work \cite{Lesniewska2022}), and $\kappa$ is the bulk elastic constant of the liquid crystal.

The dynamic viscosity $\eta$ is calculated as an apparent viscosity, defined as the ratio, $\eta = \mu \Phi_0 / \Phi$ of the volumetric flux $\Phi_0$ of a simple Newtonian fluid and the volumetric flux of the liquid crystalline system $\Phi$, through a plane perpendicular to the flow in the $z$-direction, namely  

\begin{equation}
\label{volumetric_flow_rate}
\Phi = \int_0^{L_x}\int_0^{L_y} u_z(x)\; dx \,dy, \\
\end{equation}
with the flow being driven through the pressure gradient $\Psi = \Delta p/L_z$ with $\Delta p$ being the pressure difference between inlet and outlet. The volumetric flow rate $\Phi_0$ of a Newtonian fluid with dynamic viscosity $\mu$ through a gap $L_x$ driven by a pressure gradient $\Psi$ in plane Poiseuille flow can be calculated as 
\begin{eqnarray}
\label{volumetric_flow_rate_0}
\Phi_0 &=& \int_0^{L_x} \int_0^{L_y} \frac{L_x^2}{2\mu}\Psi \left(\frac{x}{L_x} - \left(\frac{x}{L_x}\right)^2\right) dx \,dy\\
&=& \frac{L_x^3 \,L_y \,\Psi}{12\,\mu}.
\end{eqnarray}

\begin{figure}[htbp!]
\centering
\includegraphics[trim={1.5cm 2cm 2.5cm 3cm},clip,width=0.95\linewidth]{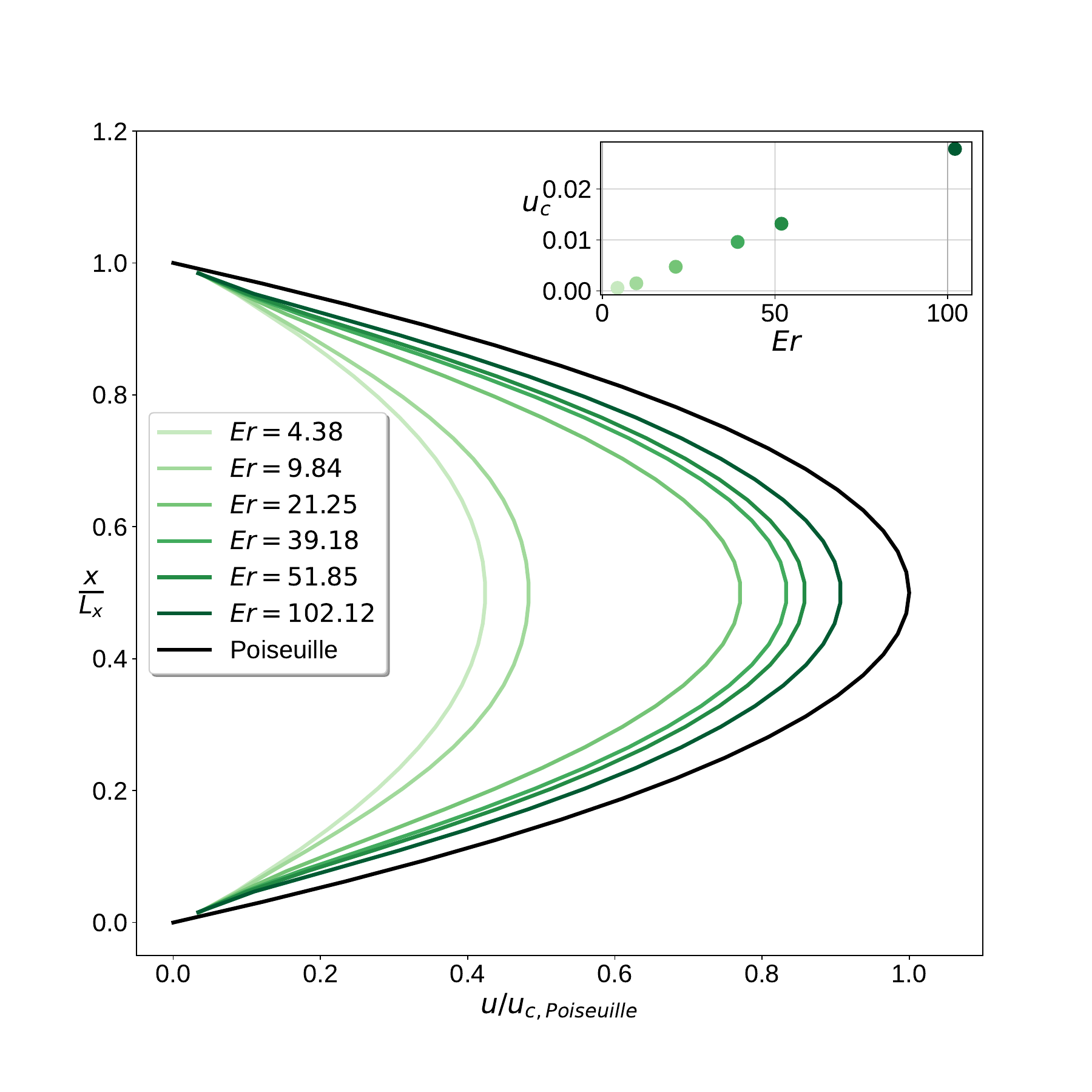}
\caption{\rev{Scaled magnitude of the fluid velocity $u(x)=|\bm{u}(x)|$ normalised against the peak flow velocity $u_{c, Poiseuille}$ of a simple Newtonian fluid in Poiseuille flow at the centre line of the duct at $x=L_x/2$. 
The image shows representative results for the confinement ratio $2R/L_x=0.6$.}
The black line is the \rev{parabolic flow profile of Poiseuille flow. Away from the walls the velocity profiles of the flowing nematic are parabolic and deviations from the parabolic profile occur only close to the walls.} The inset shows the dependence of the centre line fluid velocity $u_c$ on the Ericksen number \rev{Er}. Profiles for the other confinement ratios are not shown as they look very similar once normalised.}
\label{fig2}
\end{figure}

Fig.~\ref{fig2} shows \rev{fluid velocity profiles} for a \rev{representative} confinement ratio of $2 R/L_x= 0.6$ \rev{that have been normalised to the peak flow velocity of a simple Newtonian fluid in Poiseuille flow at the same pressure gradient and scaled using the $x$-dimension of the duct.}
The apparent viscosity $\eta$ is the ratio of the areas under the Poiseuille curve and the curves at finite Ericksen numbers.
\rev{More specifically, in Fig.~\ref{fig2} the flow velocities have been normalised against the maximum flow velocity of the Poiseuille flow $u_{c, Poiseuille}(x=L_x/2) = L_x^2 \Delta p/8\mu L_z$ at the centre line of the duct with $\mu$ as dynamic viscosity and $\Delta p$ as pressure difference between inlet and outlet, respectively. Away from the walls at $x/L_x=0$ and $x/L_x=1$ the velocity profiles of the flowing nematic are parabolic and deviations form the parabolic profile occur only close to the walls. This is a result of shear thinning as the director field flow-aligns further away from the walls, which is prevented by the normal wall anchoring in the vicinity of the walls.}

\rev{In all simulations that contain a colloidal particle the fluid flow velocity $u_c$ at the centre line was taken at $x=L_x/2$ and a point in a distance $L_z/2$ upstream/downstream from the particle, which is the point farthest away from the particle in the $z$-direction due to the periodic boundary conditions. However, owing to this large distance the values we obtained for $u_c$ in this manner are virtually identical to those of a pure liquid crystal without particle.}
Profiles for other confinement ratios are not shown as they look very similar.

\begin{table}[ht]
\begin{tabular}{|cc|cc|cc|}
\hline
\multicolumn{2}{|c|}{$2R/L_x= 0.4$}     & \multicolumn{2}{c|}{$2R/L_x= 0.6$}     & \multicolumn{2}{c|}{$2R/L_x= 0.8$}     \\ \hline
\multicolumn{1}{|c|}{Er (Lx)} & Er (2R) & \multicolumn{1}{c|}{Er (Lx)} & Er (2R) & \multicolumn{1}{c|}{Er (Lx)} & Er (2R) \\ \hline
\multicolumn{1}{|c|}{1.65}    & 0.658   & \multicolumn{1}{c|}{4.38}    & 2.63    & \multicolumn{1}{c|}{6.15}    & 4.92    \\ \hline
\multicolumn{1}{|c|}{8.30}    & 3.32    & \multicolumn{1}{c|}{9.84}    & 5.90    & \multicolumn{1}{c|}{10.37}   & 8.30    \\ \hline
\multicolumn{1}{|c|}{18.10}   & 7.24    & \multicolumn{1}{c|}{21.25}   & 12.75   & \multicolumn{1}{c|}{17.95}   & 14.36   \\ \hline
\multicolumn{1}{|c|}{35.55}   & 14.22   & \multicolumn{1}{c|}{39.18}   & 23.51   & \multicolumn{1}{c|}{22.32}   & 17.86   \\ \hline
\multicolumn{1}{|c|}{52.72}   & 21.09   & \multicolumn{1}{c|}{51.86}   & 31.11   & \multicolumn{1}{c|}{64.85}   & 52.16   \\ \hline
\multicolumn{1}{|c|}{69.77}   & 27.91   & \multicolumn{1}{c|}{102.12}  & 61.27   & \multicolumn{1}{c|}{86.06}   & 69.10   \\ \hline
\end{tabular}
\caption{Conversion of Ericksen numbers for different confinement ratios using different characteristic length scales, namely the size of the channel (odd columns) or as in Ref. \cite{Lesniewska2022} the diameter of the particle (even columns).}
\label{tab2}
\end{table}

\section{Results and Discussion}\label{sec4}

We study the \rev{advection} behaviour of a single particle moving in a nematic host phase in highly confining ducts and investigate the effect that varying pressure gradient $\Psi$, confinement ratio $2R/L_x$ and homeotropic anchoring strength have. In a simple Newtonian fluid, or in a liquid crystal at temperatures above the isotropic-nematic transition point, the motion of a freely suspended spherical particle between two parallel plane walls has been studied previously theoretically \cite{Ganatos1980, Staben2003}, with simulations \cite{Nikoubashman2013} and experimentally \cite{Staben2005}. The main effect is that the retardation of the particle motion to the fluid motion is primarily independent of the applied pressure gradient, but greater for particles closer to either of the walls, and \rev{therefore more so} for highly confined particles due to the proximity to the walls. 

In a nematic liquid crystal with homeotropic anchoring conditions at the walls the director orientation is forced to be parallel to the wall normals. The degree of alignment depends on the strength of the anchoring, but also on the velocity gradient, and therefore the pressure gradient. At low pressure gradients, the nematic order will be enforced throughout the duct. But for higher pressure gradients the director field flow-aligns at the Leslie angle. Two conformations are persistent in flowing nematics, namely the so-called bend state or H-state and the splay state or V-state. For both H- and V-state the director flow-aligns to a positive (negative) Leslie angle in the lower (upper) half of the channel. The difference between the two states is determined by the way the director rotates between the positive and negative Leslie angles at the centre: In the bend state, the director at the centre is perpendicular to the walls, whereas in the splay state the director is almost parallel to the walls at the centre. The bend state is generally adopted at low flow velocities, whereas the nematic transitions to the splay state at higher flow velocities.


\begin{figure}[htbp]
\centering
\begin{tabular}{ccc}
Er=4.38 \quad \includegraphics[trim={3.5cm 3.5cm 37.5cm 25.5cm},clip,scale=0.25]{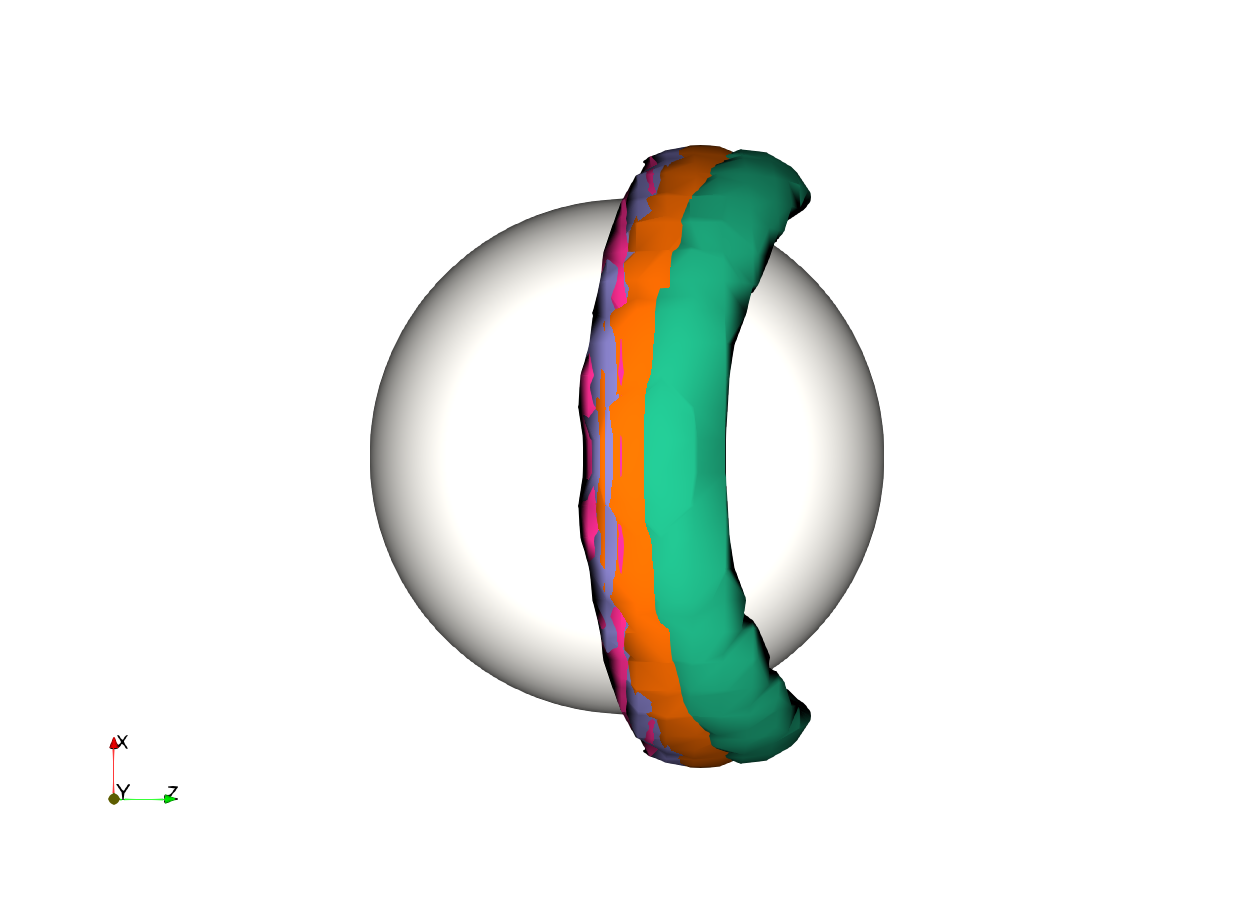} &
Er=51.86 \quad \includegraphics[trim={3.5cm 3.5cm 37.5cm 25.5cm},clip,scale=0.25]{xz_axis.png} &  \\
\includegraphics[width=0.42\linewidth]{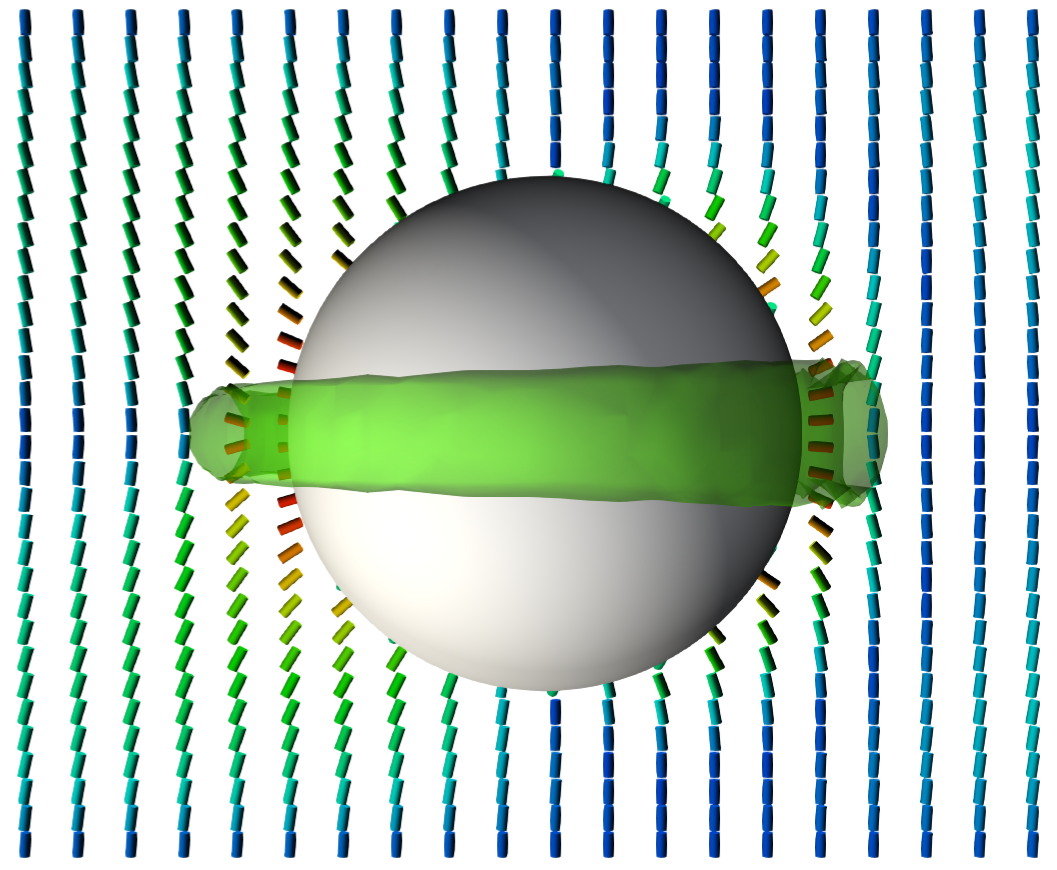}&
\includegraphics[width=0.42\linewidth]{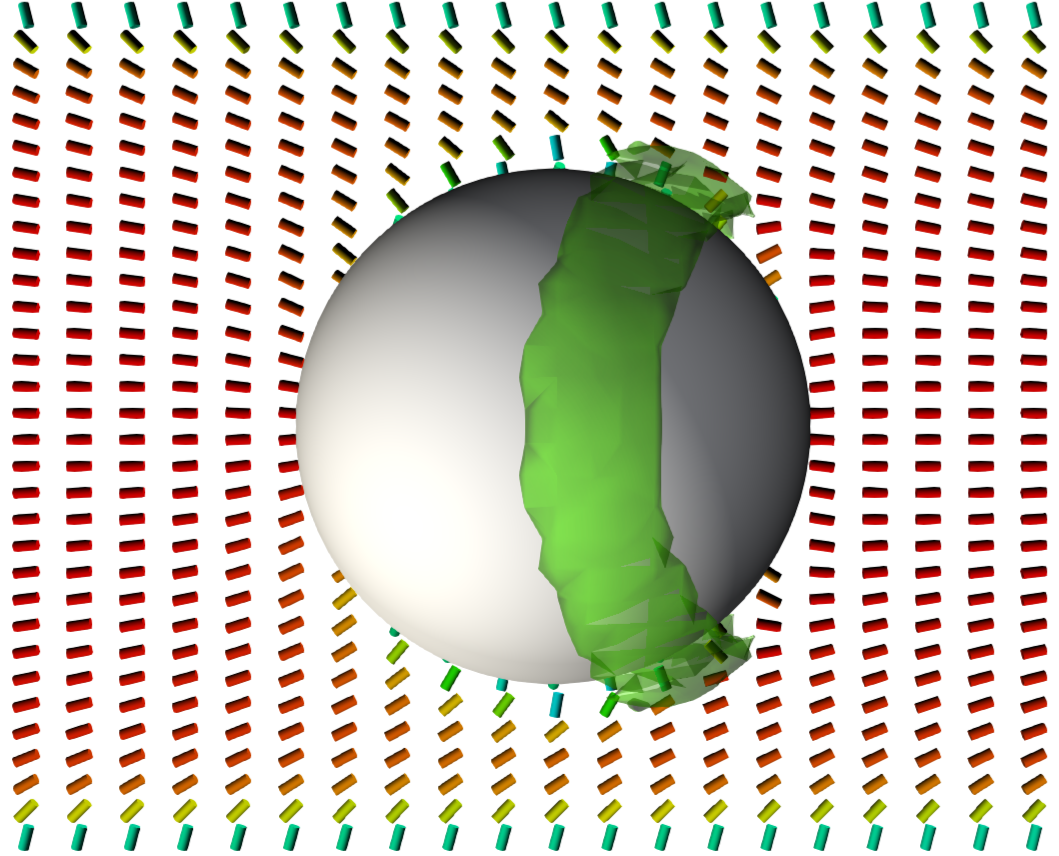}&
\hspace*{-1mm}\raisebox{1.5mm}{\includegraphics[width=0.1\linewidth]{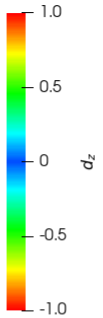}} \\
\includegraphics[width=0.42\linewidth]{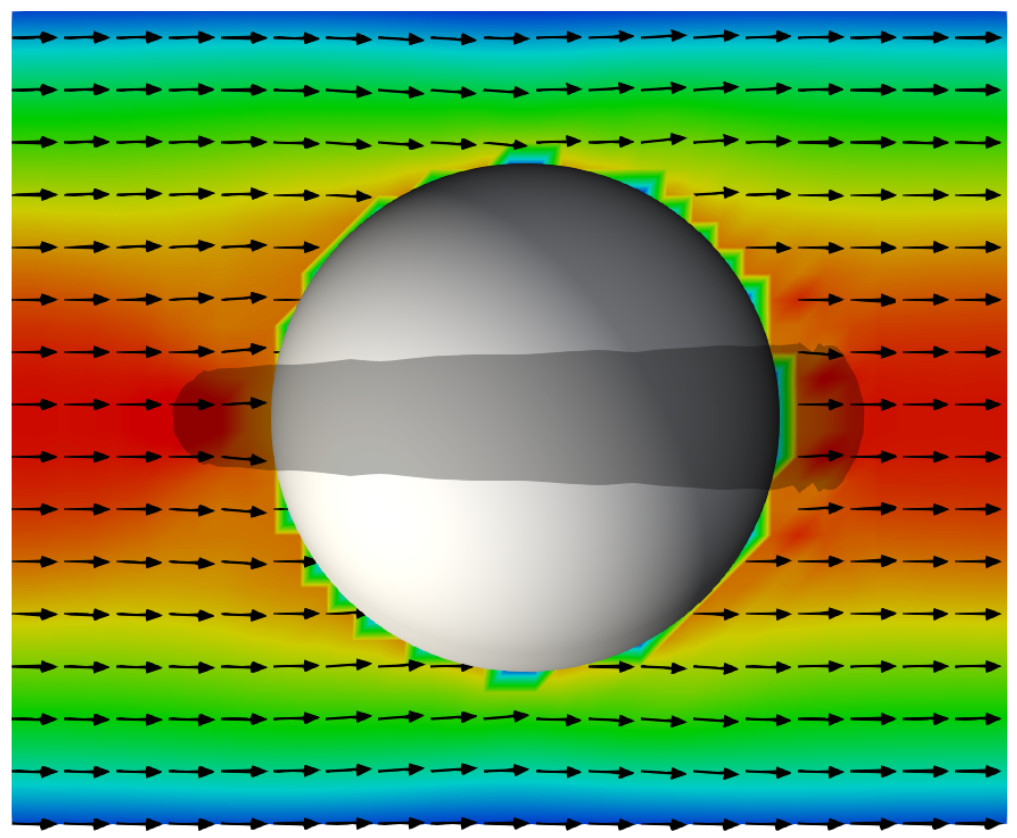}&
\includegraphics[width=0.42\linewidth]{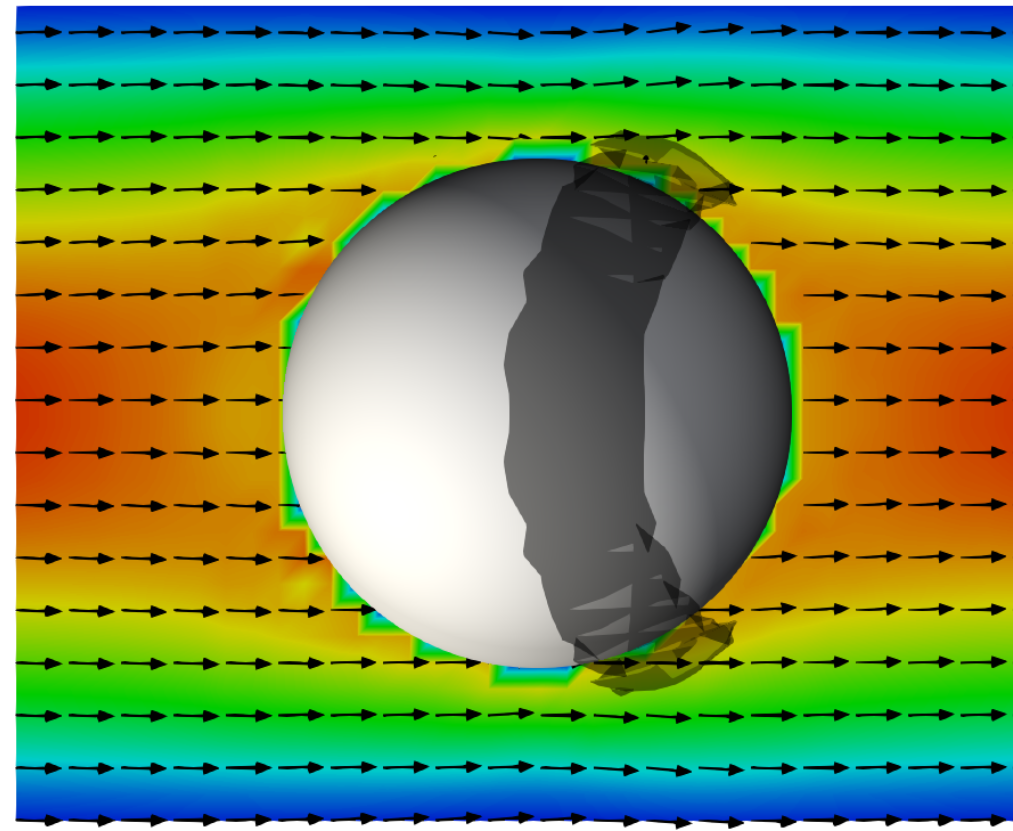}&
\raisebox{2.0mm}{\includegraphics[width=0.1\linewidth]{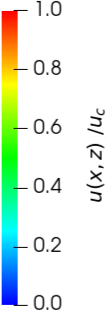}} \\
Er=4.38 \quad \includegraphics[trim={3.5cm 3.5cm 37.5cm 25.5cm},clip,scale=0.25]{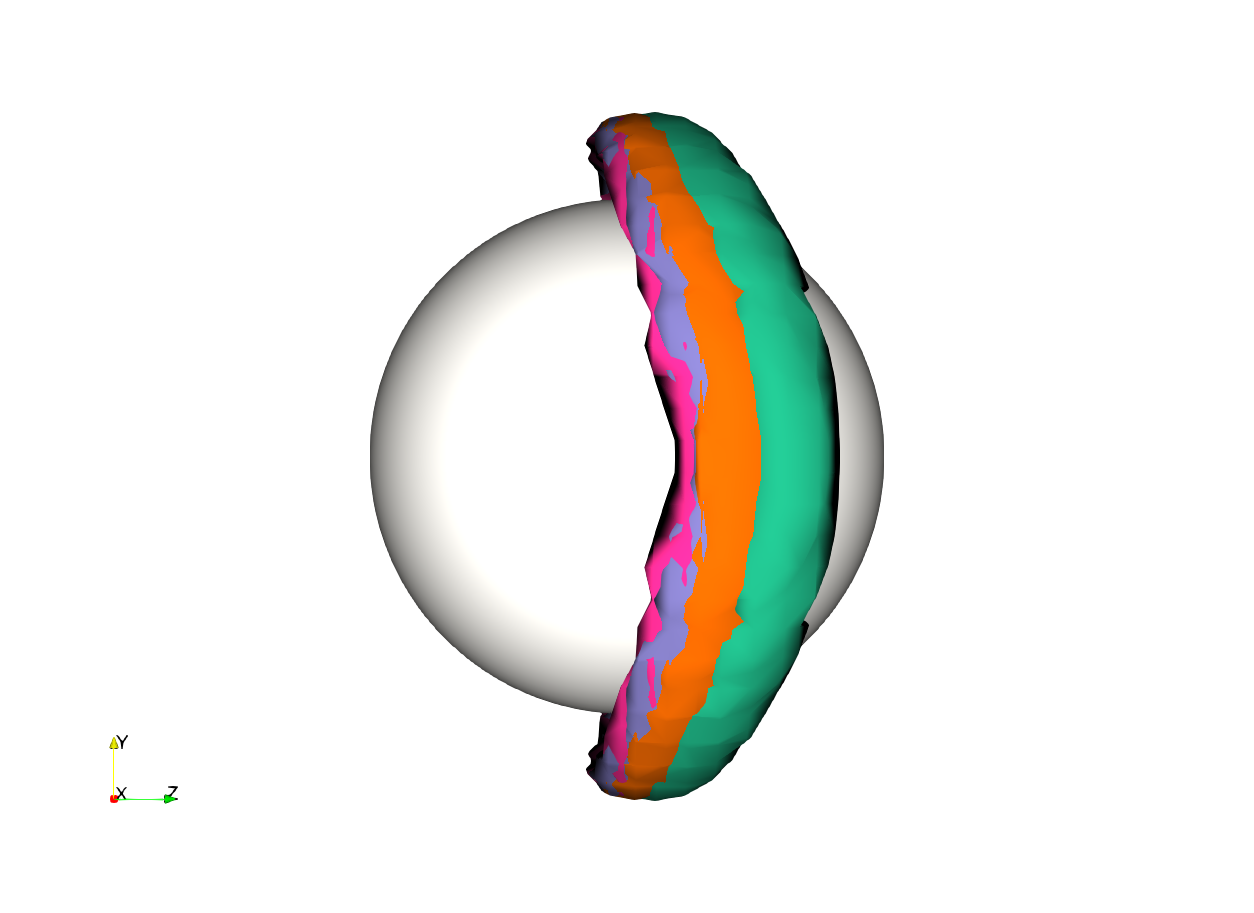}  &
Er=51.86 \quad \includegraphics[trim={3.5cm 3.5cm 37.5cm 25.5cm},clip,scale=0.25]{yz_axis.png} & \\
\includegraphics[width=0.42\linewidth]{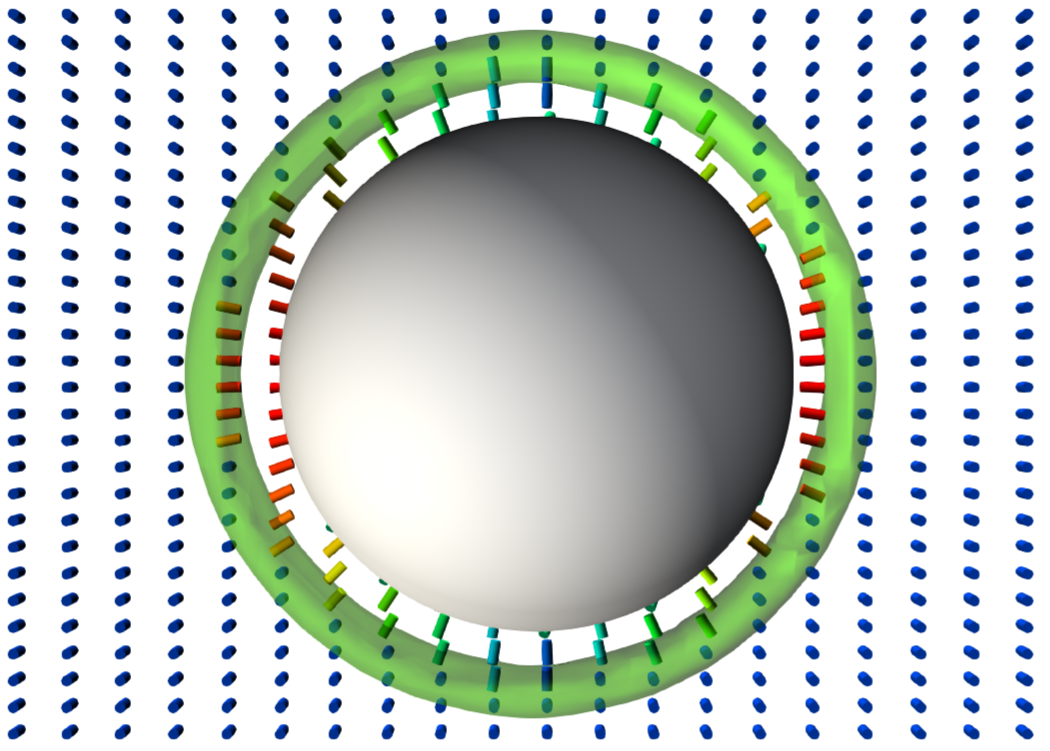}&
\includegraphics[width=0.42\linewidth]{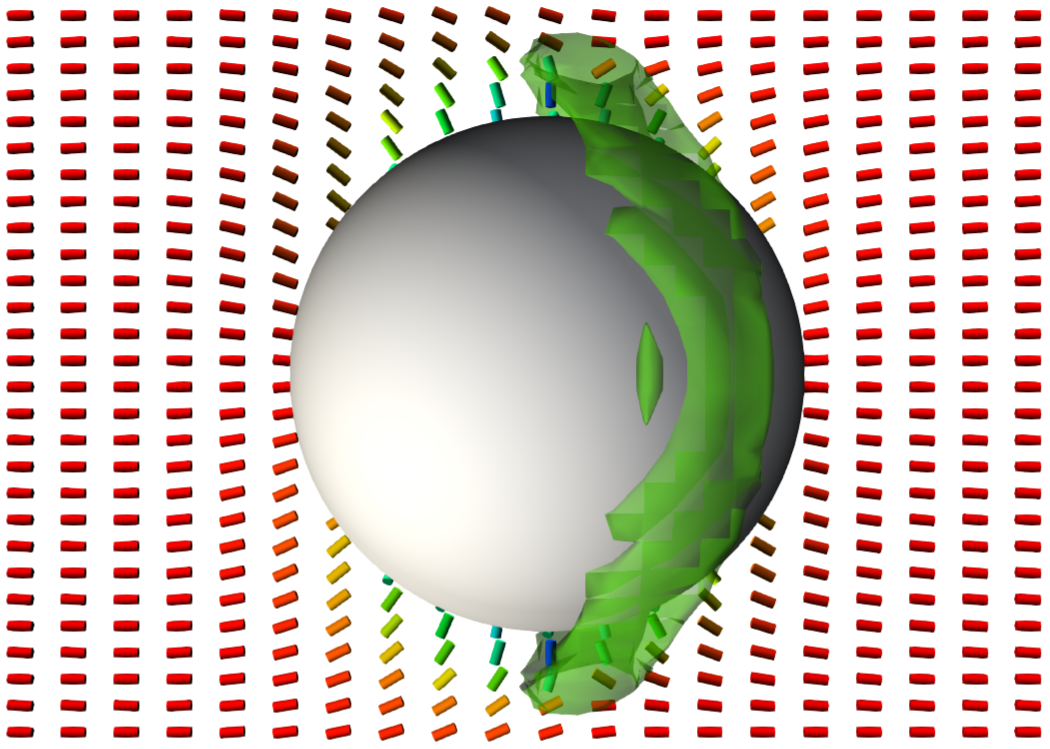}&
\hspace*{-1mm}\includegraphics[width=0.1\linewidth]{colorbar_dir.png} \\
\includegraphics[width=0.42\linewidth]{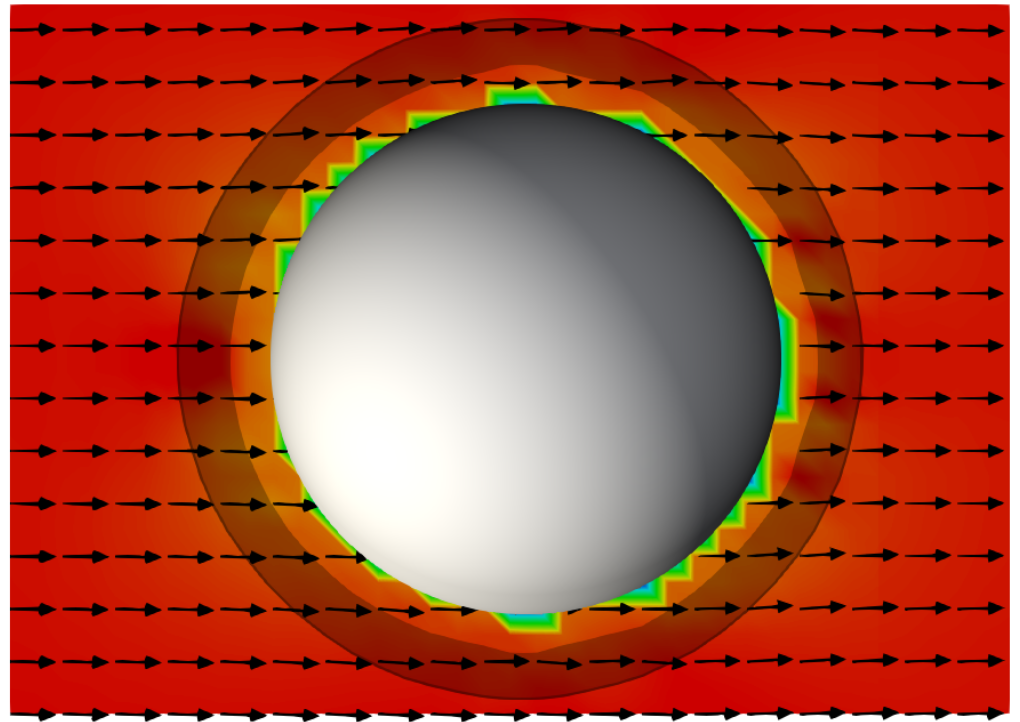}&
\includegraphics[width=0.42\linewidth]{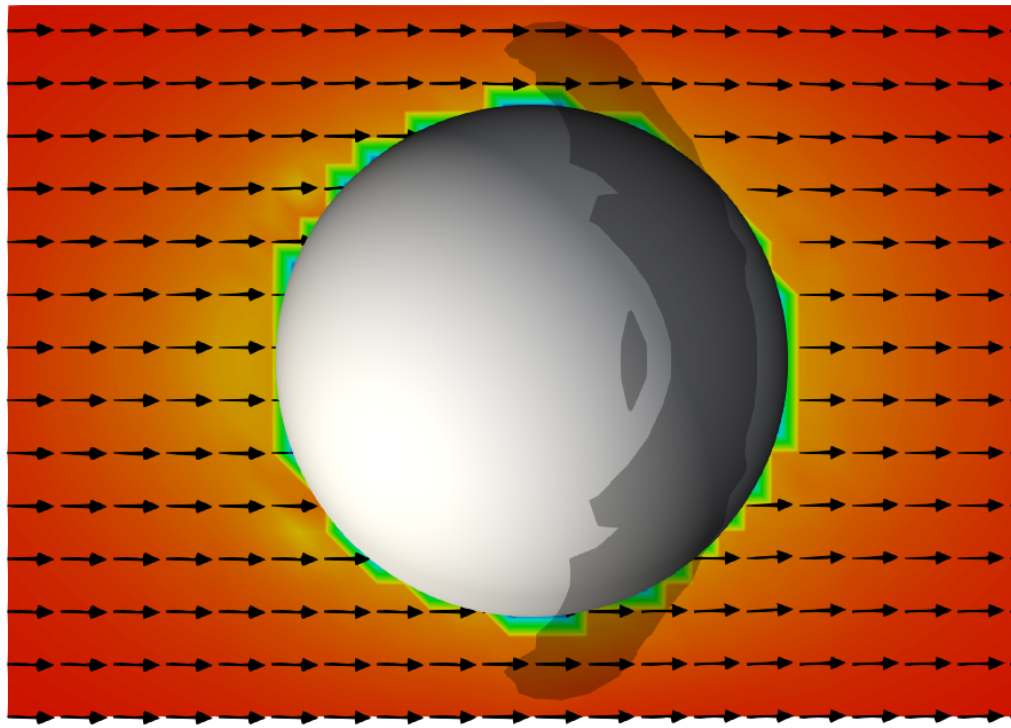}&
\raisebox{0.5mm}{\includegraphics[width=0.1\linewidth]{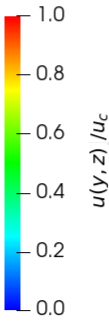}} \\
\end{tabular}
\caption{Director field, defect structure and \rev{fluid velocity profiles} for confinement ratio $2R/L_x= 0.6$ and anchoring parameter $\omega=48$ before and after the bend-to-splay transition. The \rev{left column} shows the bend state (H-state), while the \rev{right column} shows the splay state (V-state). 
\rev{The first and third row show the director field $\bm{d}$ with the magnitude $d_z$ of its $z$-component indicated through the colour code. The second and fourth row show the magnitude of the fluid velocity $u(x,z)$ and $u(y,z)$ through the centre of the particle, normalised to the maximum velocity $u_c$ at the centre line of the duct, where arrows give a sense of the vectorial dependence of the fluid velocity field.} 
The images in the \rev{two top rows} represent slices through the middle of the channel in the $xz$-plane (narrowest duct dimension and flow direction) and have the view along the negative $y$-dimension. Those in the \rev{two bottom rows} show slices in the $yz$-plane (widest and narrowest duct dimension) and have the view in positive $x$-direction. The flow direction is from left to right in positive $z$-direction. The opacity of the defect rings (green isosurfaces) has been slightly reduced to enhance the visibility of the local director field.}
\label{fig3}
\end{figure}

Fig.~\ref{fig3} shows the director field, defect structure \rev{and magnitude of the fluid velocity} at a medium confinement ratio $2R/L_x=0.6$ and different Ericksen numbers. The \rev{left column} displays the bend state at Er=$4.38$ prior to the transition to the splay state, whereas the \rev{right column} shows the splay state at Er=$51.86$ after transitioning from the bend state. The \rev{two top rows} contain slices in the $xz$-plane (narrowest duct dimension and flow direction) at $y=L_y/2$ with walls at the $x$-boundaries at the top and bottom, whereas the \rev{two bottom rows} show slices in the $yz$-plane (widest duct dimension and flow direction) at $x=L_x/2$ cropped to the vicinity of the colloidal particle.

\begin{figure*}[htbp]
\centering
\begin{tabular}{||ll||ll||ll||}
\hline
\multicolumn{2}{||c||}{ \includegraphics[trim={3.5cm 3.5cm 37.5cm 25.5cm},clip,scale=0.3]{xz_axis.png} \quad $2R/L_x= 0.4$ \quad  \includegraphics[trim={3.5cm 3.5cm 37.5cm 25.5cm},clip,scale=0.3]{yz_axis.png}   } &
\multicolumn{2}{c||}{ \includegraphics[trim={3.5cm 3.5cm 37.5cm 25.5cm},clip,scale=0.3]{xz_axis.png} \quad $2R/L_x= 0.6$ \quad  \includegraphics[trim={3.5cm 3.5cm 37.5cm 25.5cm},clip,scale=0.3]{yz_axis.png}   } &
\multicolumn{2}{c||}{ \includegraphics[trim={3.5cm 3.5cm 37.5cm 25.5cm},clip,scale=0.3]{xz_axis.png} \quad $2R/L_x= 0.8$ \quad  \includegraphics[trim={3.5cm 3.5cm 37.5cm 25.5cm},clip,scale=0.3]{yz_axis.png}   }
\\ \hline
\rev{Er}$=1.65 $  & &\rev{Er}$= 4.38$ && \rev{Er}$=6.15$ & \\
\includegraphics[trim={9cm 3cm 9cm 3cm},clip,width=0.13\linewidth]{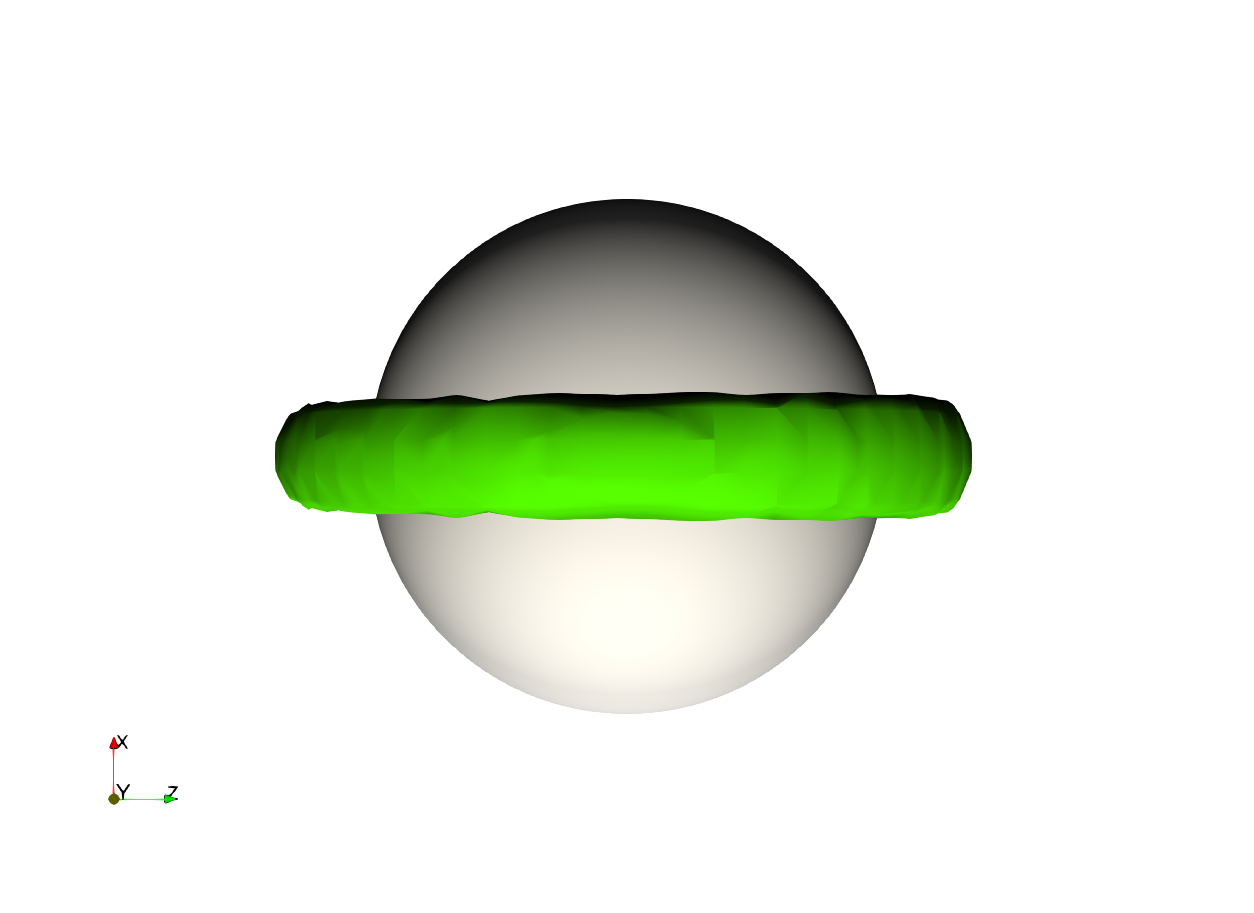} & 
\includegraphics[trim={9cm 3cm 9cm 3cm},clip,width=0.13\linewidth]{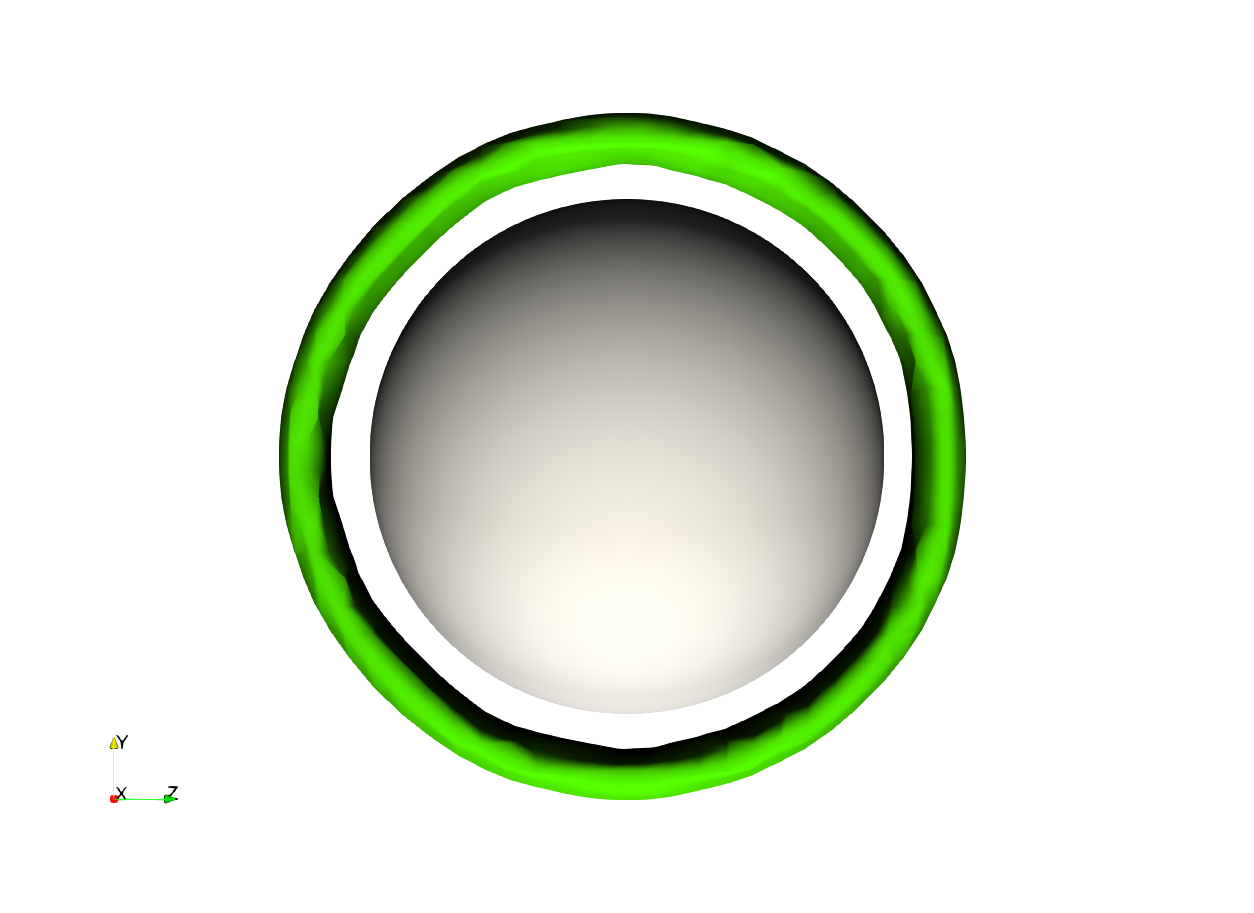}&
\includegraphics[trim={9cm 3cm 9cm 3cm},clip,width=0.13\linewidth]{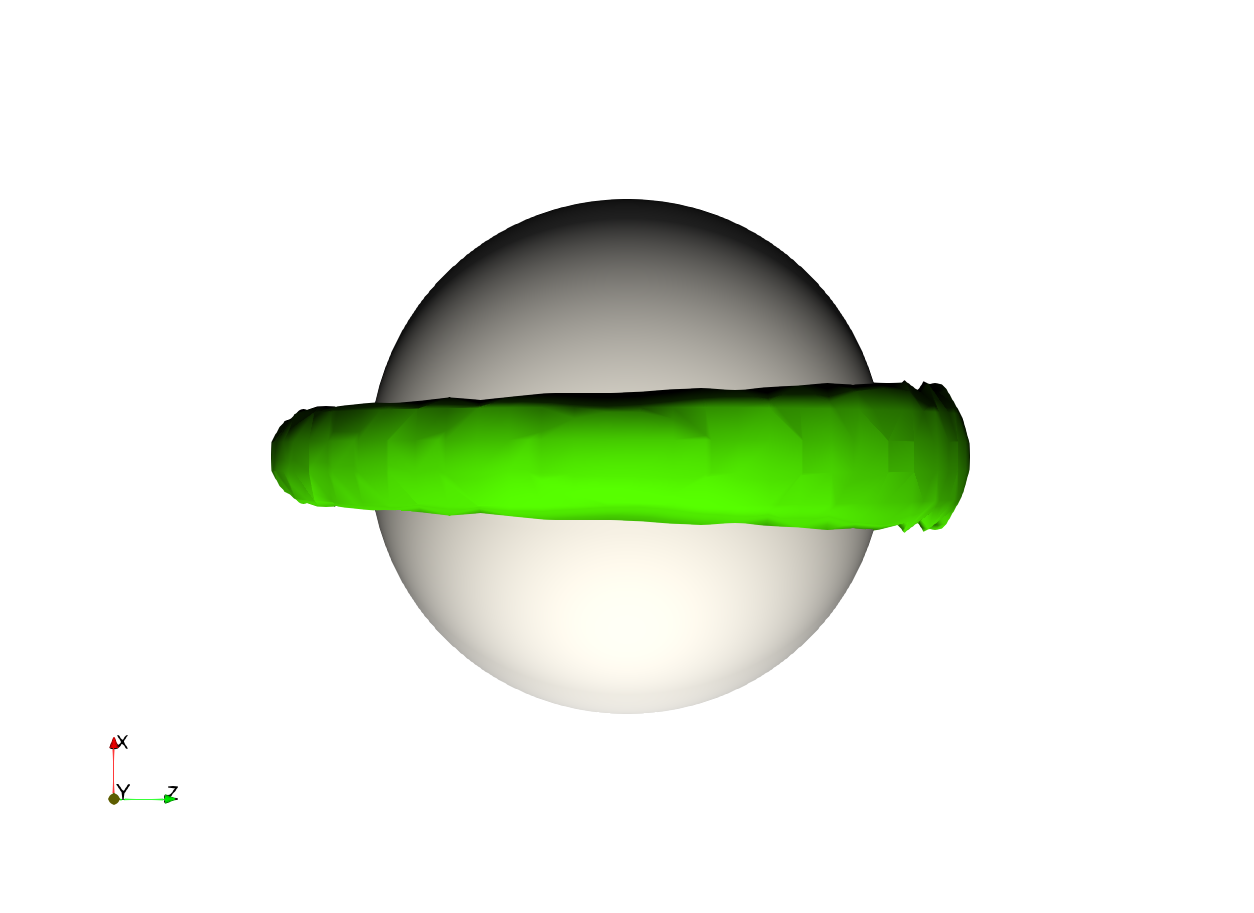} & 
\includegraphics[trim={9cm 3cm 9cm 3cm},clip,width=0.13\linewidth]{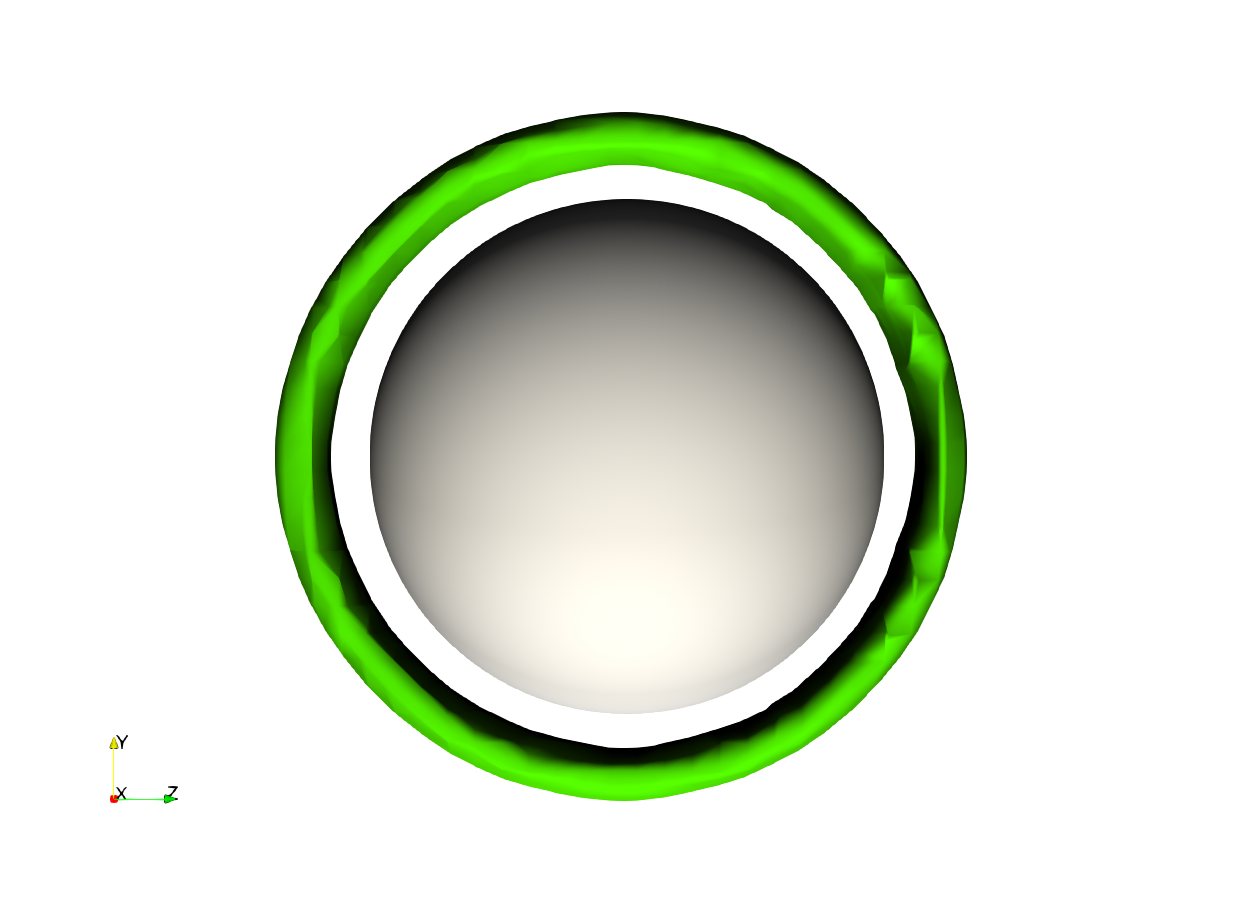} & 
\includegraphics[trim={9cm 3cm 9cm 3cm},clip,width=0.13\linewidth]{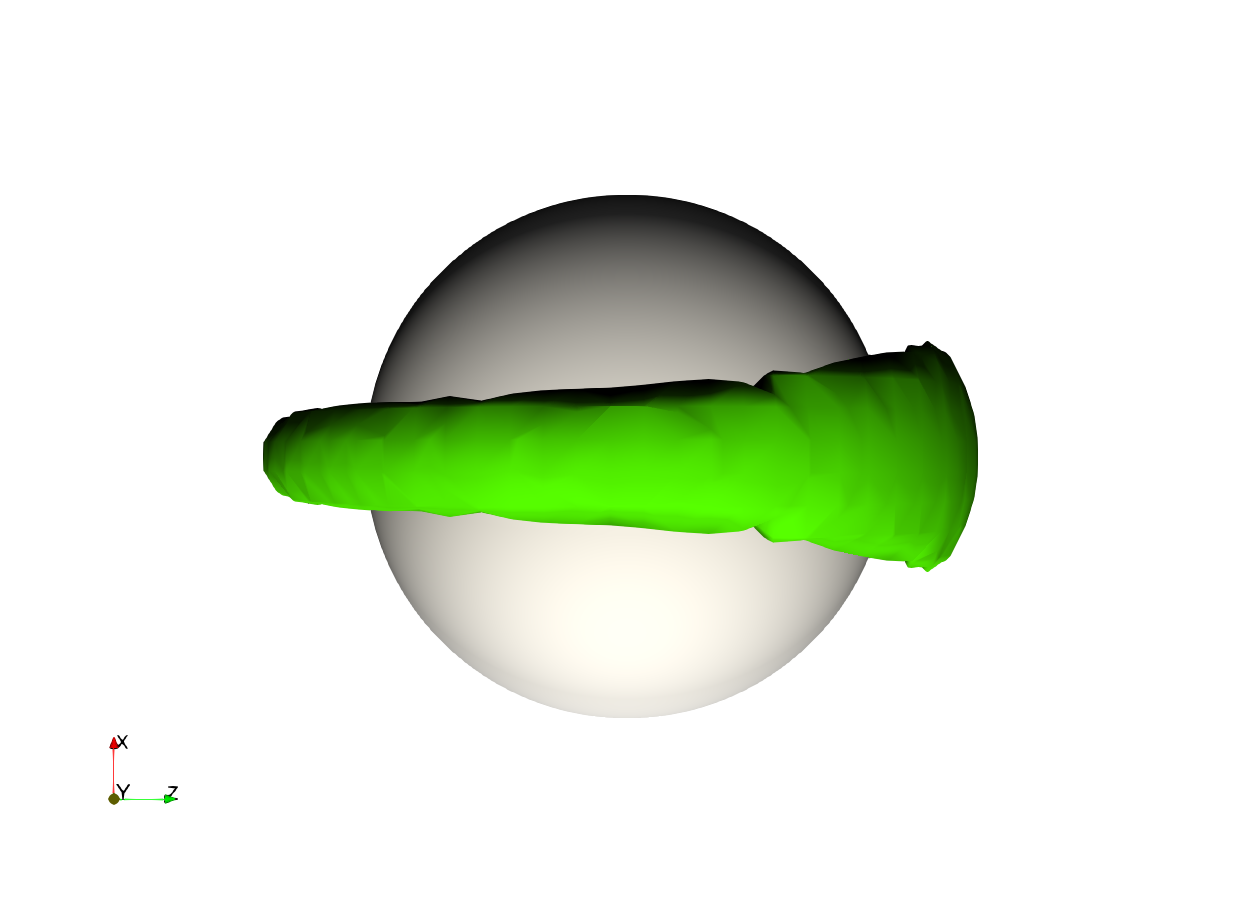} &
\includegraphics[trim={9cm 3cm 9cm 3cm},clip,width=0.13\linewidth]{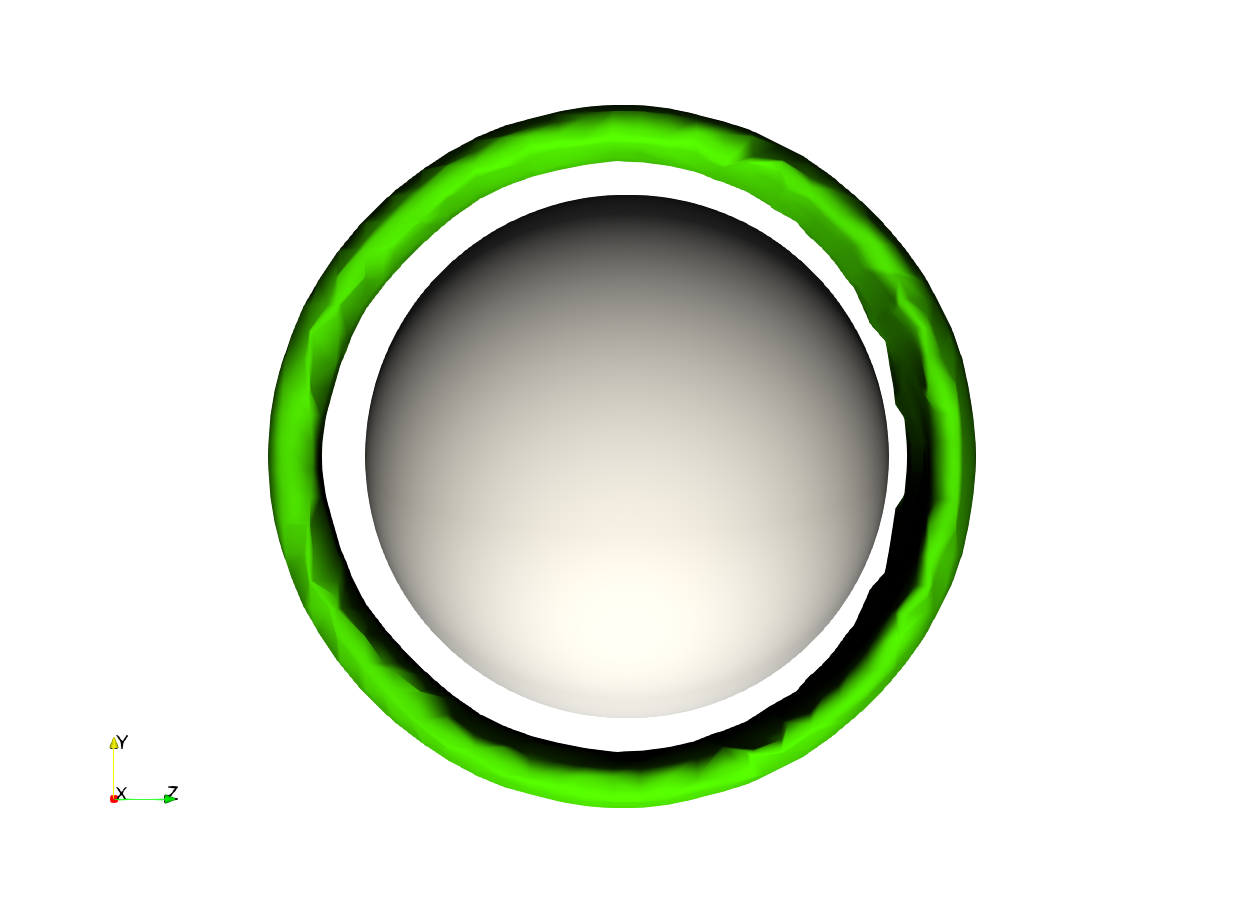}\\ \hline
\rev{Er}$=8.30 $  && \rev{Er}$= 9.84$& & \rev{Er}$=10.37$ &  \\
\includegraphics[trim={9cm 3cm 9cm 3cm},clip,width=0.13\linewidth]{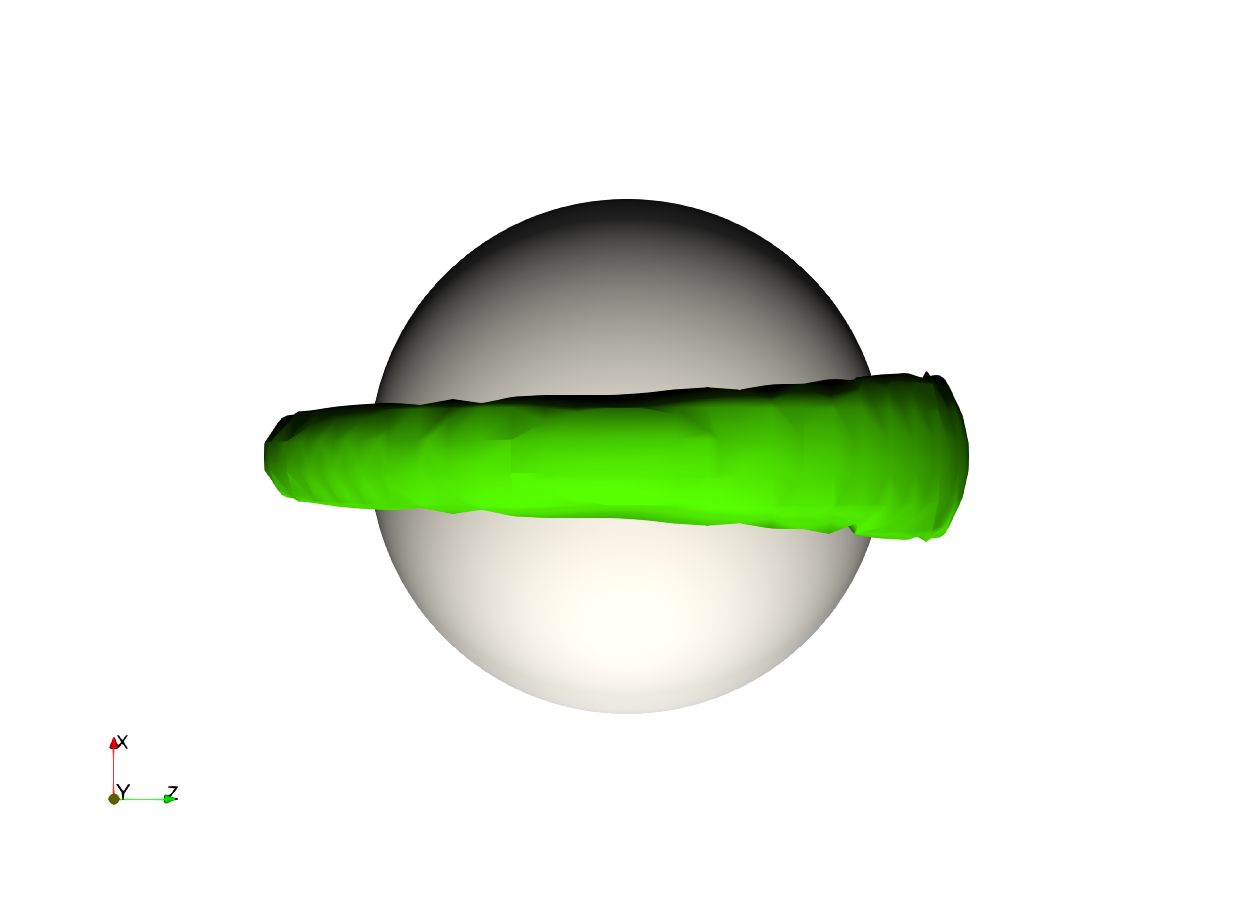} & 
\includegraphics[trim={9cm 3cm 9cm 3cm},clip,width=0.13\linewidth]{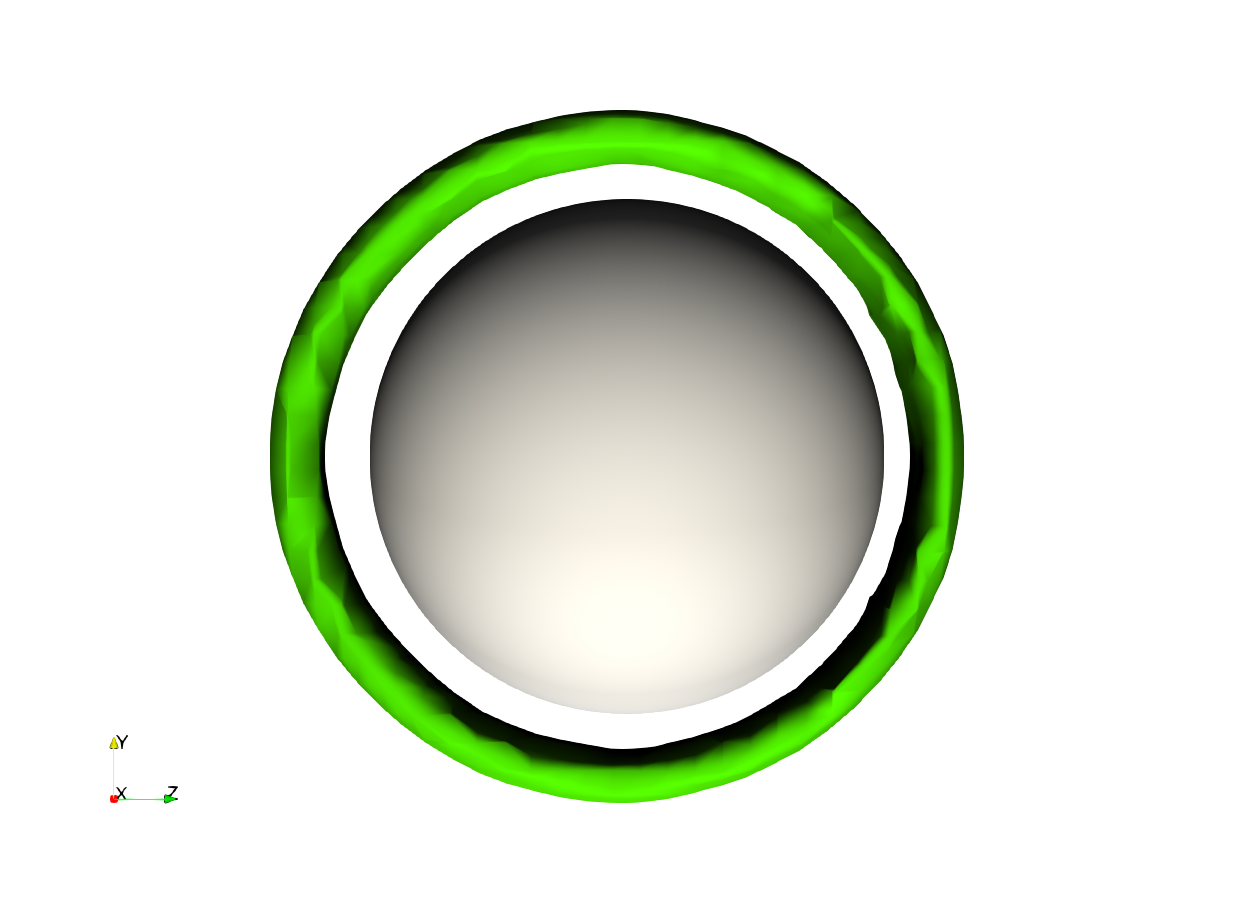} & 
\includegraphics[trim={11cm 5cm 12cm 5cm},clip,width=0.13\linewidth]{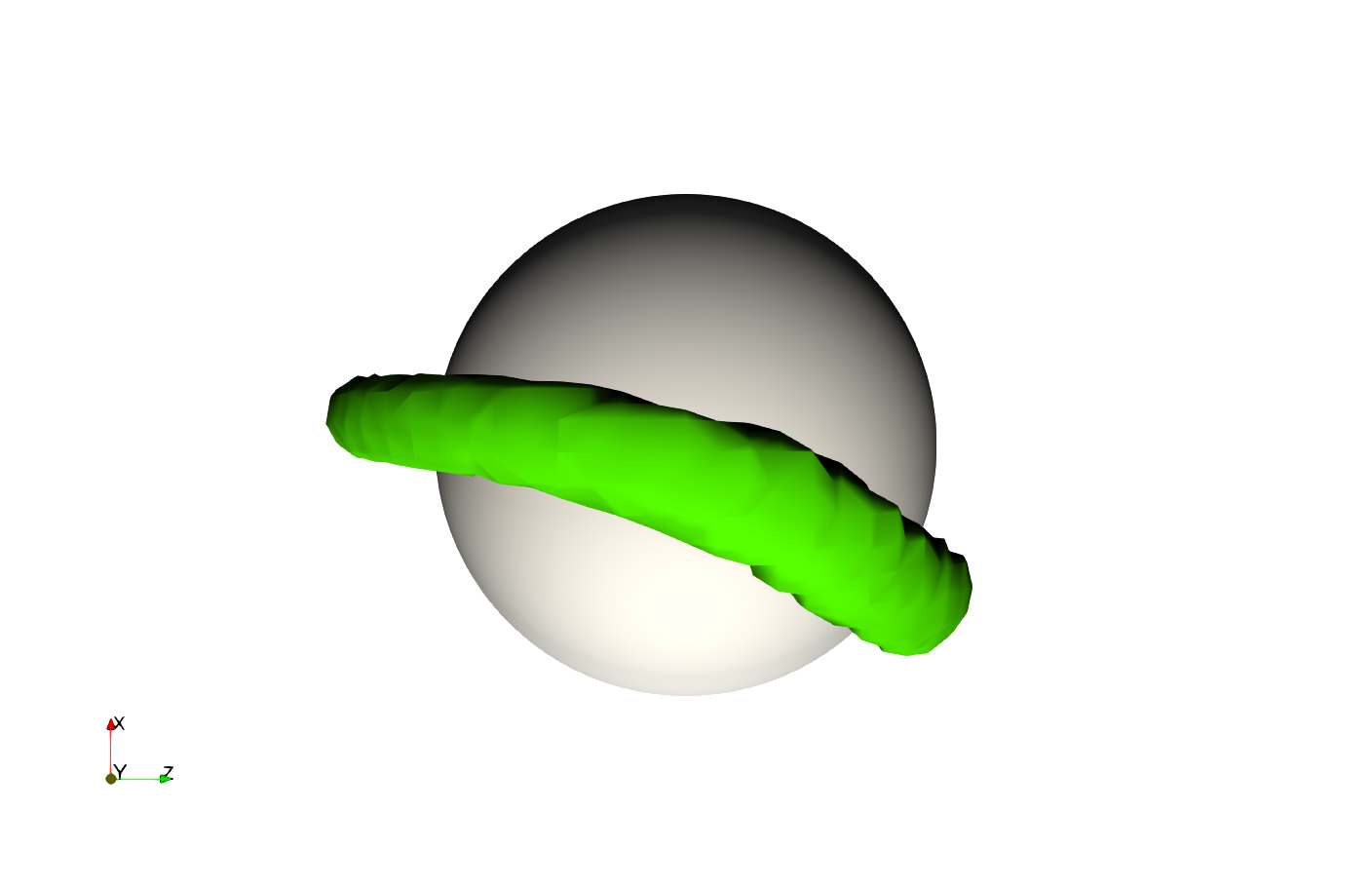} &  
\includegraphics[trim={10cm 3cm 11cm 3cm},clip,width=0.13\linewidth]{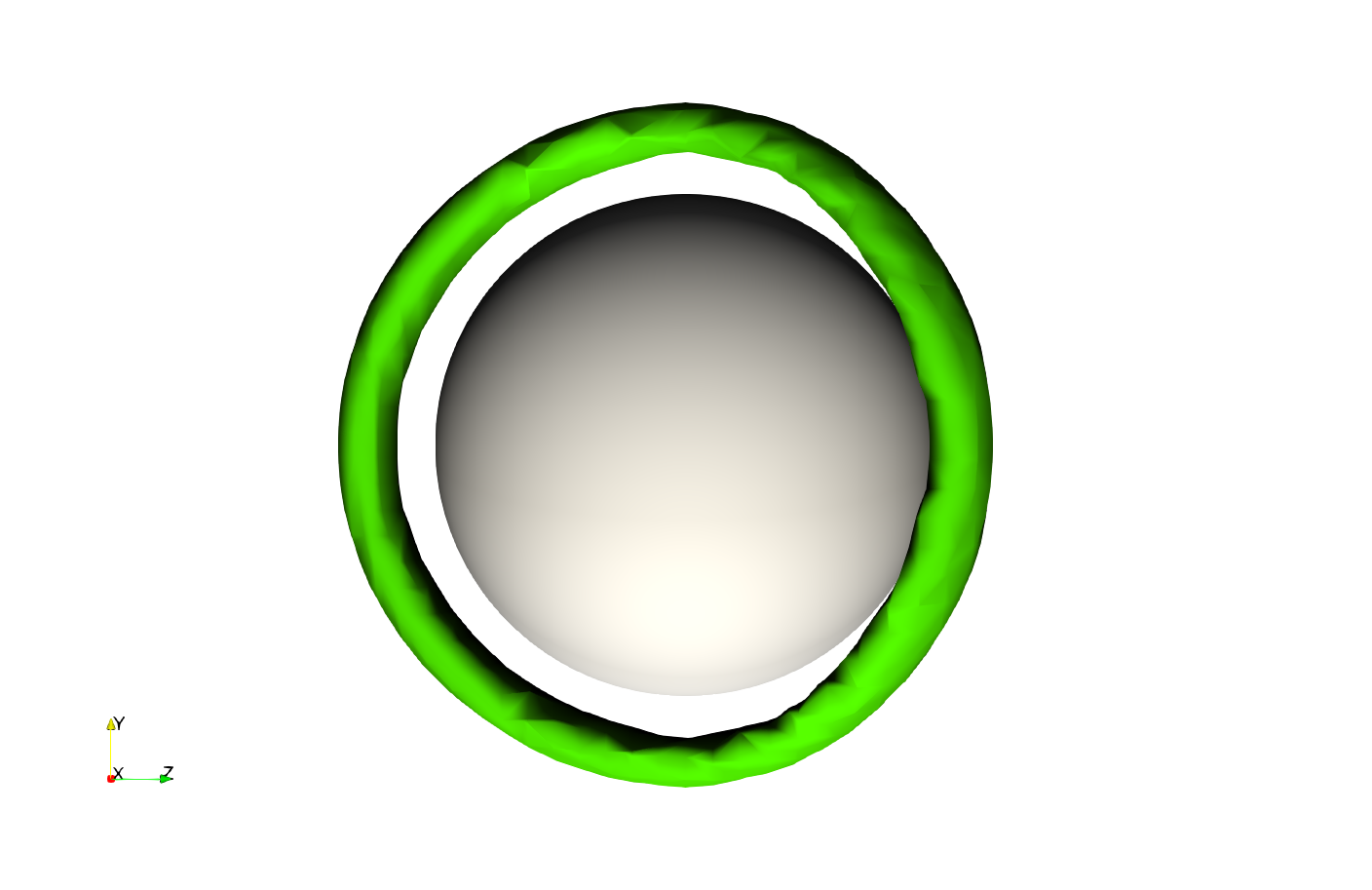}& 
\includegraphics[trim={11cm 5cm 12cm 5cm},clip,width=0.13\linewidth]{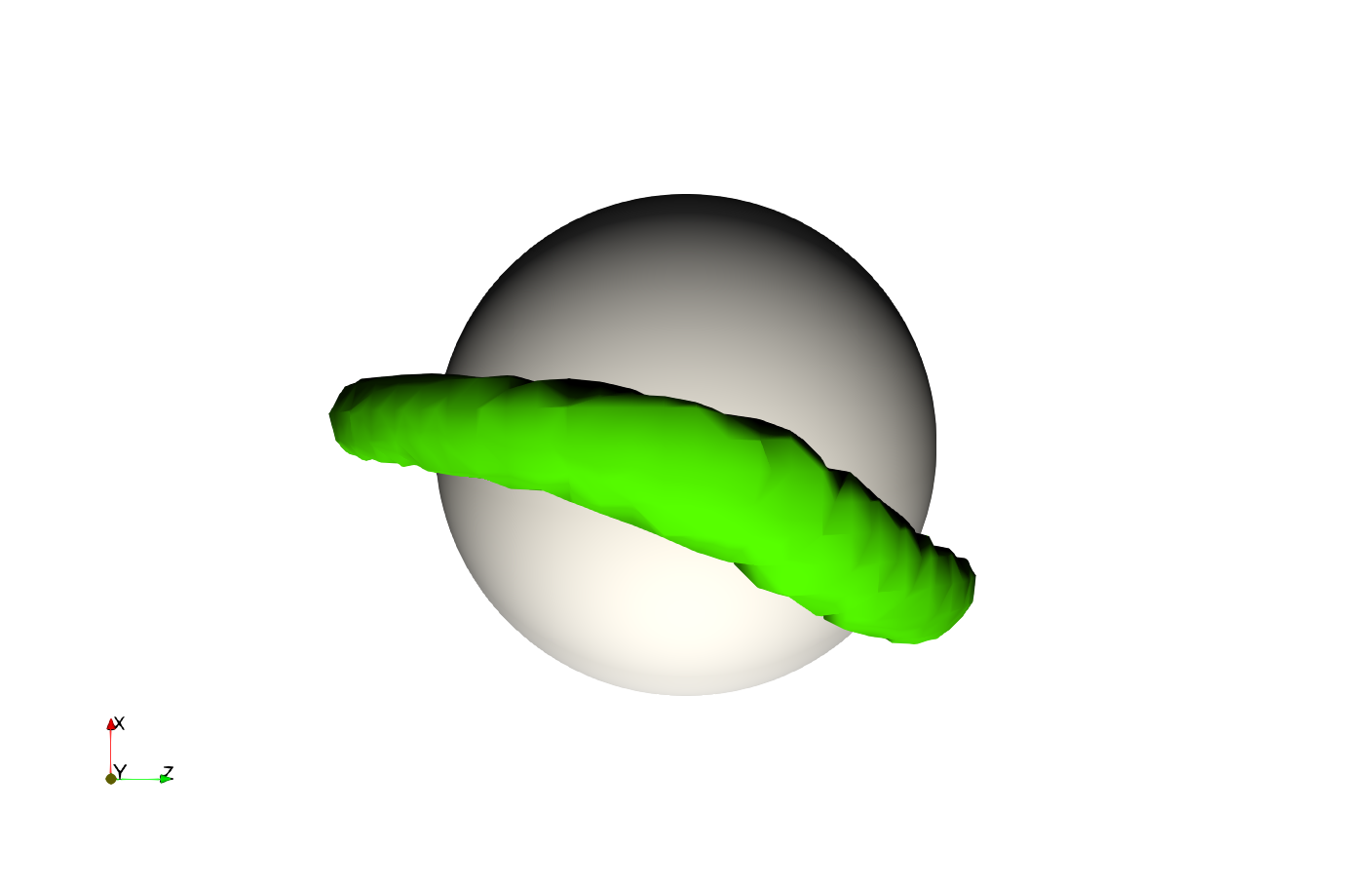} & 
\includegraphics[trim={11cm 3cm 11cm 3cm},clip,width=0.13\linewidth]{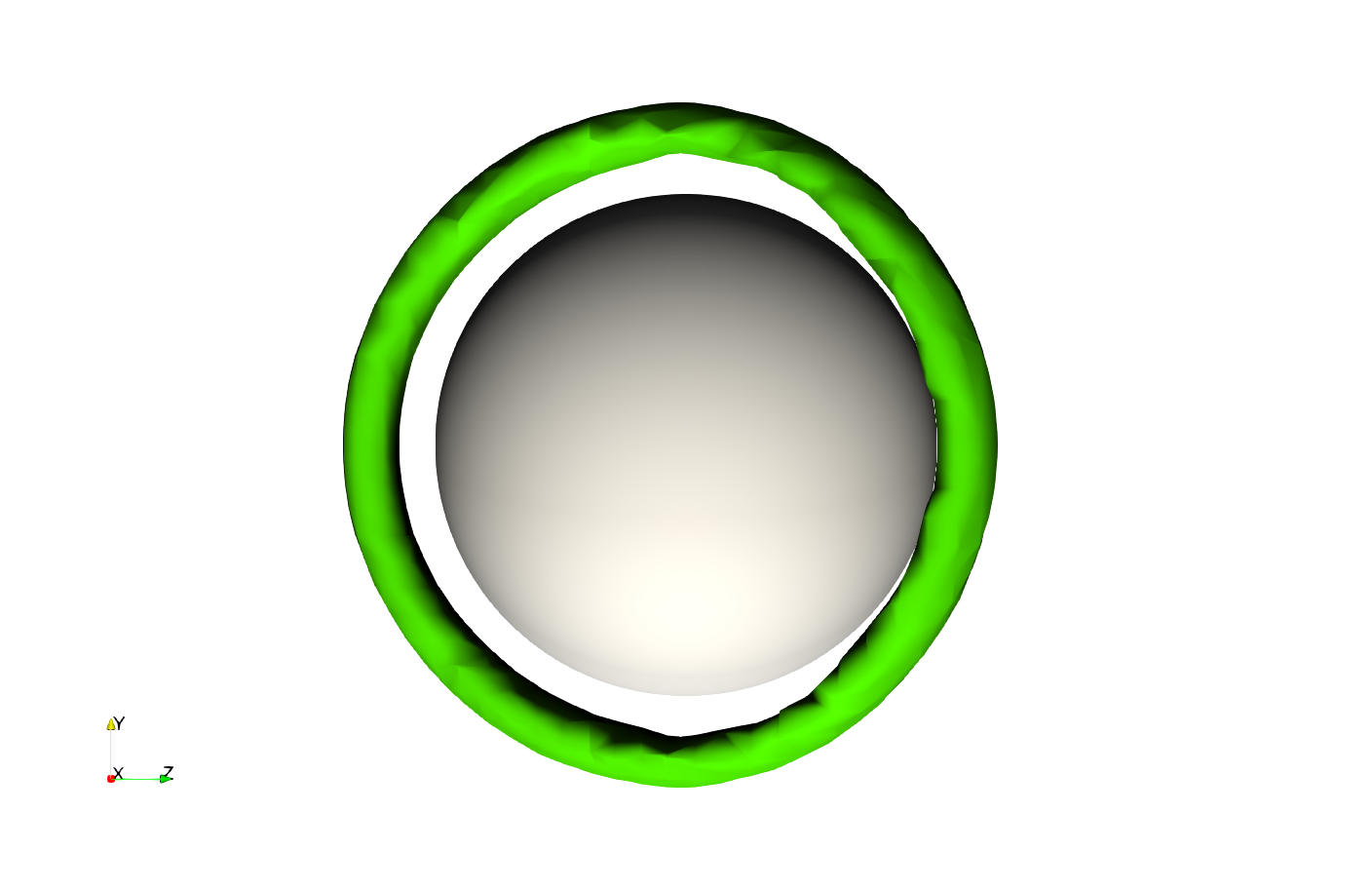}\\  \hline
\rev{Er}$=18.10 $ & & \rev{Er}$= 21.25$ && \rev{Er}$=22.32$  & \\
\includegraphics[trim={12cm 3cm 13cm 3cm},clip,width=0.13\linewidth]{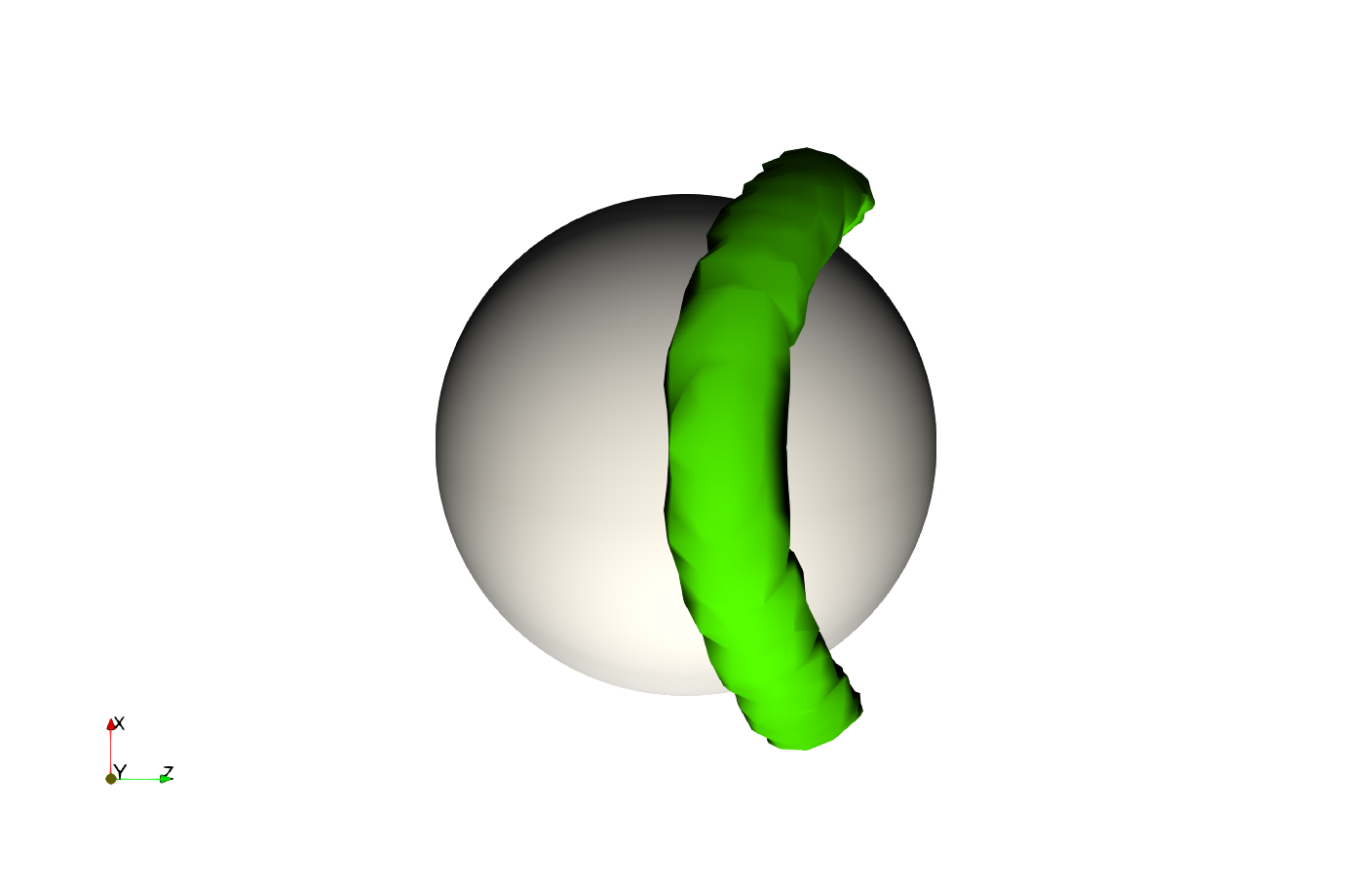} &
\includegraphics[trim={12cm 3cm 12cm 3cm},clip,width=0.13\linewidth]{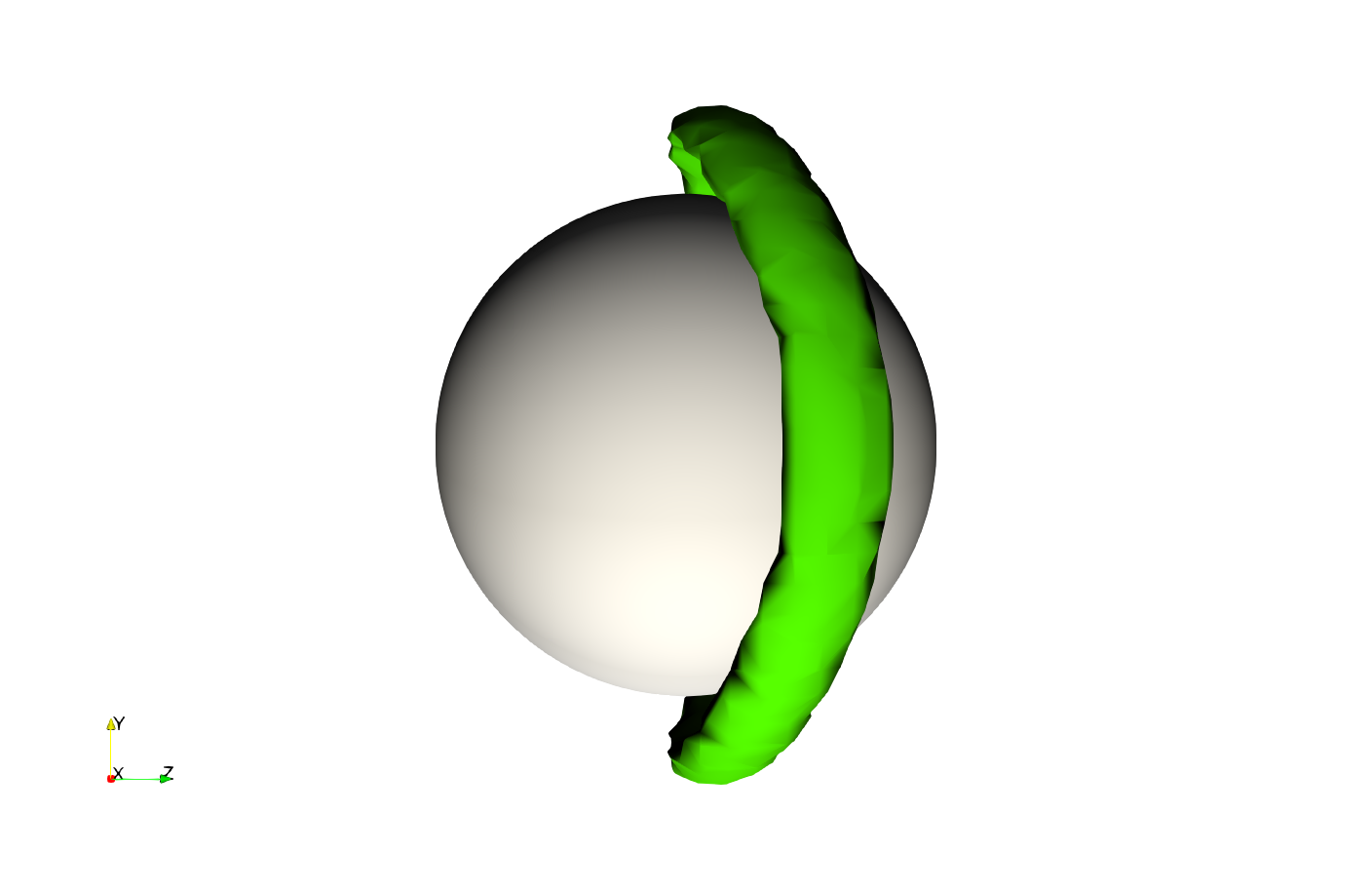} &
\includegraphics[trim={9cm 3cm 9cm 3cm},clip,width=0.13\linewidth]{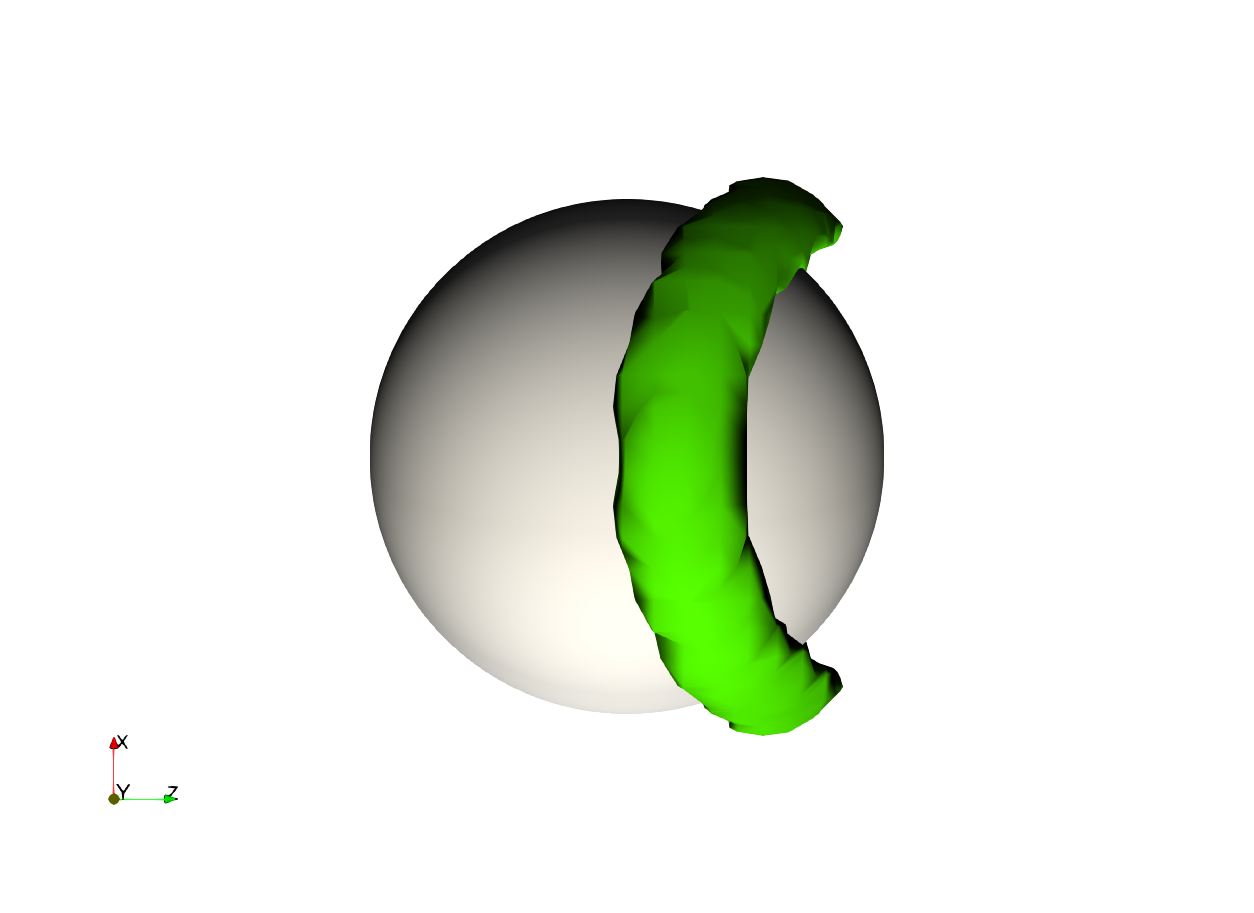} &  
\includegraphics[trim={9cm 3cm 9cm 3cm},clip,width=0.13\linewidth]{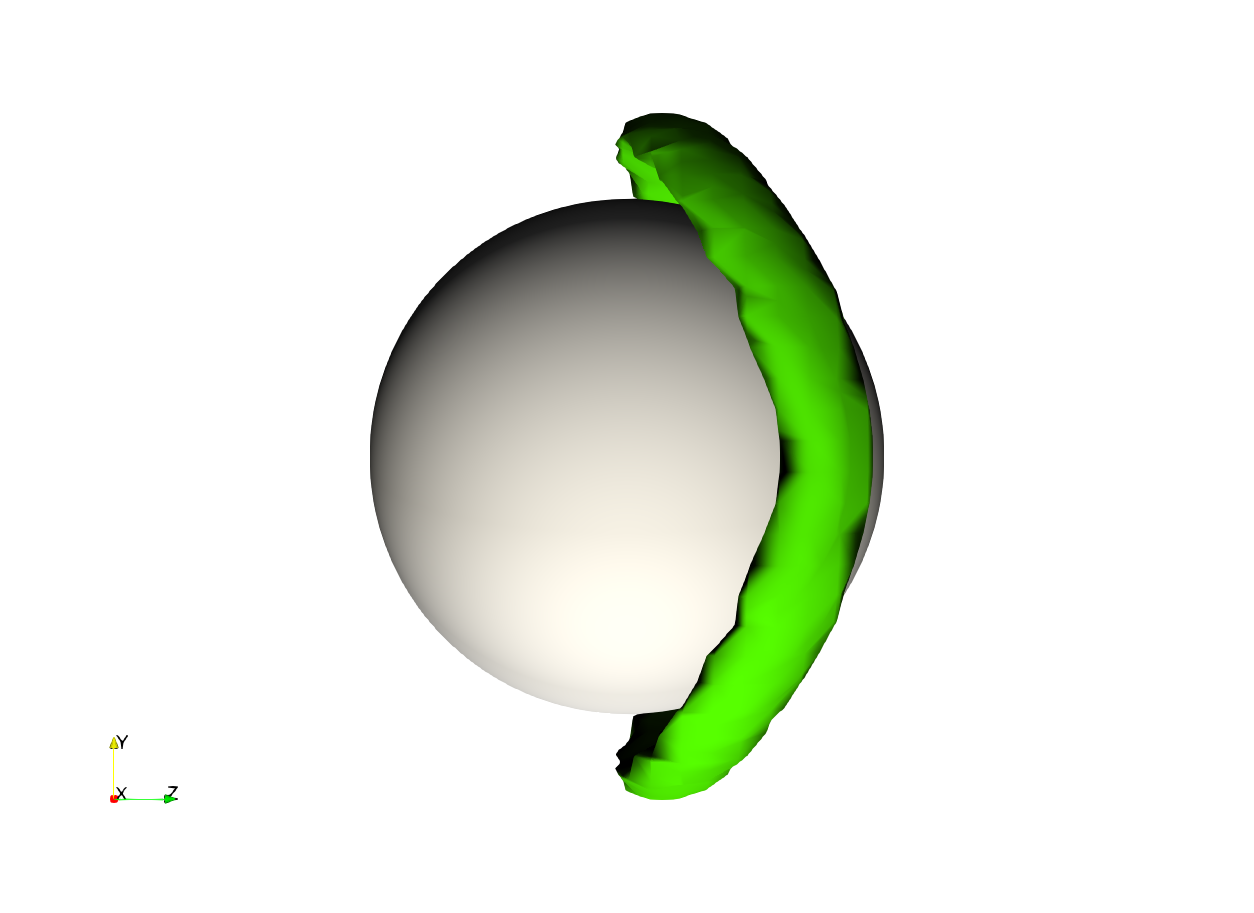} & 
\includegraphics[trim={9cm 3cm 9cm 3cm},clip,width=0.13\linewidth]{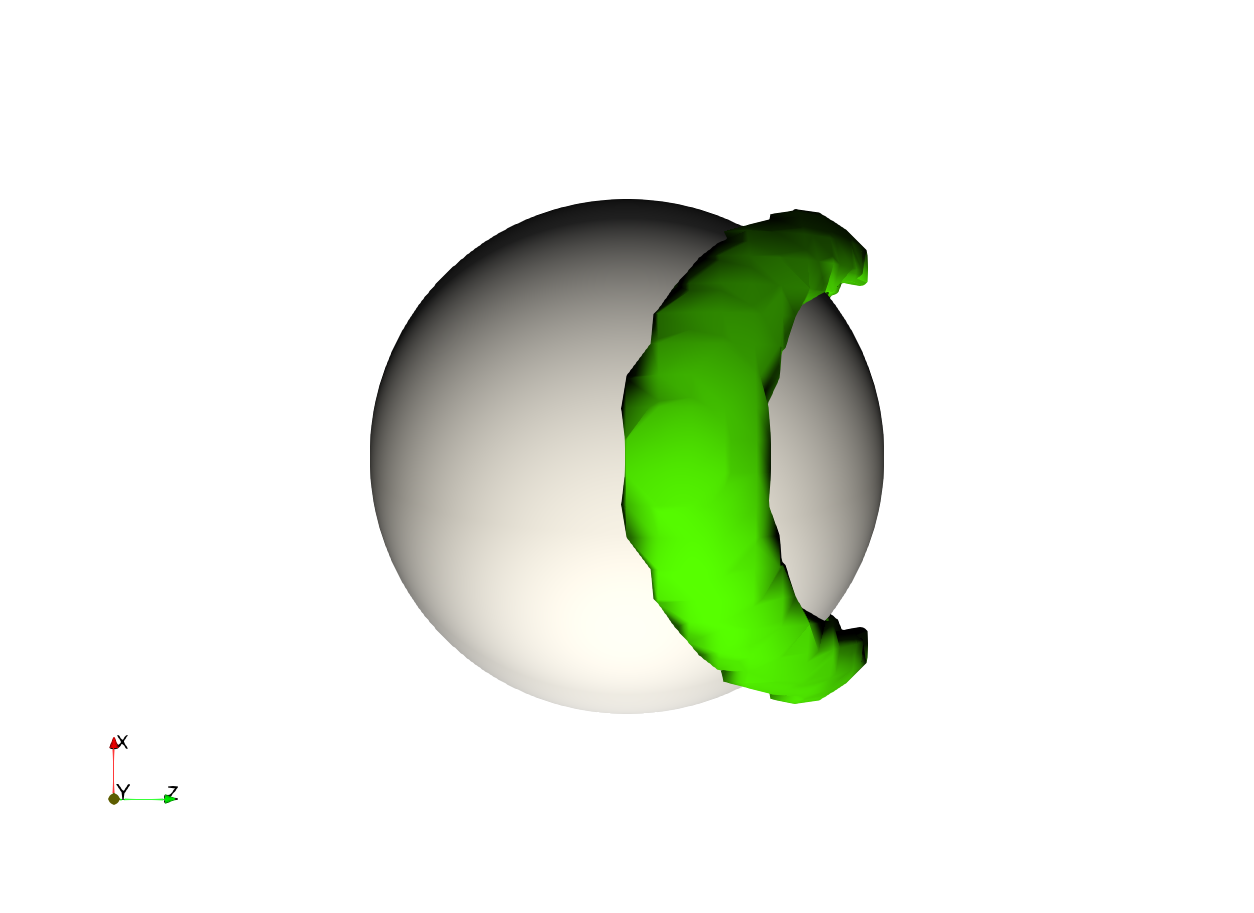} & 
\includegraphics[trim={9cm 3cm 9cm 3cm},clip,width=0.13\linewidth]{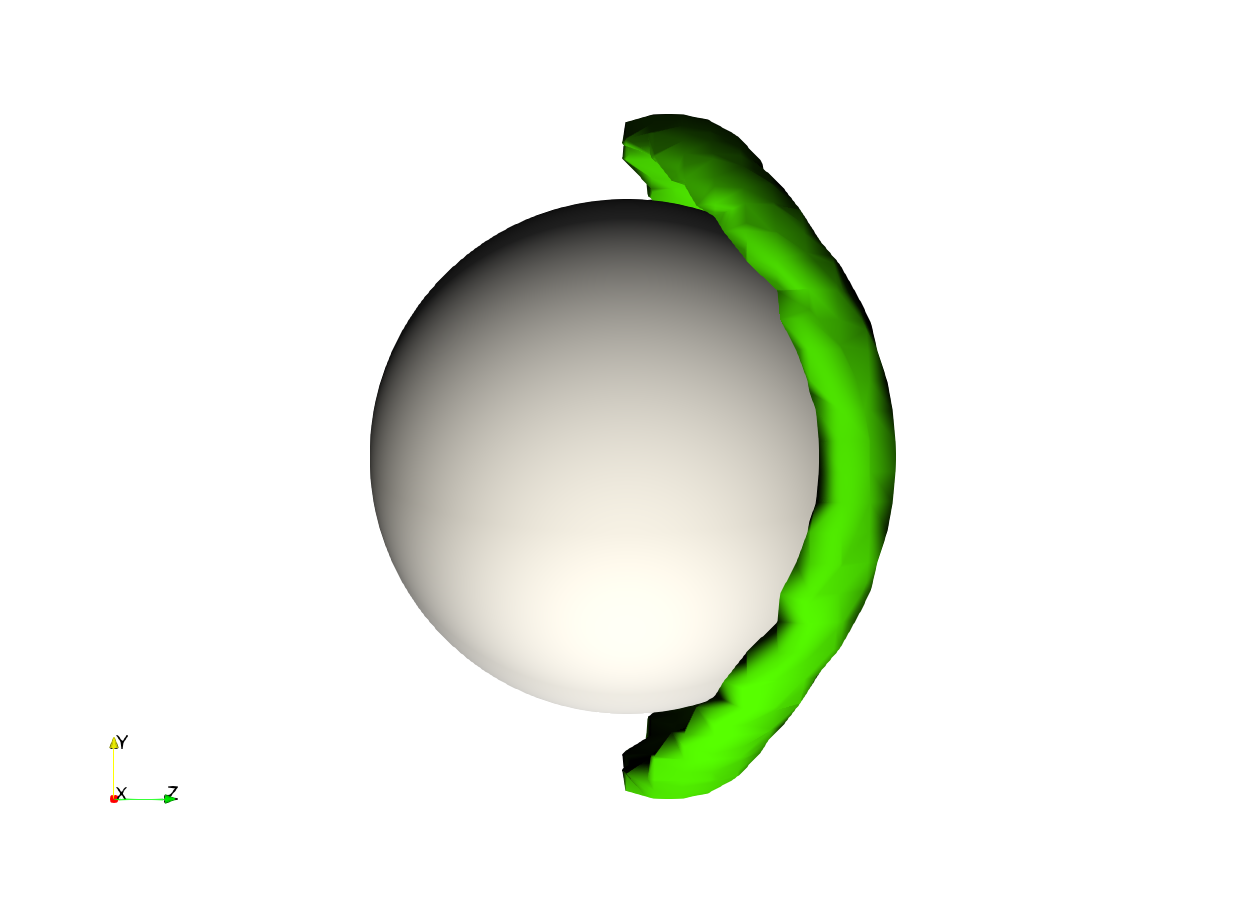}\\ \hline
\rev{Er}$=52.72 $  && \rev{Er}$= 51.86$ && \rev{Er}$=64.85$ &   \\
\includegraphics[trim={9cm 3cm 9cm 3cm},clip,width=0.13\linewidth]{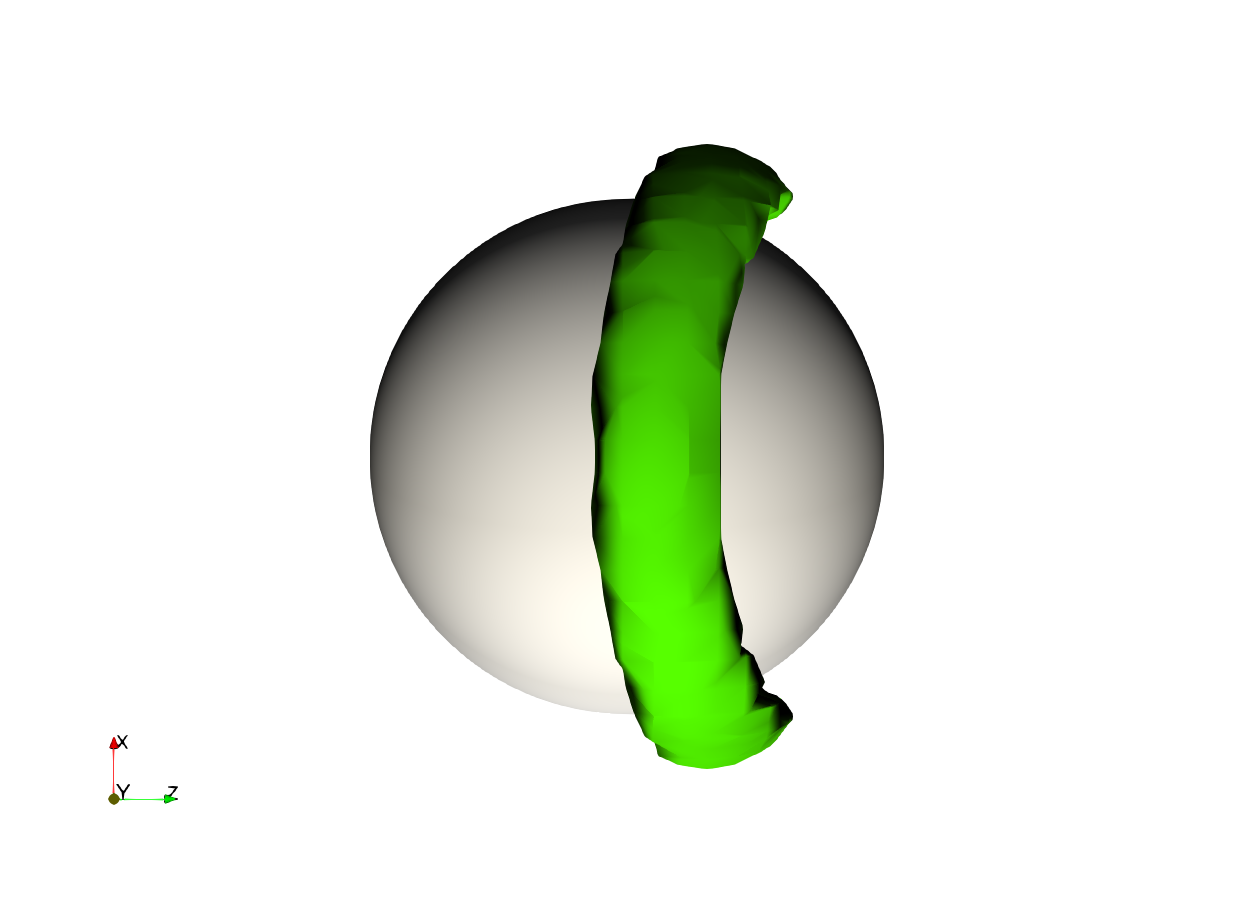} & 
\includegraphics[trim={9cm 3cm 9cm 3cm},clip,width=0.13\linewidth]{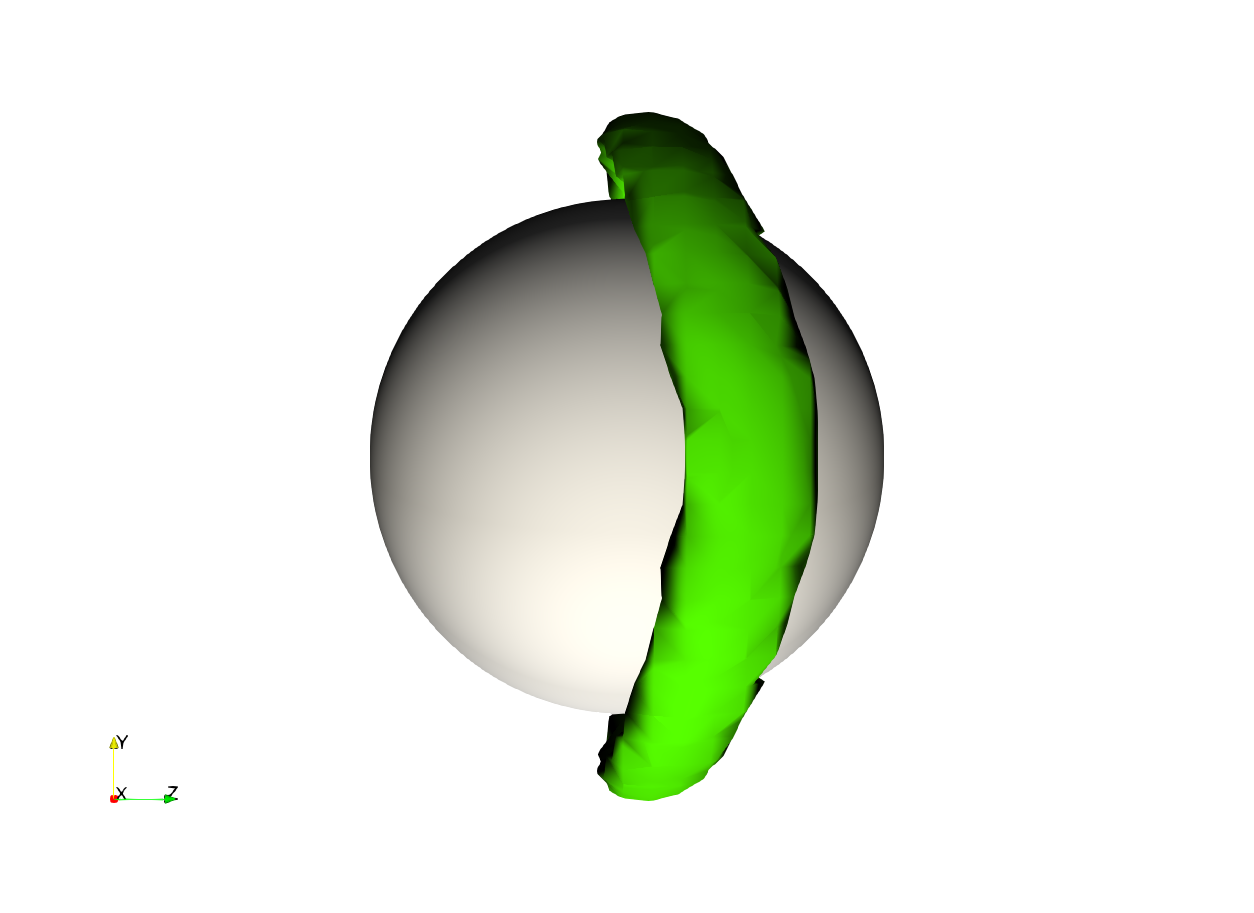} &  
\includegraphics[trim={9cm 3cm 9cm 3cm},clip,width=0.13\linewidth]{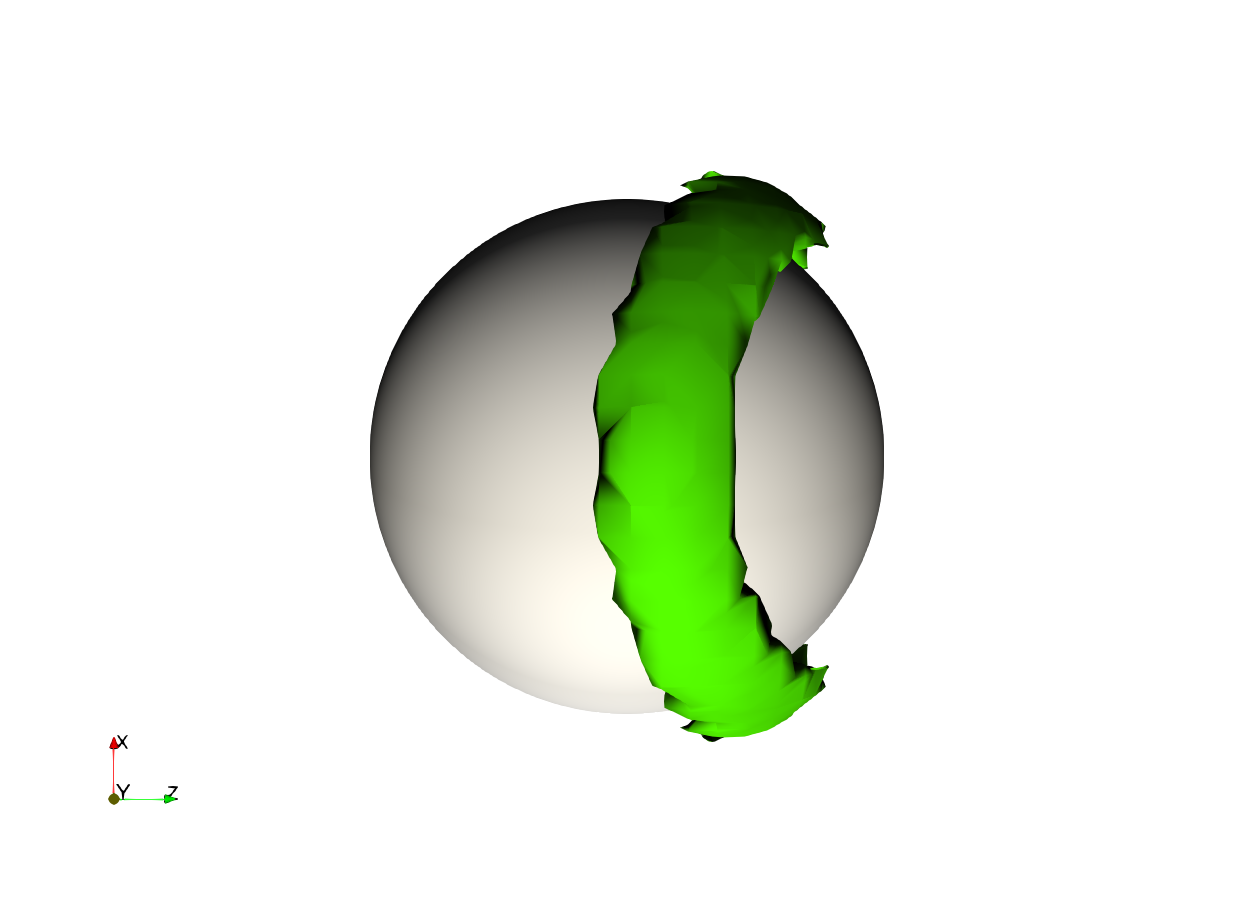} &  
\includegraphics[trim={9cm 3cm 9cm 3cm},clip,width=0.13\linewidth]{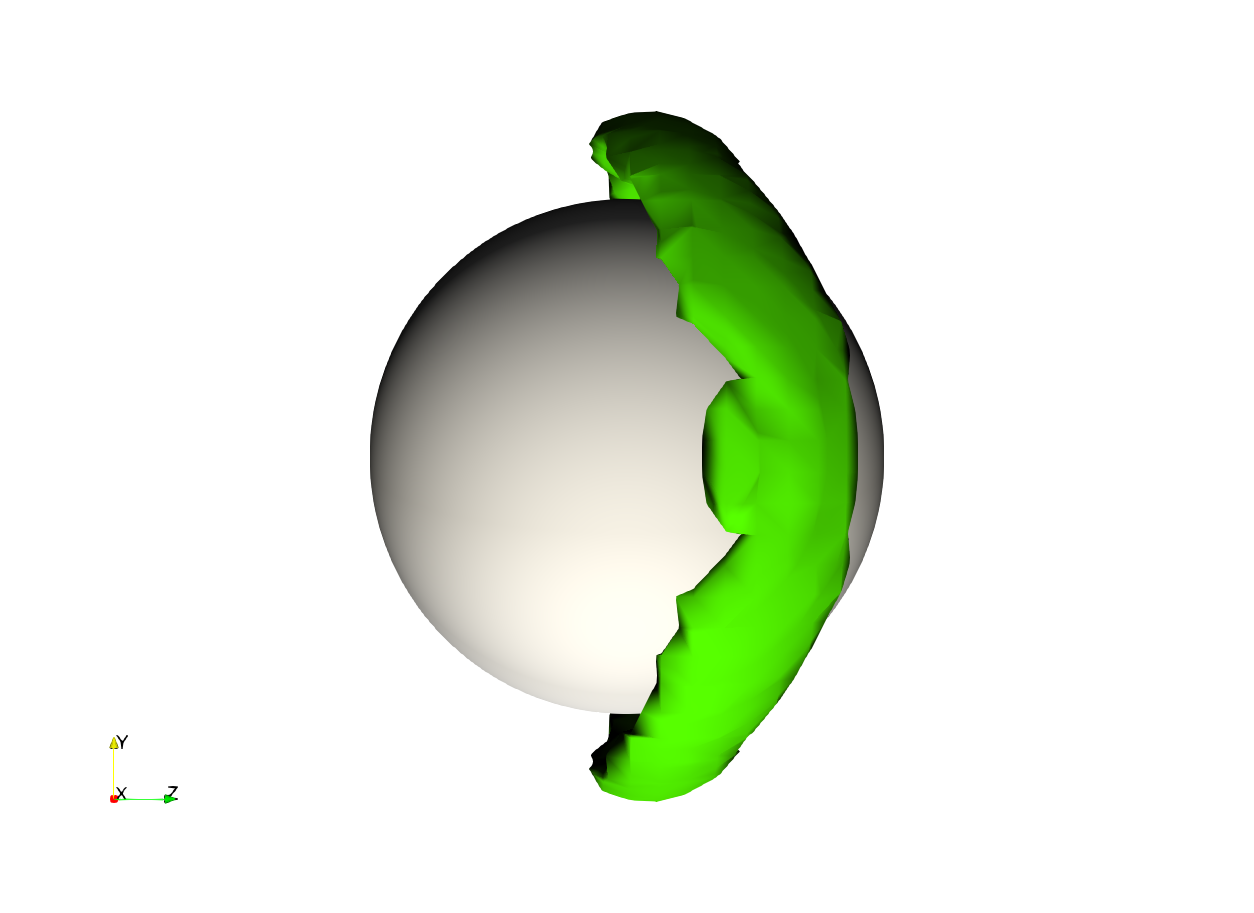} & 
\reflectbox{\rotatebox[origin=c]{180}{\includegraphics[trim={12cm 4cm 12cm 4cm},clip,width=0.13\linewidth]{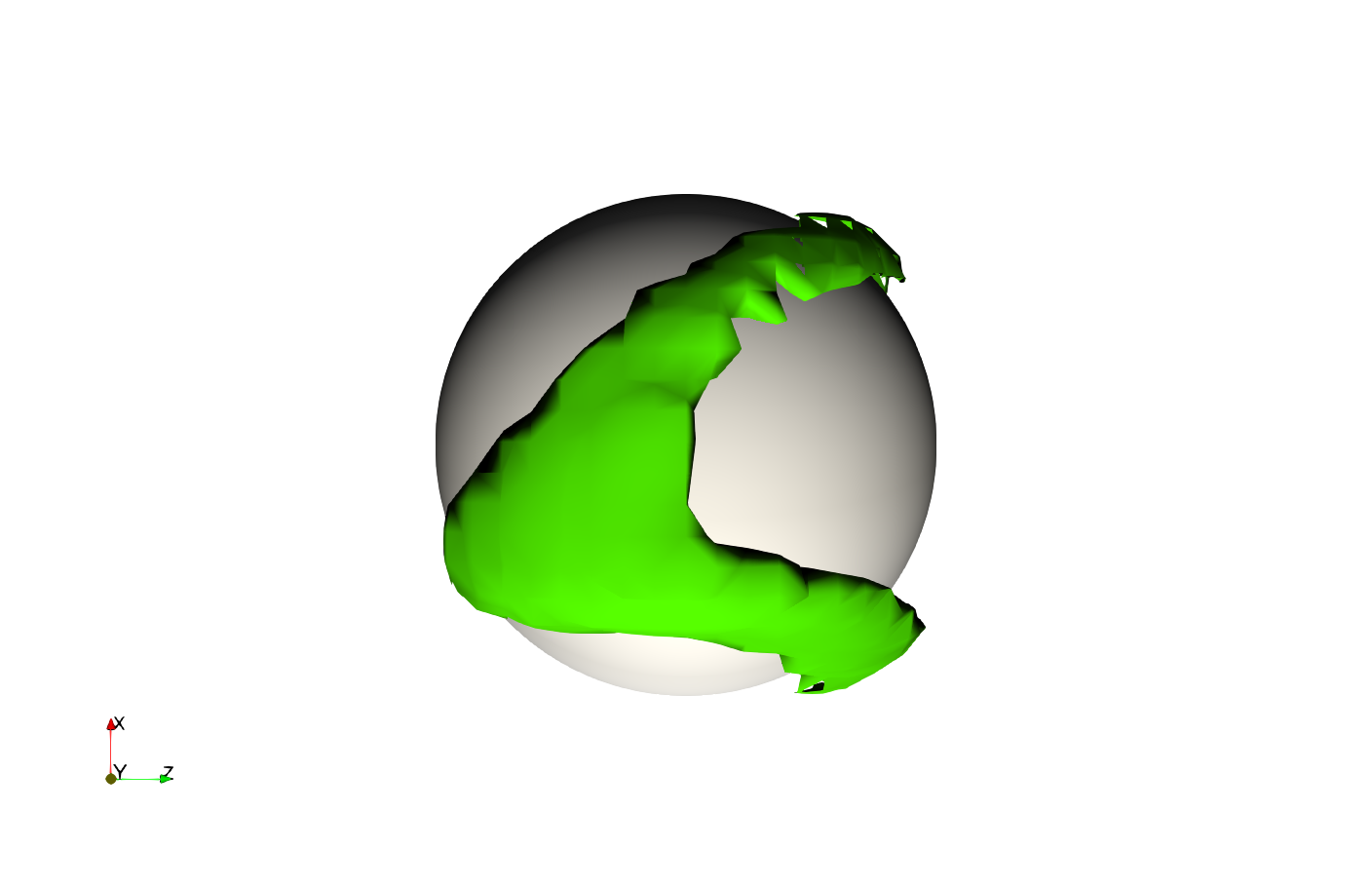}}}& 
\reflectbox{\rotatebox[origin=c]{180}{\includegraphics[trim={12cm 4cm 12cm 4cm},clip,width=0.13\linewidth]{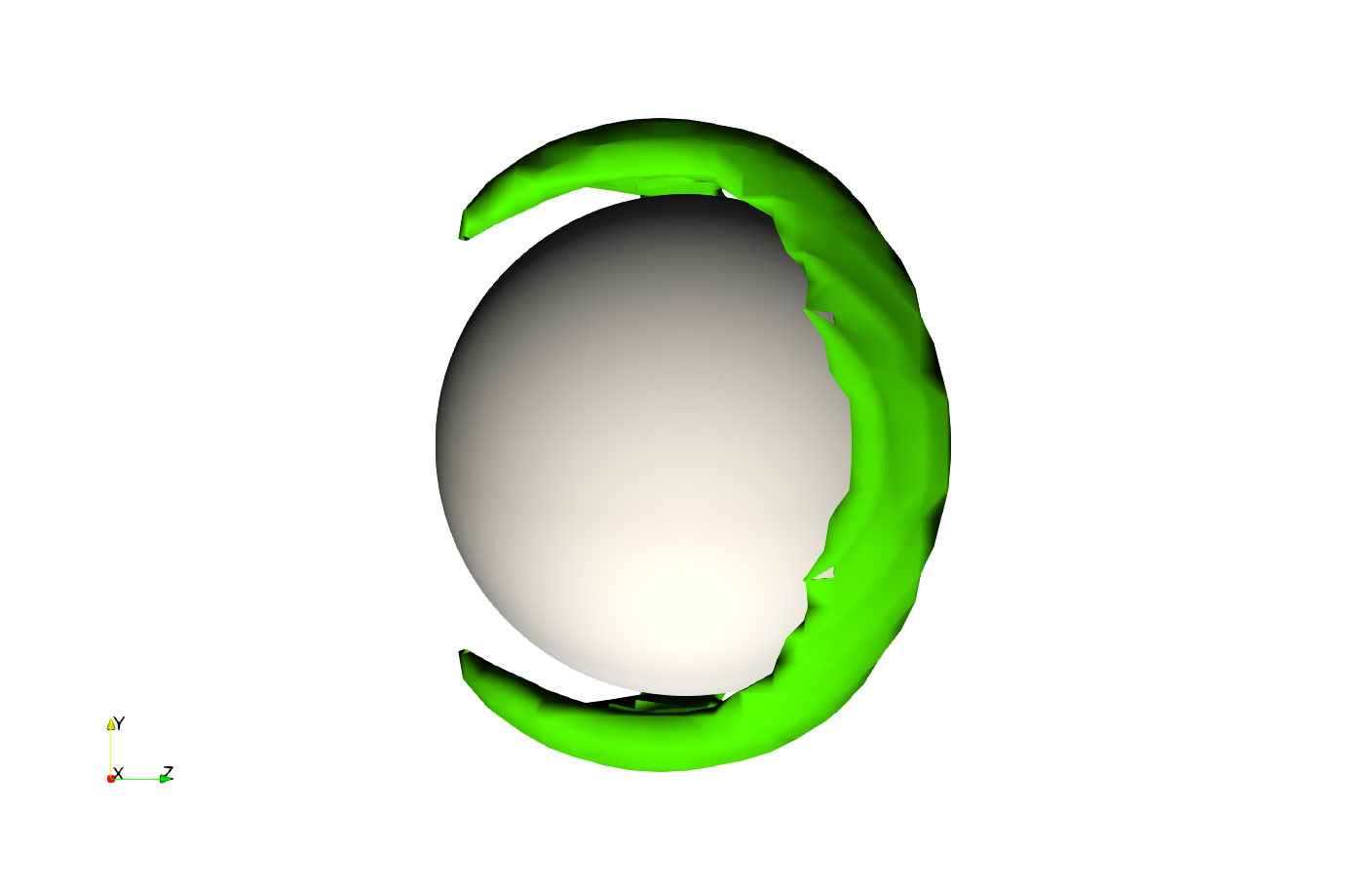}}}\\  \hline
\rev{Er}$=69.77 $ & & \rev{Er}$= 102.12$ && \rev{Er}$=86.06$ 
&  \\
\includegraphics[trim={9cm 3cm 9cm 3cm},clip,width=0.13\linewidth]{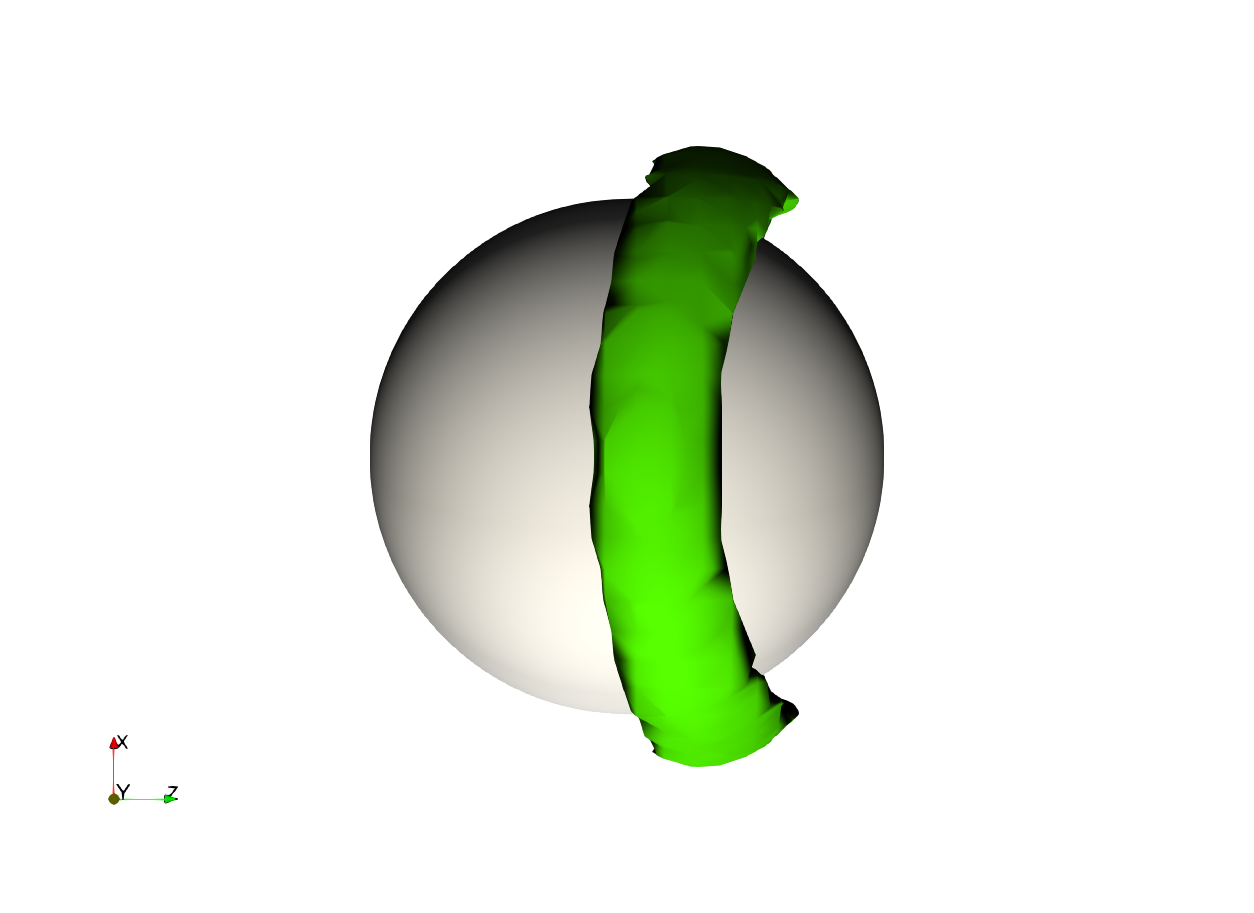} &
\includegraphics[trim={9cm 3cm 9cm 3cm},clip,width=0.13\linewidth]{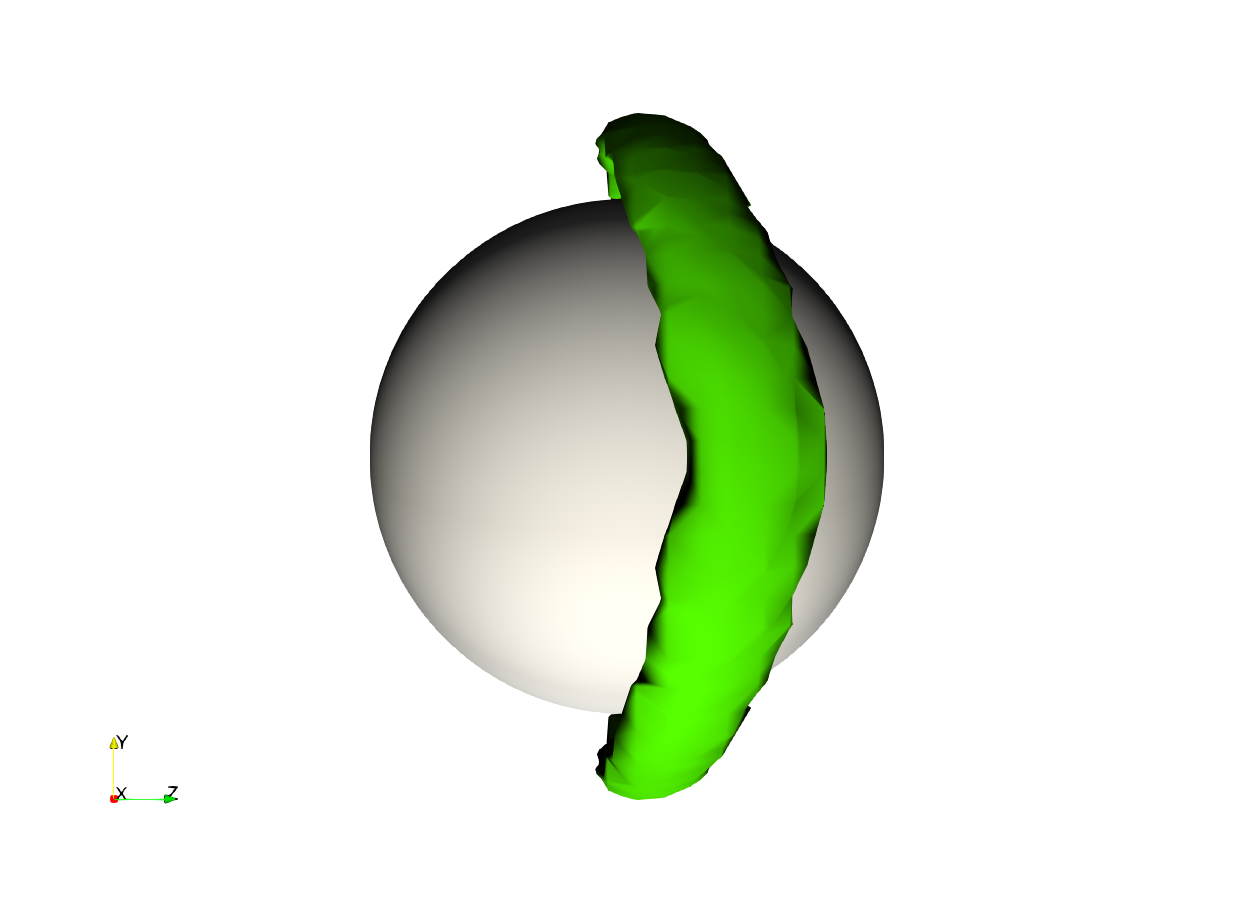} &
\includegraphics[trim={9cm 3cm 9cm 3cm},clip,width=0.13\linewidth]{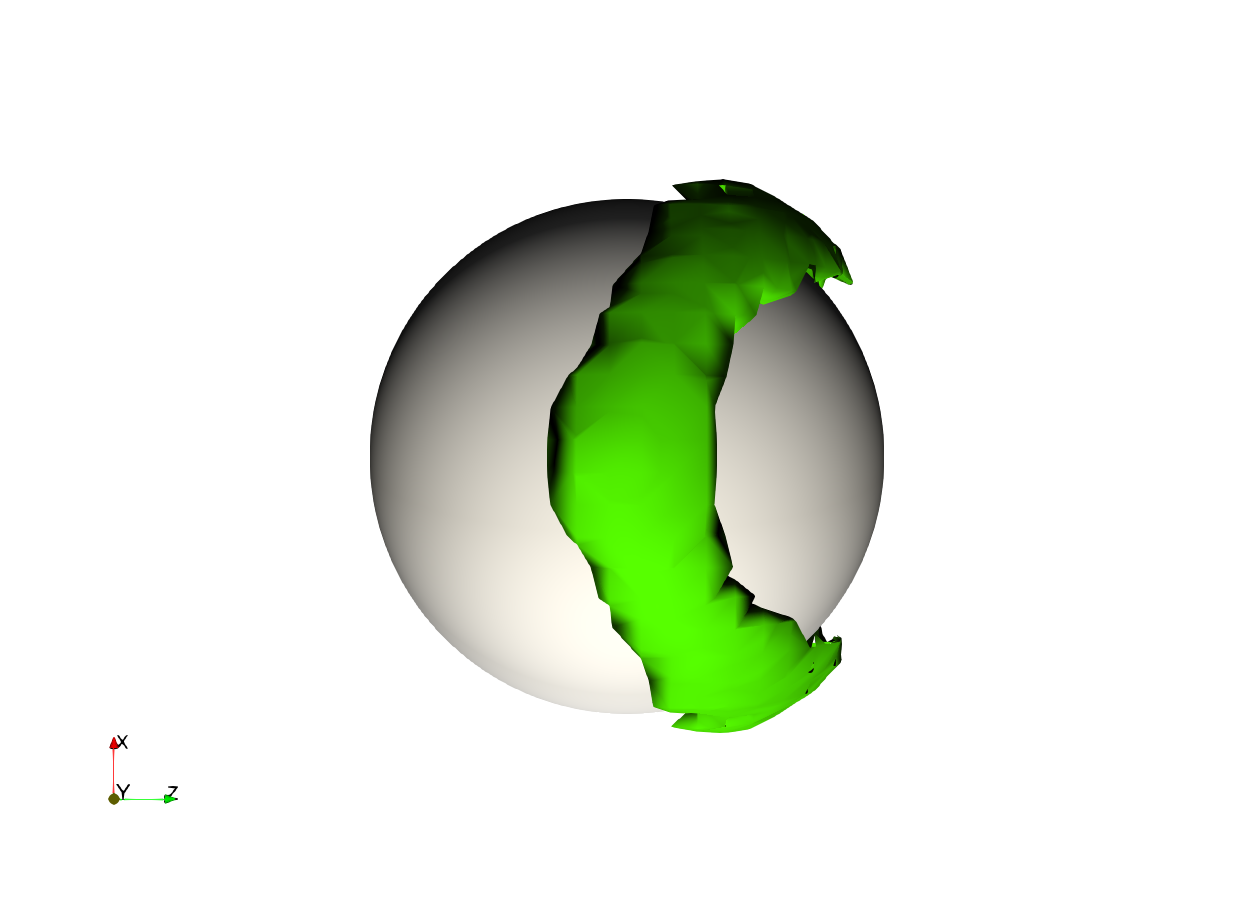}&  
\includegraphics[trim={9cm 3cm 9cm 3cm},clip,width=0.13\linewidth]{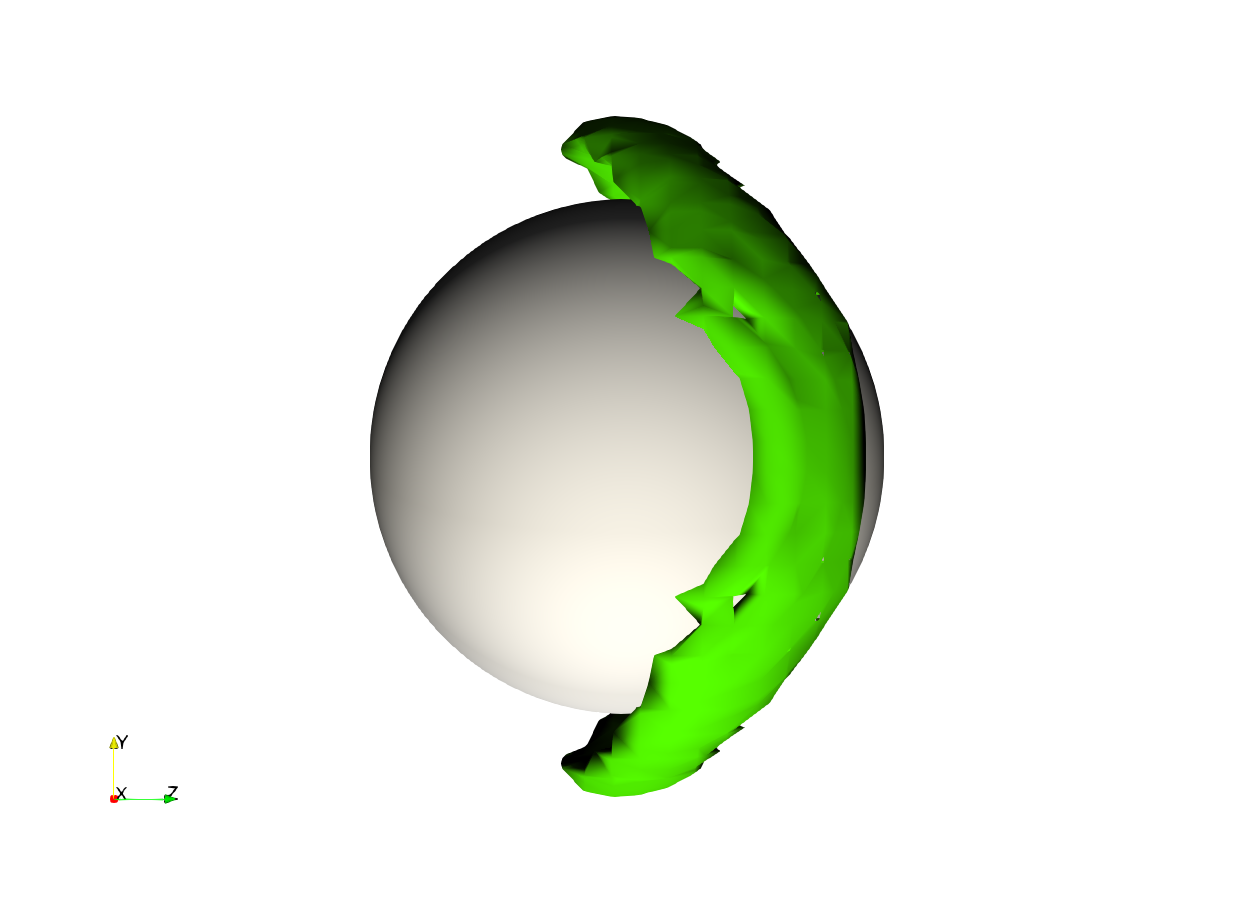}&  
\includegraphics[trim={8cm 4cm 14cm 4cm},clip,width=0.13\linewidth]{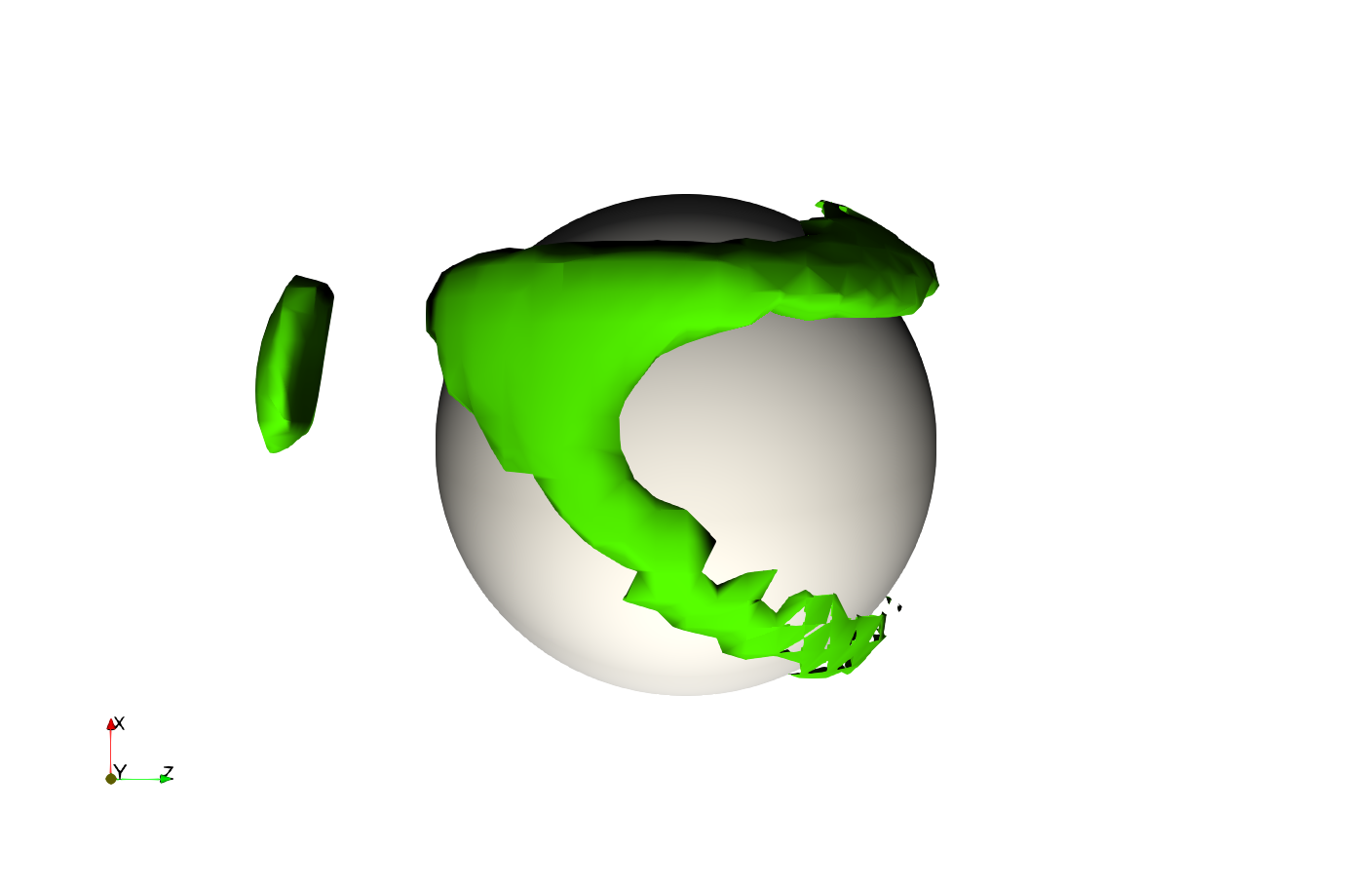}& 
\includegraphics[trim={9cm 4cm 14cm 4cm},clip,width=0.13\linewidth]{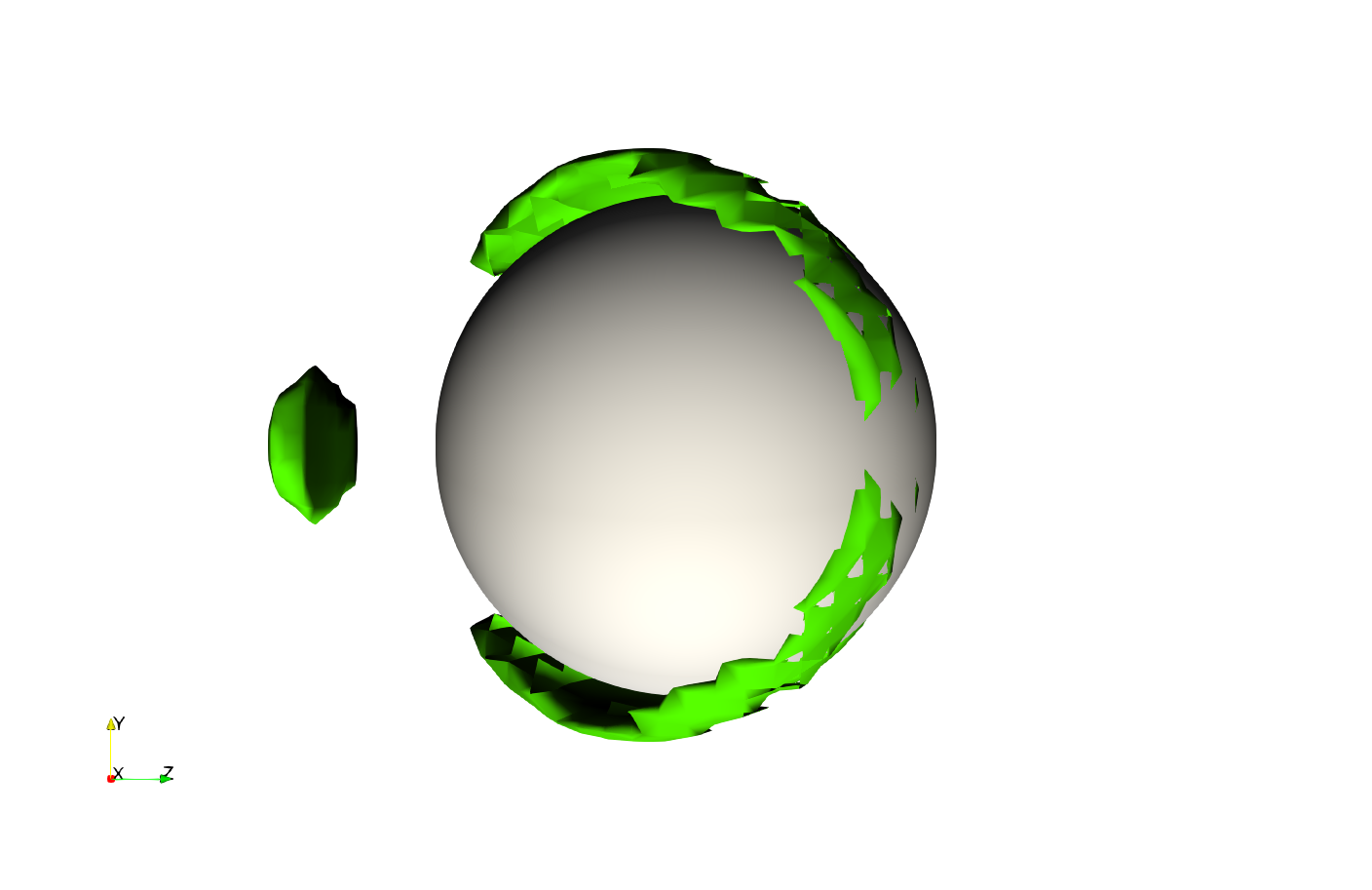} \\ \hline
\end{tabular}
\caption{Snapshots of the director field and defect structure in the steady state at various Ericksen numbers and strongest particle anchoring parameter $\omega=48$. The bright green region corresponds to the defect where liquid crystalline order is reduced. In each cell the left images in each cell show the side view looking in the negative $y$-direction with walls at the top and bottom. The images on the right in each cell show the view from the top looking in the positive $x$-direction. \rev{The flow is in the horizontal positive $z$-direction from left to right.}
The confinement increases from left to right, and Ericksen numbers increase from top to bottom. While in most cases the particle stays at the centre of the duct throughout the simulation, there are a small number of cases where they migrate away from it. Specifically, there are two cases where the particle migrates fully to a wall (for the two highest Ericksen number and confinement ratio $2R/L_x= 0.8$), and three cases where the particle migrates to a stable position between the wall and the centre (for \rev{Er}$=18.10$ and $2R/L_x= 0.4$, for \rev{Er}$=9.84$ and $2R/L_x= 0.8$, and for \rev{Er}$=10.37$ and $2R/L_x= 0.8$).}
\label{fig4}
\end{figure*}

The director field in the first and third row is colour-coded with red indicating an orientation parallel to the flow or $z$-direction and blue indicating an orientation perpendicular to the flow direction or in $xy$-plane. The bend state at low Ericksen number shows the Saturn ring defect oriented parallel to the walls with only very minor deformations, while the splay state has the Saturn ring defect oriented approximately perpendicular to the walls \rev{and displaced slightly downstream from the meridian of the particle in positive $z$-direction.}

\rev{The second and fourth row show the magnitude of the fluid velocity $u(x,z)=|\bm{u}(x,z)|$ in the $xz$-plane and $u(y,z)=|\bm{u}(y,z)|$ in the $yz$-plane normalised to the maximum velocity $u_c$ at the centre line of the duct. It is interesting to see that despite the striking differences in the director field structure and defect ring orientation at the two different Ericksen numbers both flow profiles are very similar. A minor exception is that at the lower Ericksen number the peak velocity is attained very close to the particle, whereas at the higher Ericksen number the relative fluid velocity is slightly reduced around the particle. This is a consequence of the different differential velocities between the colloidal particle and the fluid in both cases (see Fig.~\ref{fig8})}.  

As a quantitative overview of our findings, we include in Fig.~\ref{fig4}, snapshots of the particle and its defect in the steady state for varying confinement ratios and Ericksen numbers. In each cell the left images show the side view looking in the negative $y$-direction with walls at the top and bottom. The images on the right show the view from the top looking in the positive $x$-direction. The confinement increases from left to right from confinement ratios $2R/L_x=0.4$ to $2R/L_x=0.6$ to $2R/L_x=0.8$, and Ericksen numbers increase from top to bottom. The defect is shown as a green isosurface defined by a local order parameter $q \le 0.188$ and the particle anchoring strength and dimensionless anchoring parameter are $w_{part}=0.05$ and $\omega=48$, respectively, as lower anchoring strengths do not result in defects that could be distinctively visualised.

At low Ericksen numbers, below the bend-to-splay transition, the particle has a Saturn ring defect whose ring plane remains parallel to the walls. This is the case for all confinement ratios and Ericksen numbers below \rev{Er}$=10.37$, as shown in the first and second row of Fig.~\ref{fig4}. Two aspects are noteworthy: Firstly, there is a slight increase of the defect isosurface radius downstream of the particle, for which both Ericksen number (see images for $2R/L_x=0.4$ with \rev{Er}$=1.65$ and \rev{Er}$=8.30$) and confinement ratio (see image for $2R/L_x=0.6$, \rev{Er}$=4.38$ and image and Movie.1 for $2R/L_x=0.8$, \rev{Er}$=6.15$) seem responsible. However, confinement appears to play a more important role in this context.

Secondly, at slightly increased Ericksen numbers (see images for $2R/L_x=0.6$, \rev{Er}$=9.84$ and $2R/L_x=0.8$, \rev{Er}$=10.37$), the Saturn ring becomes angled such that the part downstream of the particle is closer to the bottom wall, while the other part upstream of the particle remains virtually unchanged. These two particular cases reached steady state positions that are offset somewhere between the centre of the duct and the walls in the $x$-direction, which contributes to this asymmetric appearance. This can be explained with the migration \rev{(i.e. lateral movement perpendicular to the flow direction)} to the weak attractor region that we observed in our previous work on controllable particle migration \cite{Lesniewska2022} in practically unconfined conditions using a much wider duct and lower confinement ratio $2R/L_x=0.15$. For direct comparison we provide in Table \ref{tab2} an approximate conversion between particle Ericksen numbers, as used our previous publication, and Ericksen numbers based on the smallest duct dimension, as used in this work. 

\begin{figure}[htbp]
\centering
\begin{tabular}{lll}
a) \rev{\quad \includegraphics[trim={3.5cm 3.5cm 37.5cm 25.5cm},clip,scale=0.3]{xz_axis.png}} & b) \rev{\quad \includegraphics[trim={3.5cm 3.5cm 37.5cm 25.5cm},clip,scale=0.3]{xz_axis.png}} & c) \rev{\quad \includegraphics[trim={3.5cm 3.5cm 37.5cm 25.5cm},clip,scale=0.3]{xz_axis.png}} \\
\includegraphics[trim={11cm 5cm 12cm 5cm},clip,scale=0.1]{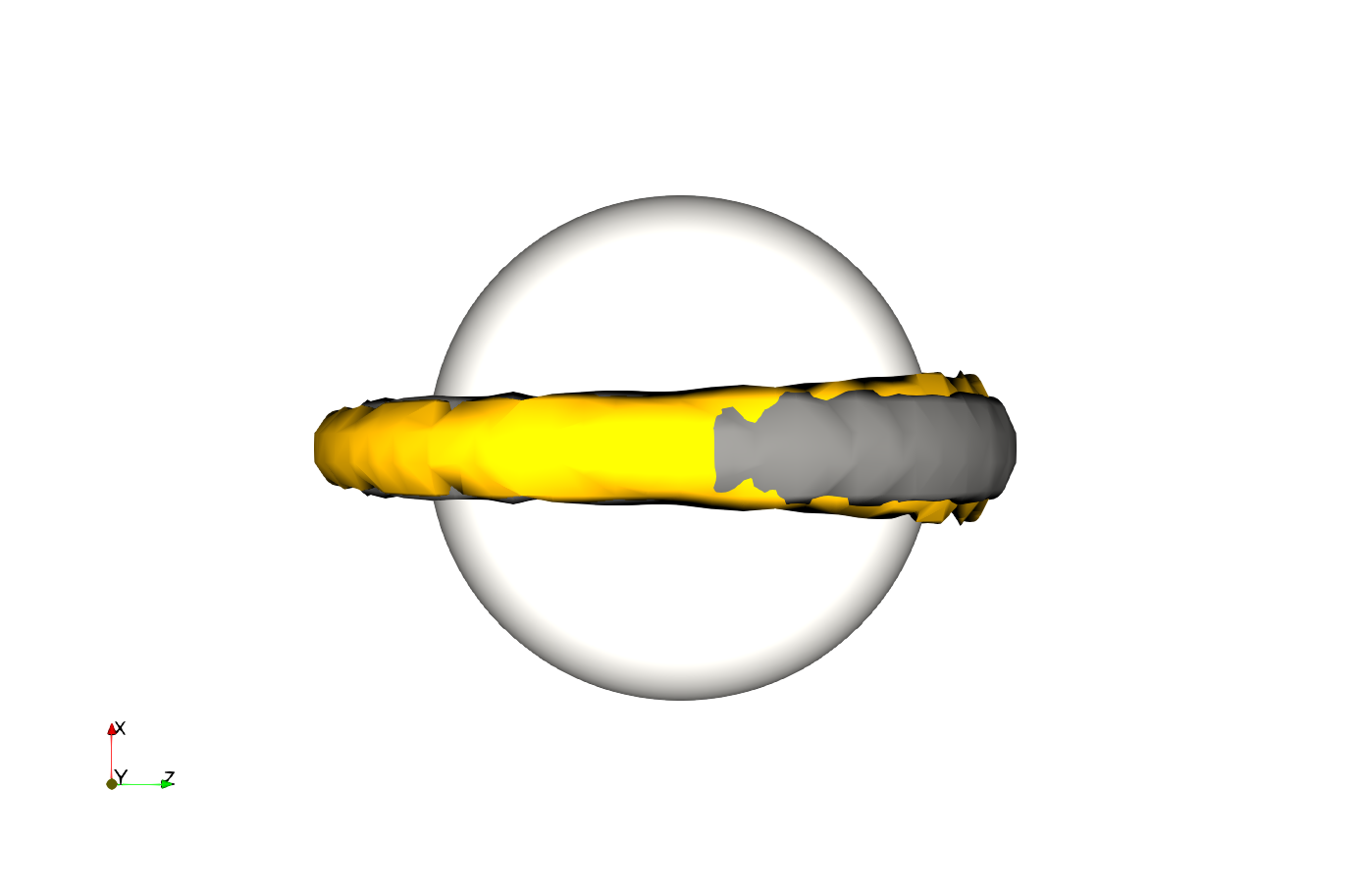}&
\includegraphics[trim={11cm 5cm 12cm 5cm},clip,scale=0.1]{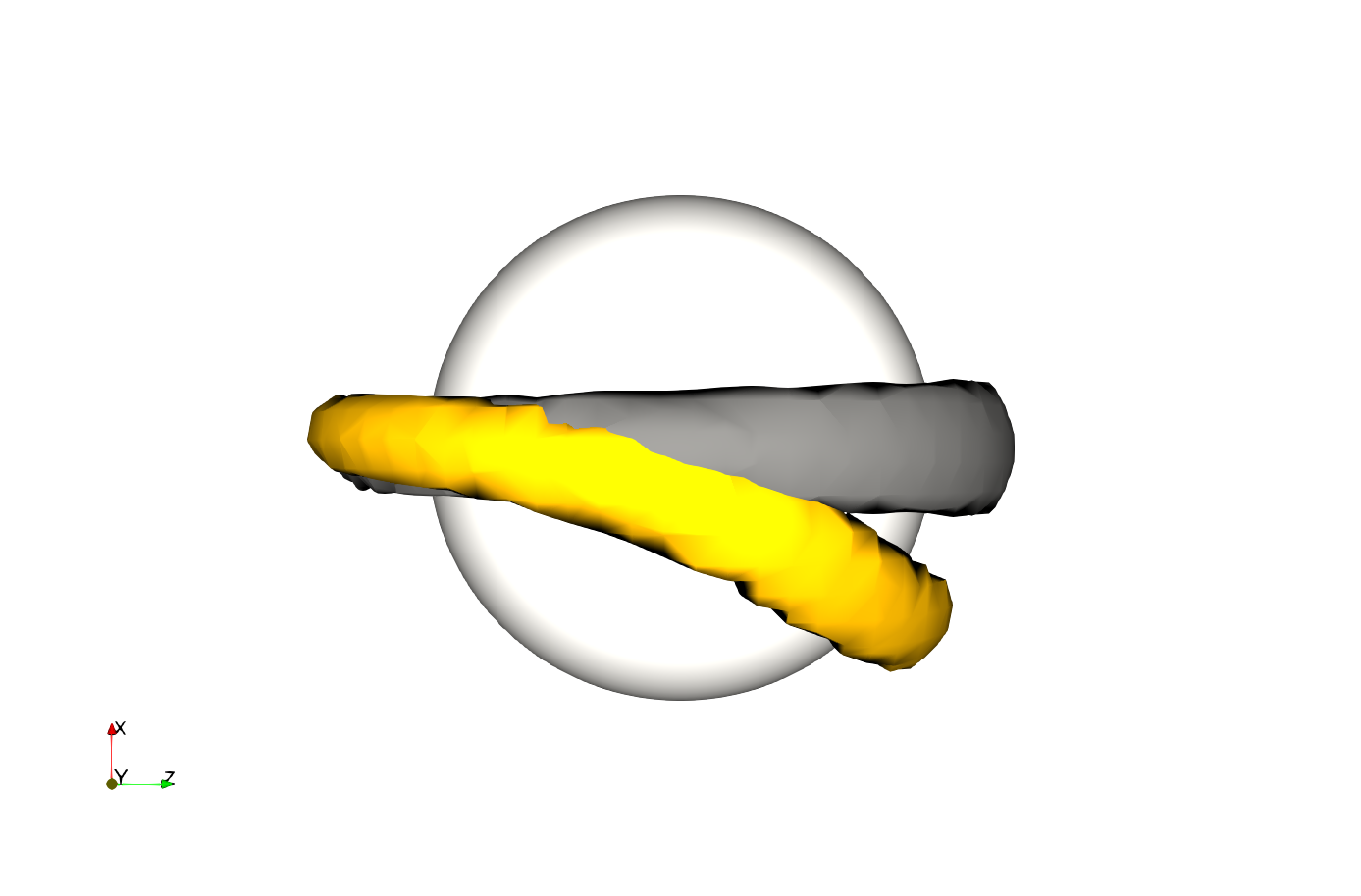}&
\includegraphics[trim={11cm 5cm 12cm 5cm},clip,scale=0.1]{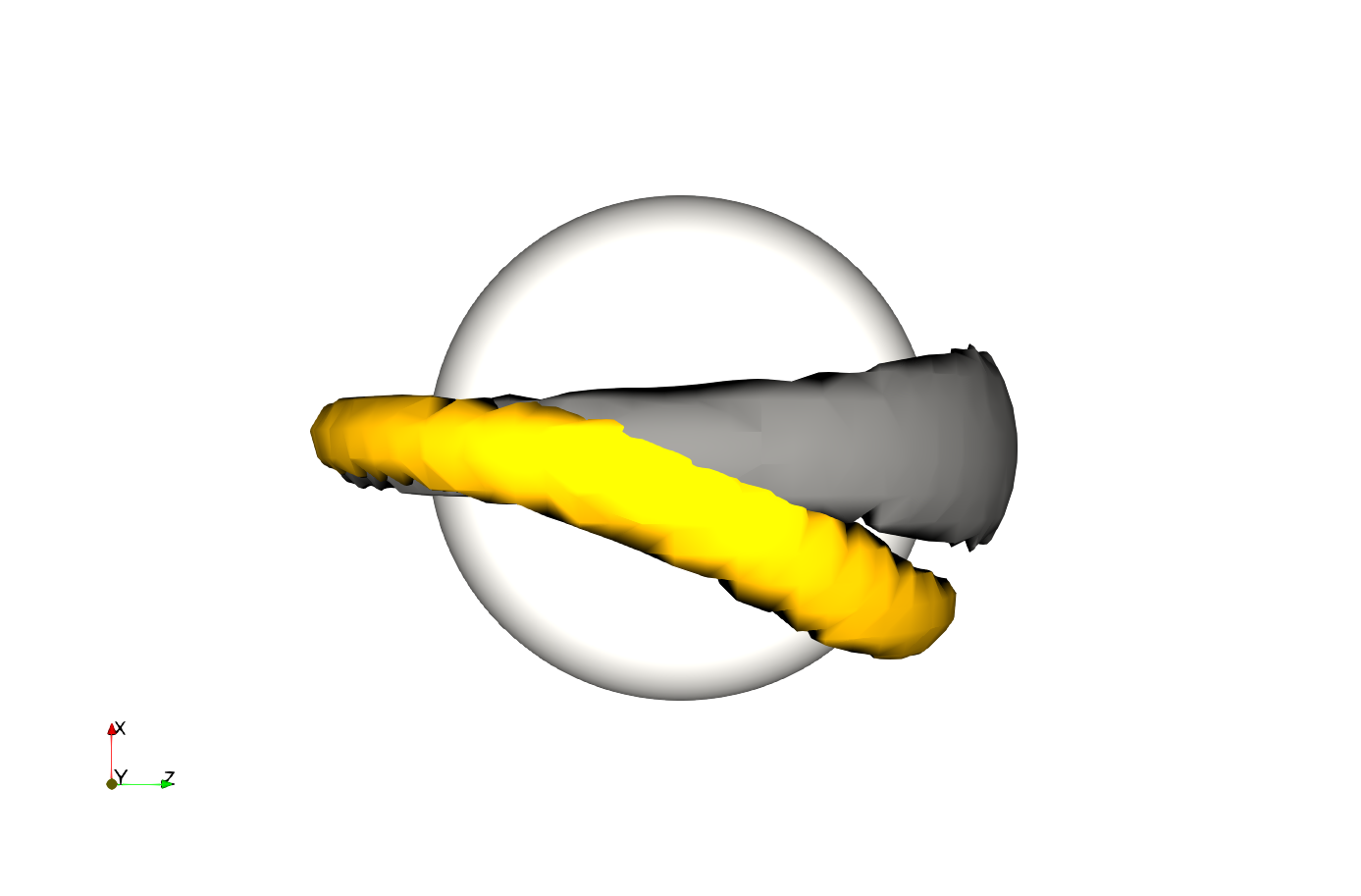}\\
\includegraphics[trim={11cm 3cm 12cm 3cm},clip,scale=0.1]{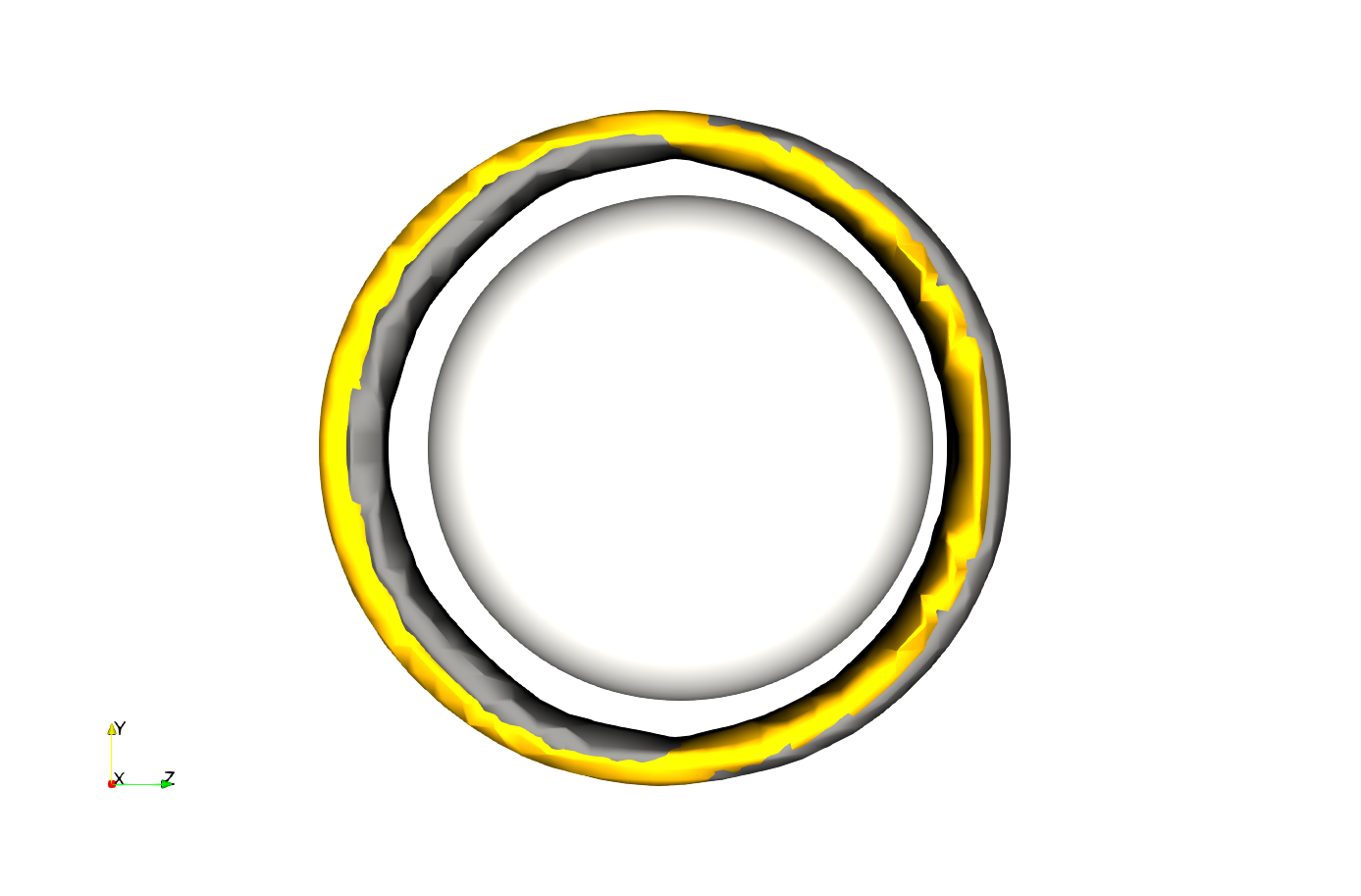}&
\includegraphics[trim={11cm 3cm 12cm 3cm},clip,scale=0.1]{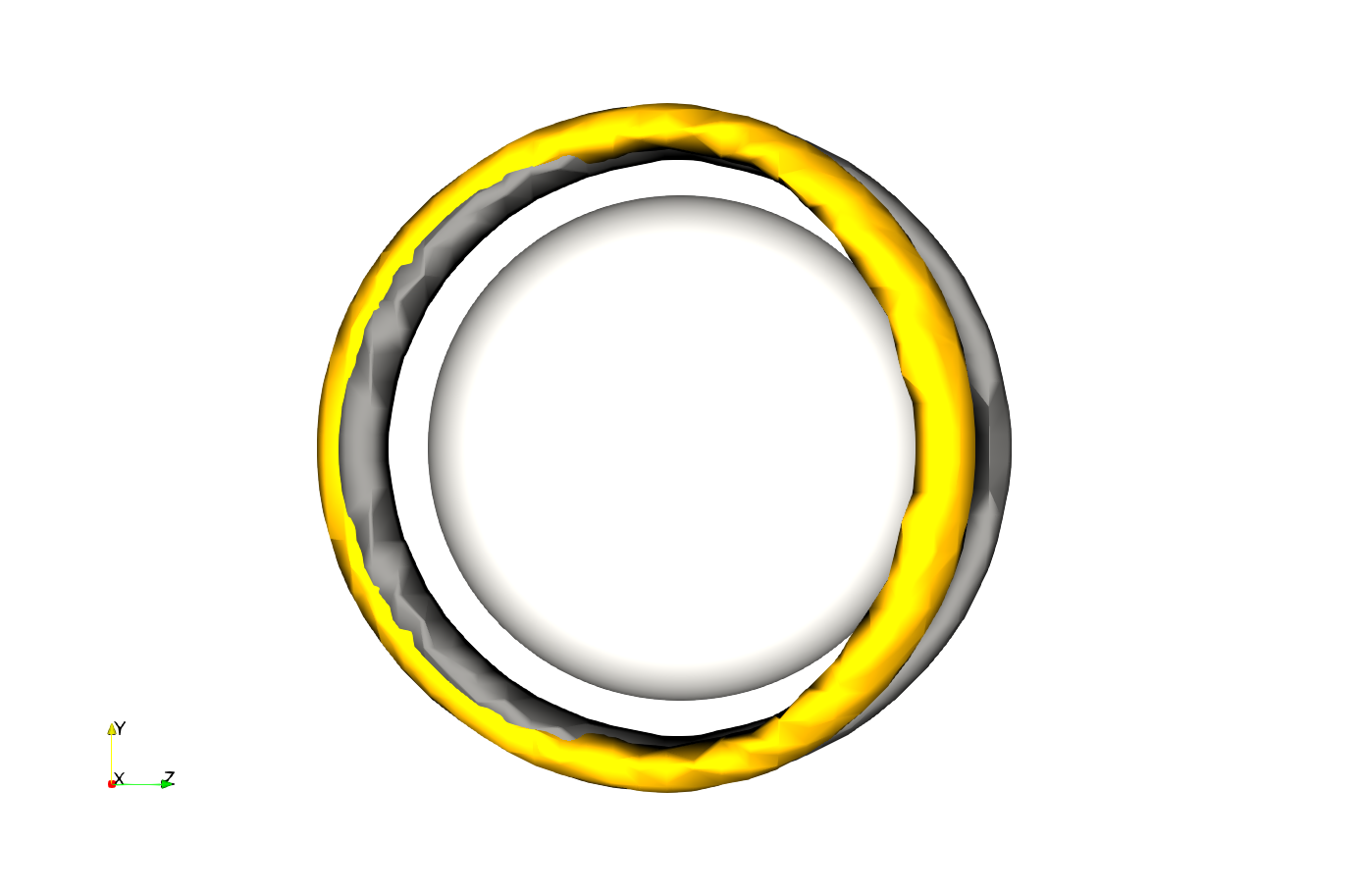}&
\includegraphics[trim={11cm 3cm 12cm 3cm},clip,scale=0.1]{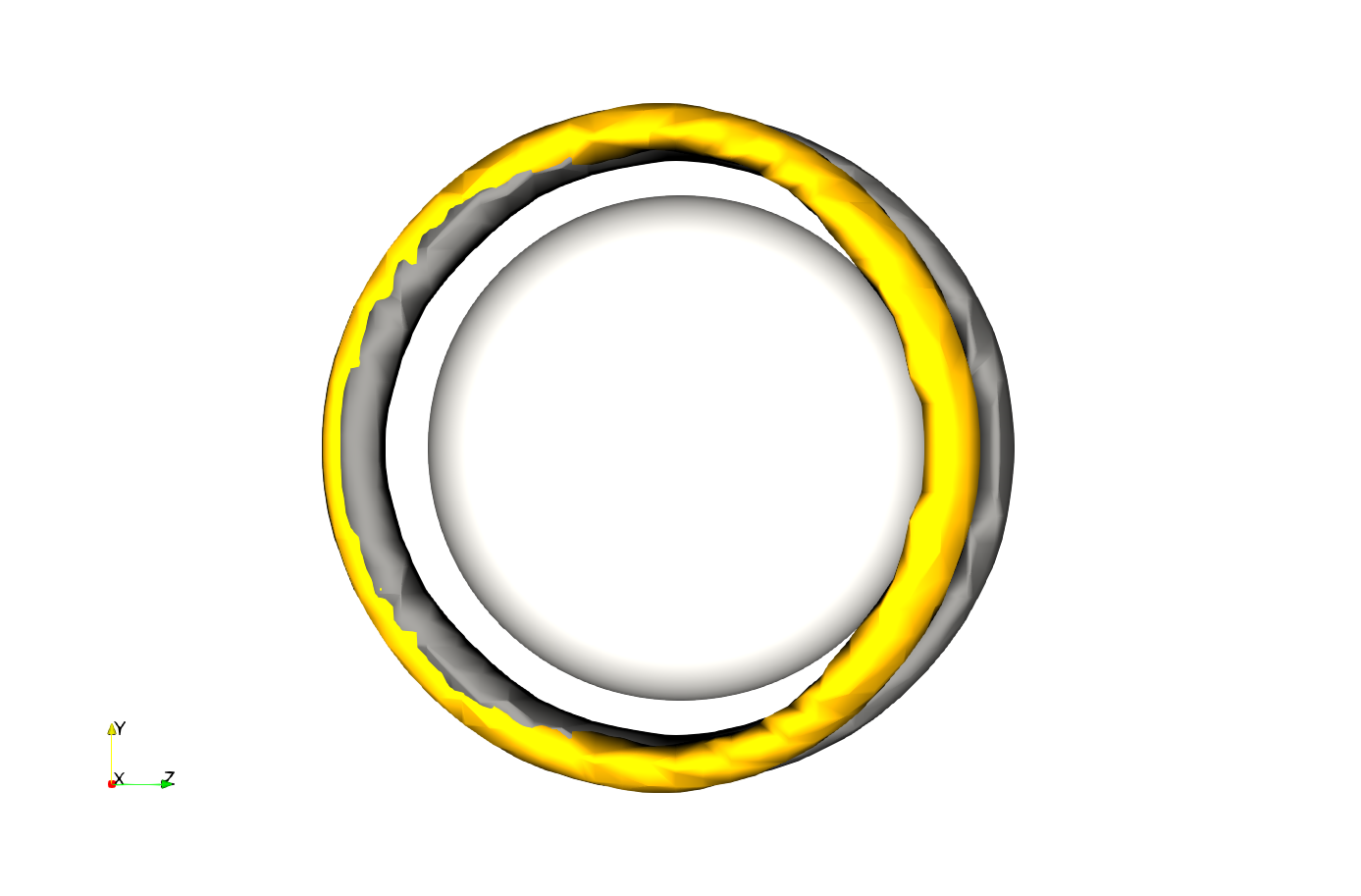}\\
{\small \textcolor{bla1_1}{Er=1.65} \textcolor{bla1_2}{Er=8.30} } 
& {\small \textcolor{bla1_1}{Er=4.38} \textcolor{bla1_2}{Er=9.84}}
& {\small \textcolor{bla1_1}{Er=6.15} \textcolor{bla1_2}{Er=10.37}}  \\
\rev{\qquad \includegraphics[trim={3.5cm 3.5cm 37.5cm 25.5cm},clip,scale=0.3]{yz_axis.png}} &  \rev{\qquad \includegraphics[trim={3.5cm 3.5cm 37.5cm 25.5cm},clip,scale=0.3]{yz_axis.png}} & \rev{\qquad \includegraphics[trim={3.5cm 3.5cm 37.5cm 25.5cm},clip,scale=0.3]{yz_axis.png}} \\
\end{tabular}
\caption{Disclination lines around the particle for the lowest \textcolor{bla1_1}{ \textemdash} and highest \textcolor{bla1_2}{ \textemdash} simulated Ericksen numbers below the bend-to-splay transition at confinement ratios a) $2R/L_x= 0.4$,  b) $2R/L_x= 0.6$ and c) $2R/L_x= 0.8$, respectively.
The top row has the view in negative $y$-direction, the widest duct dimension, with the walls in the narrowest duct dimension at the $x$-boundaries situated closely above and below the particle. The bottom row shows the view in positive $x$-direction. The flow is in the horizontal positive $z$-direction from left to right.}
\label{fig5}
\end{figure}

Fig.~\ref{fig5} shows a direct comparison of the defect rings around the particle for the lowest and highest simulated Ericksen numbers below the bend-to-splay transition, at confinement ratios (a) $2R/L_x=0.4$ (b) $0.6$ and (c) $0.8$.
The defects at the lowest Ericksen numbers, depicted in grey, are distinctive Saturn rings that are oriented parallel to the walls in the $x-$direction. As previously mentioned, increasing confinement leads to a defect ring that is thicker at the downstream side of the particle, while it remains oriented parallel to the wall at the $x$-boundary. Increasing the Ericksen number alone does not change the orientation of the defect ring, but leads to a very slight shift in position upstream (see grey and yellow defect rings in Fig. \ref{fig5} a)). However, increasing the Ericksen number and confinement ratio induces a noticeable tilt of the defect ring, shown in Fig.~\ref{fig5} b) and c), as the particle migrates into the off-centre steady state position somewhere between the centre of the duct and one of the walls. Despite the difference in confinement ratio and Ericksen number ($2R/L_x=0.6$ and $0.8$, \rev{Er}$=9.84$ and $10.37$, respectively) the shape of the defect rings is almost the same.

Upon increasing the Ericksen number, a bend-to-splay transition takes place somewhere between $8.30 < $\rev{Er}$ < 18.10$ (for $2R/L_x=0.4$), $9.84 < $\rev{Er}$ < 21.25$ (for $2R/L_x=0.6$) and $10.37 < $\rev{Er}$ < 17.95$ (for $2R/L_x=0.8$). The defect ring is now reoriented with its ring plane approximately perpendicular to the walls and flow direction, as shown in Fig.~\ref{fig4}, for instance in the third row, and retains a similar shape at higher Ericksen numbers (see images for $2R/L_x=0.4$, \rev{Er}$=18.10$, $2R/L_x=0.6$, \rev{Er}$=21.25$, and $2R/L_x=0.8$, \rev{Er}$=22.32$). The case for $2R/L_x=0.4$, \rev{Er}$=18.10$ forms an exception in that the particle moves very slightly away from the centre into a stable off-centre position, while in the other cases the particle remains at the centre of the duct, which can be also understood with the migration to the previously observed weak attractor region at similar Ericksen numbers \cite{Lesniewska2022} (see Table \ref{tab2} for conversion of Ericksen numbers).
A noticeable difference is that with increasing confinement the defect ring appears compressed in the smallest duct dimension due to the relative proximity of the walls (see image and Movie.2 for $2R/L_x=0.8$, \rev{Er}$=22.32$). 

With increasing Ericksen numbers the shape of the vertically oriented defect ring remains largely unchanged for low and medium confinement, as shown in the first and second column, forth and fifth row, of Fig.~\ref{fig4} ($2R/L_x=0.4$ and $0.6$) for Ericksen numbers \rev{Er}$=52.72,\, 69.77$ and \rev{Er}$=51.86,\,102,12$, respectively. At \rev{Er}$=102.12$ a slight change occurs such that the defect close to the mid-plane of the duct in the $x$-direction are distorted and pulled in the upstream direction, i.e.~against the flow. This effect is a precursor of the more dramatic elongation of the Saturn ring that will become even more evident as the confinement ratio increases.

At even higher Ericksen numbers \rev{Er}$=64.85$ and \rev{Er}$=86.06$ and confinement $2R/L_x=0.8$, shown in the third column forth and fifth row of Fig.~\ref{fig4}, we observe defects that differ substantially from those discussed before. 
In these cases the particle migrates fully to one of the walls. This has also been previously observed for similar Ericksen numbers in much lower confinement \cite{Lesniewska2022}. \rev{But there it occurred when the particle was within a distance of one and a half to two diameters from the walls, depending on the Ericksen number (see Table~\ref{tab2} for a conversion of Er). Given the proximity of the walls in the present work with increased confinement, this means that attraction to the walls should occur in practically all situations. This, however, is not the case as we observe attraction to the walls only for the highest Ericksen numbers and the largest confinement. Thus, increased confinement prevents particle migration to the walls and stabilises trajectories around the centre of the duct.}
The migration to one of the walls results in a different defect shape such that there is a pronounced elongation of the Saturn ring defect in the upstream direction. There is also the indication of  a small satellite region of low order, upstream of the particle that never merges up with the rest of the defect (see image and Movie.3 for $2R/L_x=0.8$ and \rev{Er}$=86.06$). 

\begin{figure}[htbp]
\centering
\begin{tabular}{lll}
a) \rev{\quad \includegraphics[trim={3.5cm 3.5cm 37.5cm 25.5cm},clip,scale=0.3]{xz_axis.png}} & b) \rev{\quad \includegraphics[trim={3.5cm 3.5cm 37.5cm 25.5cm},clip,scale=0.3]{xz_axis.png}} & c) \rev{\quad \includegraphics[trim={3.5cm 3.5cm 37.5cm 25.5cm},clip,scale=0.3]{xz_axis.png}} \\
\includegraphics[trim={13cm 3cm 13cm 3cm},clip,scale=0.1]{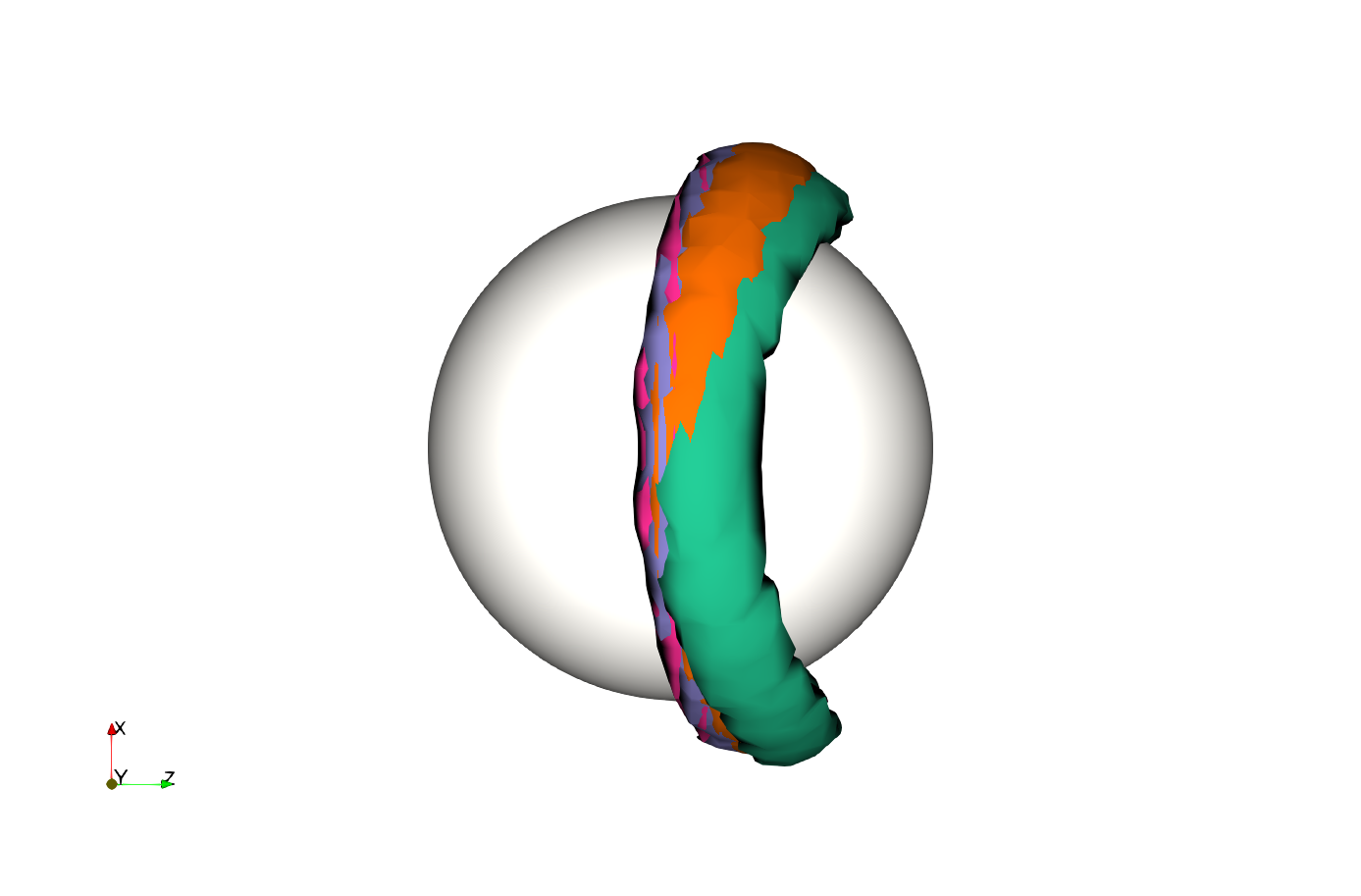} &
\includegraphics[trim={13cm 3cm 13cm 3cm},clip,scale=0.1]{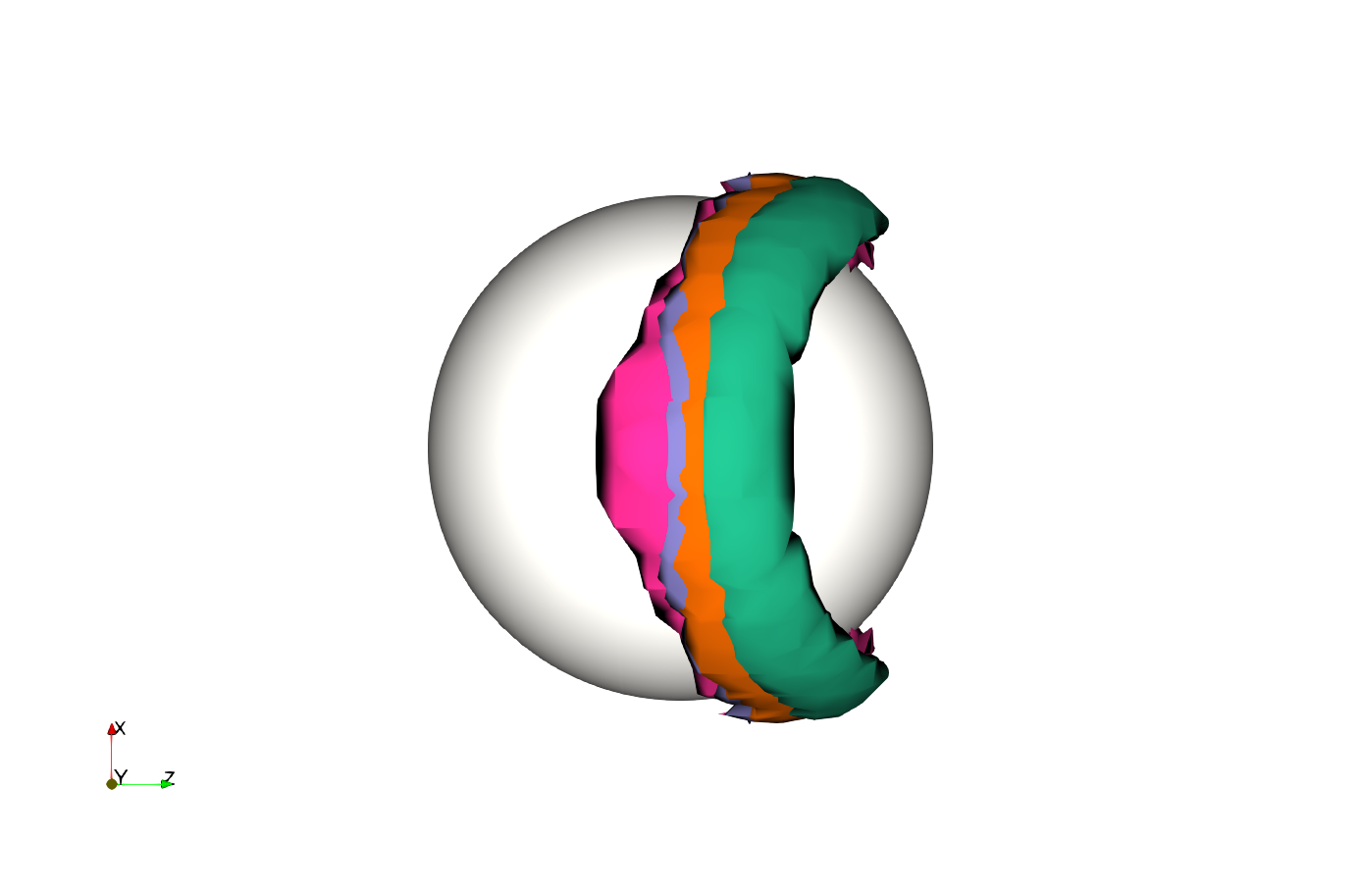} &
\includegraphics[trim={13cm 3cm 13cm 3cm},clip,scale=0.1]{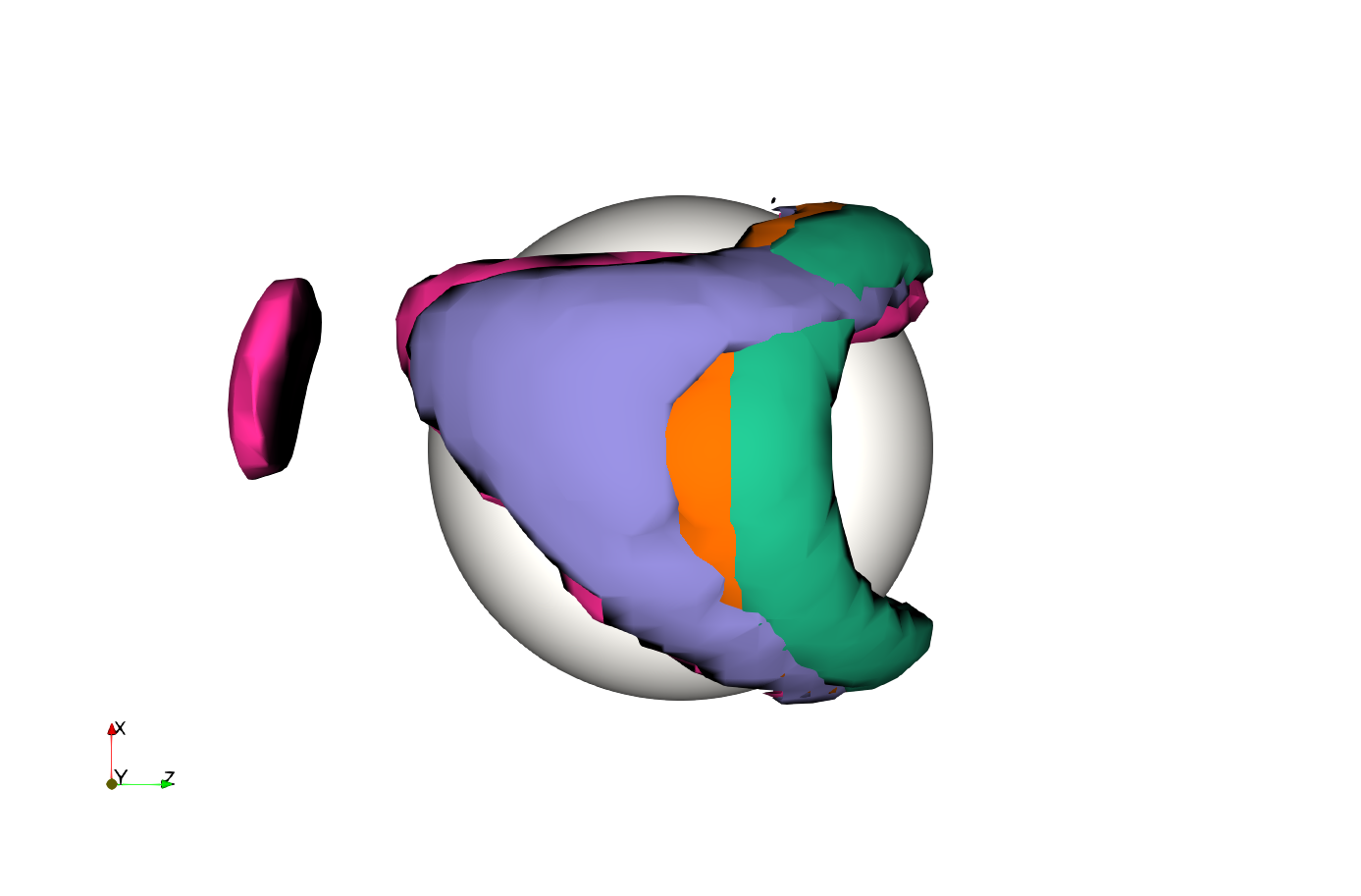} \\
\includegraphics[trim={13cm 3cm 13cm 3cm},clip,scale=0.1]{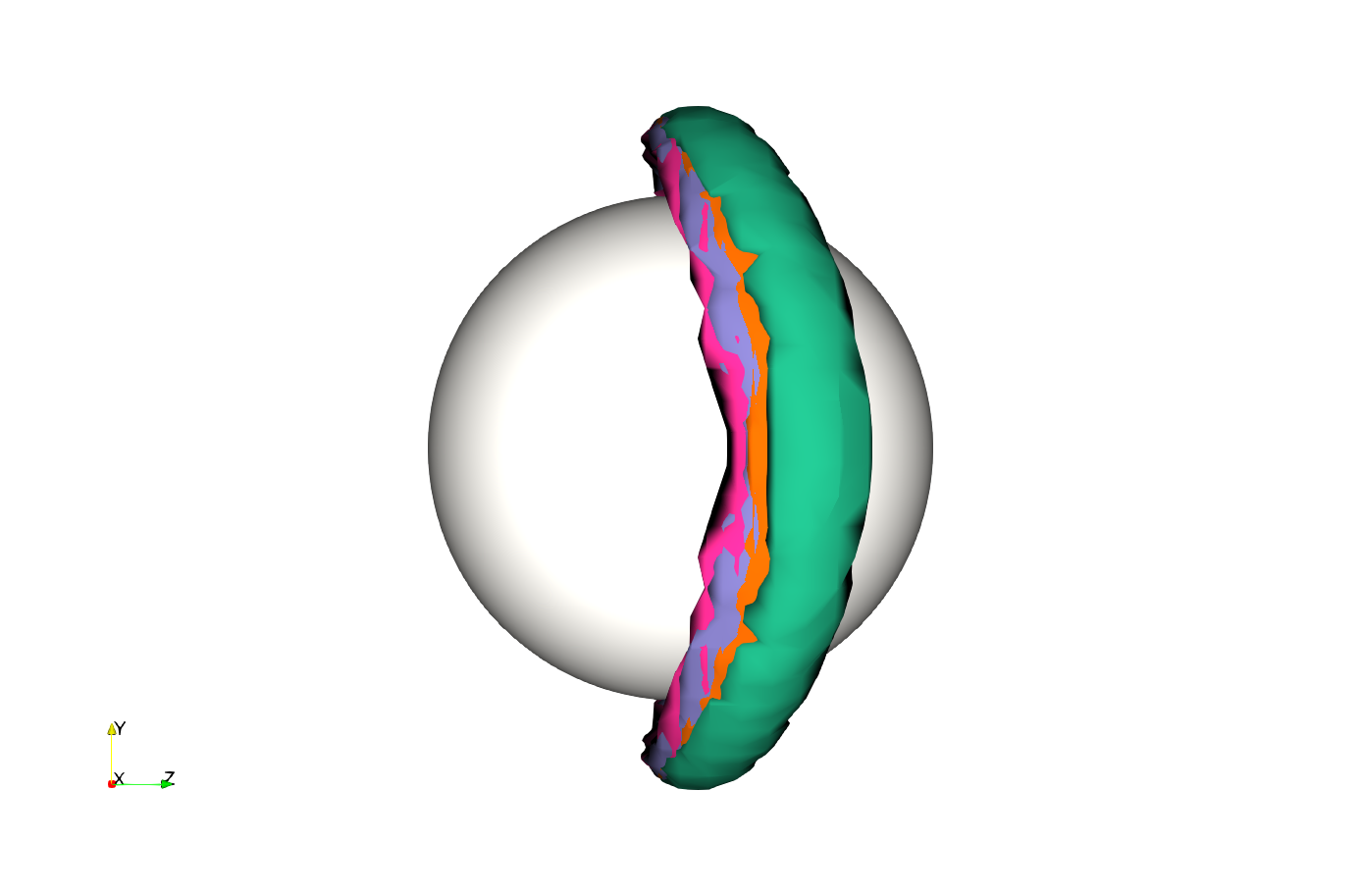} &
\includegraphics[trim={13cm 3cm 13cm 3cm},clip,scale=0.1]{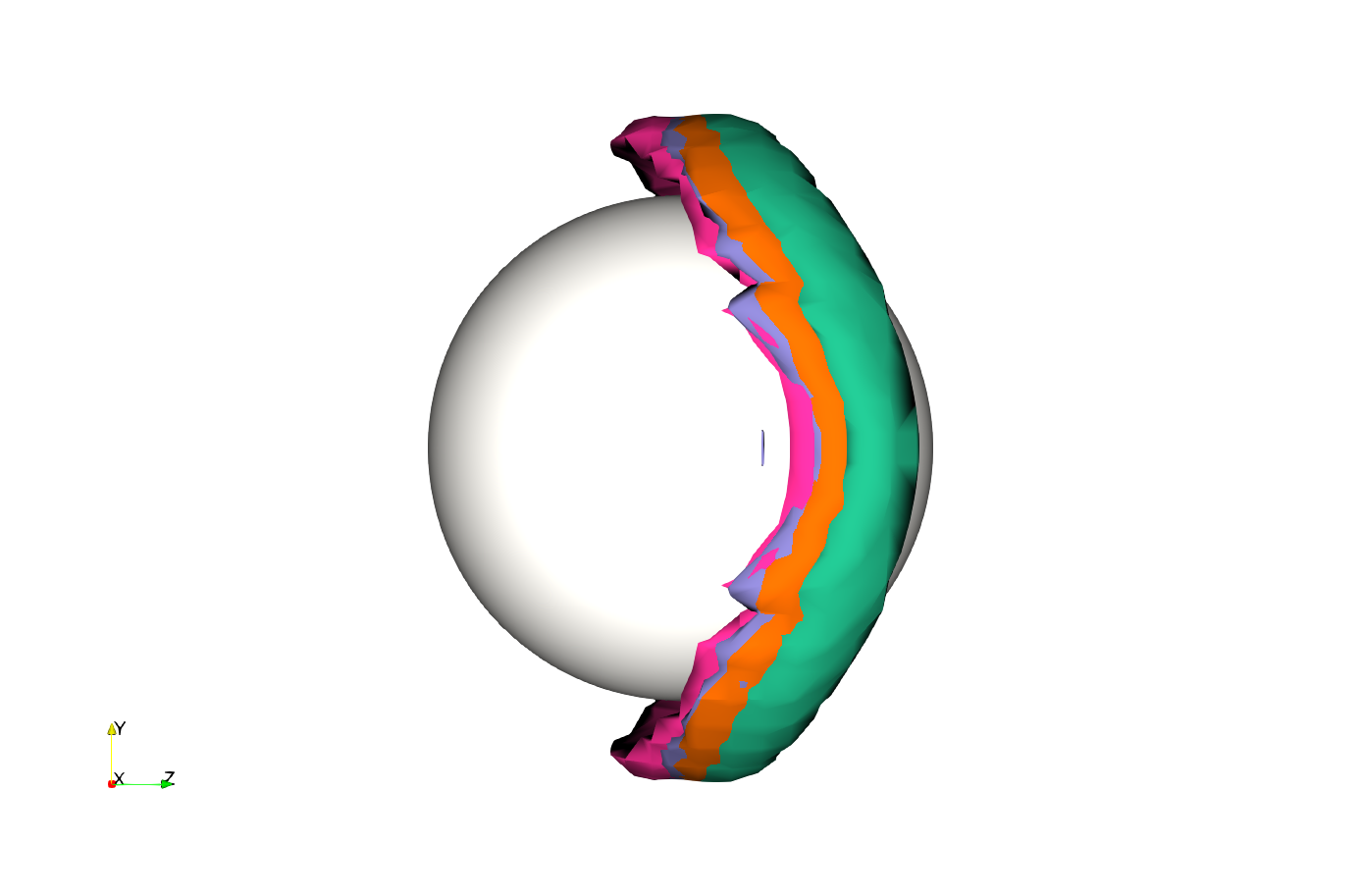} &
\includegraphics[trim={13cm 3cm 13cm 3cm},clip,scale=0.1]{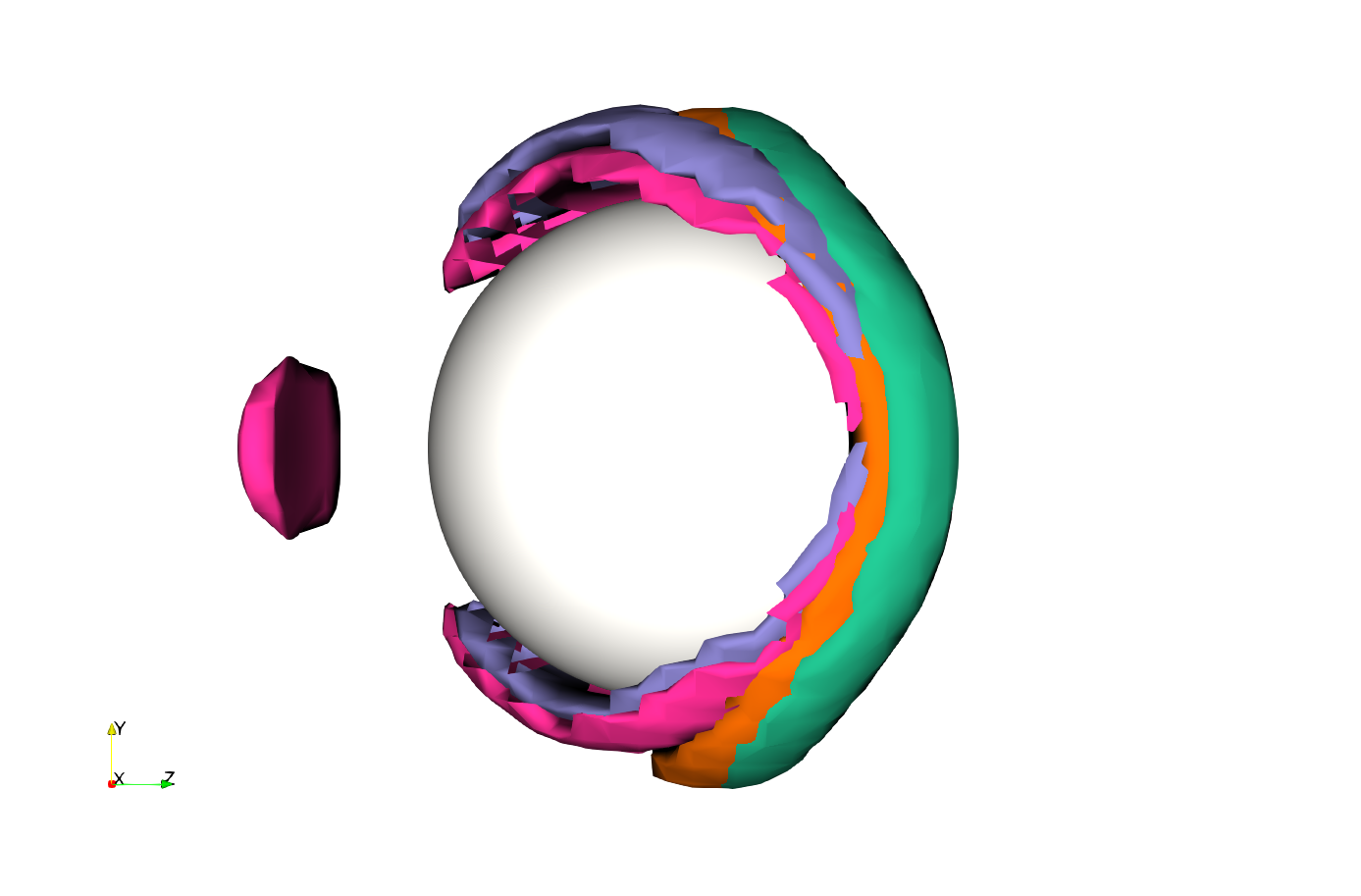} \\
{\small \textcolor{bla1}{Er=18.10} \textcolor{bla2}{Er=35.55 }}
& {\small \textcolor{bla1}{Er=21.25} \textcolor{bla2}{Er=39.18 }}
& {\small \textcolor{bla1}{Er=17.95} \textcolor{bla2}{Er=22.32 }} \\
{\small \textcolor{bla3}{Er=52.72} \textcolor{bla4}{Er=69.77}} 
& {\small \textcolor{bla3}{Er=51.86} \textcolor{bla4}{Er=102.12}}
& {\small \textcolor{bla3}{Er=64.85}  \textcolor{bla4}{Er=86.06}}   \\
\rev{\qquad \includegraphics[trim={3.5cm 3.5cm 37.5cm 25.5cm},clip,scale=0.3]{yz_axis.png}} &  \rev{\qquad \includegraphics[trim={3.5cm 3.5cm 37.5cm 25.5cm},clip,scale=0.3]{yz_axis.png}} & \rev{\qquad \includegraphics[trim={3.5cm 3.5cm 37.5cm 25.5cm},clip,scale=0.3]{yz_axis.png}} \\
\end{tabular}
\caption{Saturn ring disclination lines around the particle for various  Ericksen numbers after the bend-to-spay transition has taken place. The confinement ratios are a) $2R/L_x= 0.4$, b) $2R/L_x= 0.6$ and c) $2R/L_x= 0.8$.
In the top row the view is along the negative $y$-direction, the widest duct dimension, with the walls in the narrowest duct dimension at the $x$-boundaries situated above and below the particle. The bottom row is the view in the positive $x$-direction. The flow is in the horizontal $z$-direction from left to right.}
\label{fig6}
\end{figure}

\begin{figure*}[htbp]
\centering
\begin{tabular}{lllllll}
\rev{Er}$= 22.32$ & \multicolumn{1}{r}{\includegraphics[trim={3.5cm 3.5cm 37.5cm 25.5cm},clip,scale=0.2]{xz_axis.png}} & \rev{Er}$=64.85 $ & \multicolumn{1}{r}{\includegraphics[trim={3.5cm 3.5cm 37.5cm 25.5cm},clip,scale=0.2]{xz_axis.png}} & \rev{Er}$= 86.06$ & \multicolumn{1}{r}{\includegraphics[trim={3.5cm 3.5cm 37.5cm 25.5cm},clip,scale=0.2]{xz_axis.png}} & \\
\multicolumn{2}{l}{\includegraphics[trim={4.5cm 4.5cm 4.5cm 4.5cm }, clip, width=0.29\linewidth]{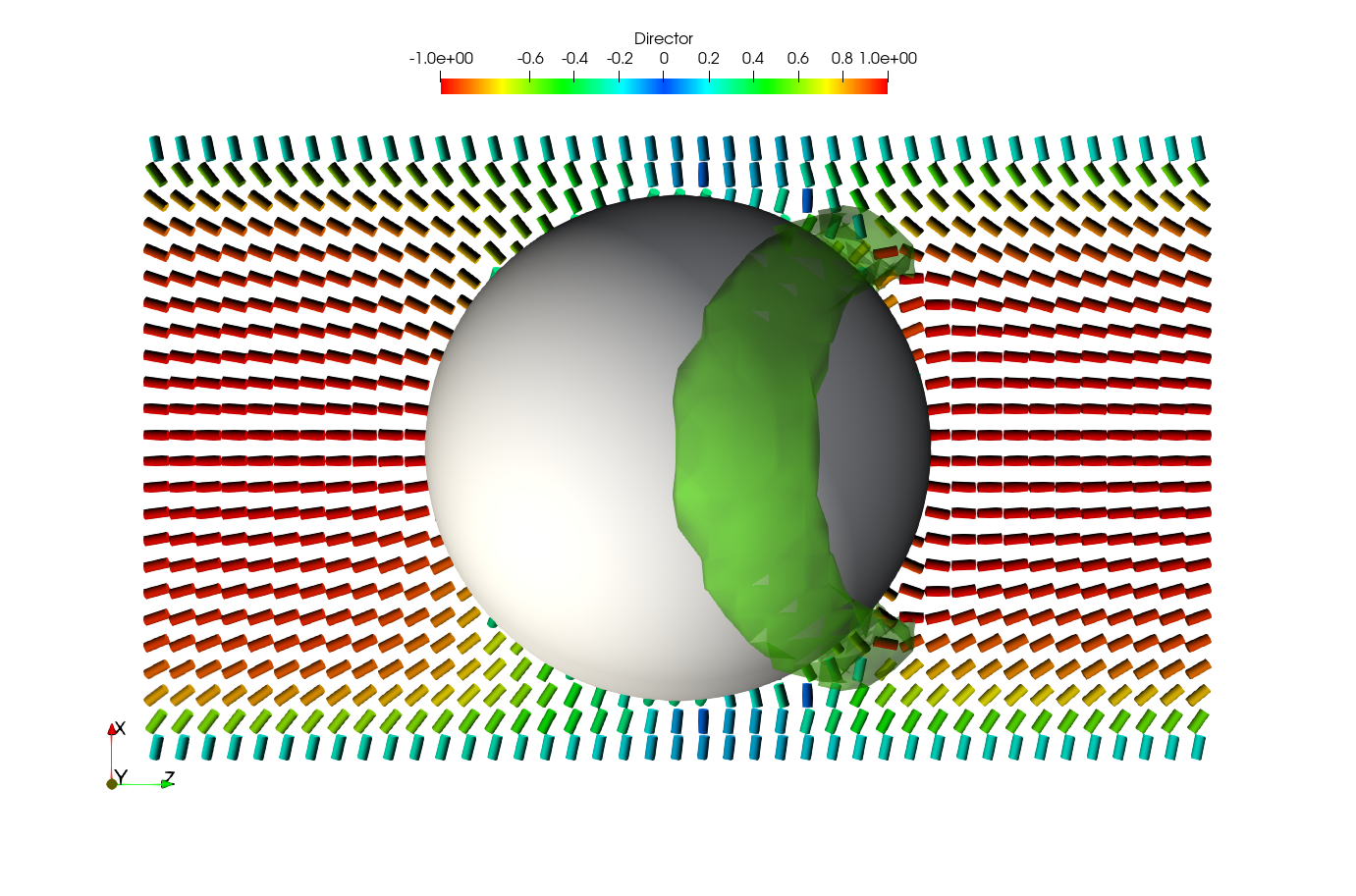}} & 
\multicolumn{2}{l}{\includegraphics[trim={3cm 6.5cm 5cm 2cm},clip, width=0.29\linewidth]{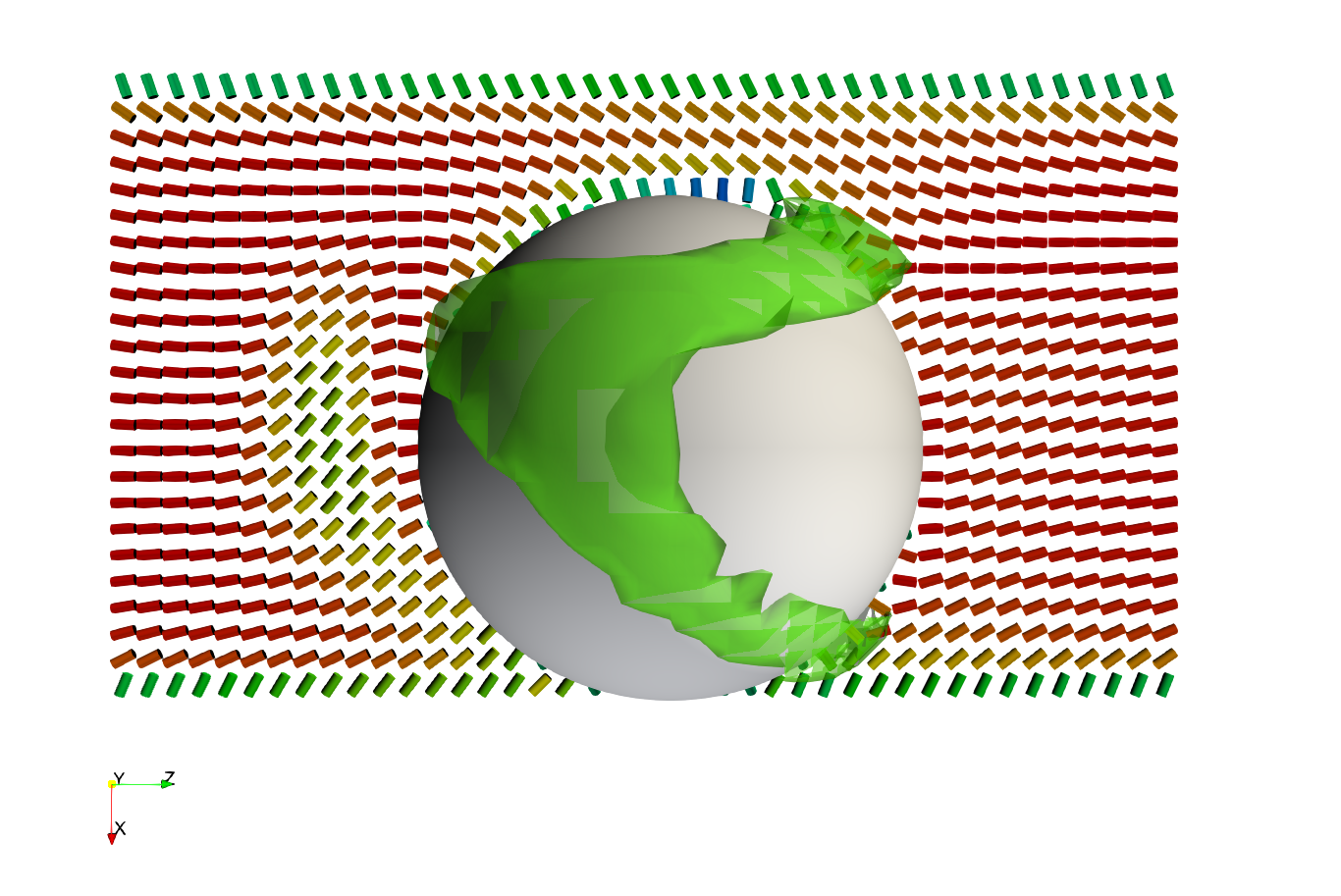}} & 
\multicolumn{2}{l}{\includegraphics[trim={3cm 6.5cm 5cm 2cm},clip,width=0.29\linewidth]{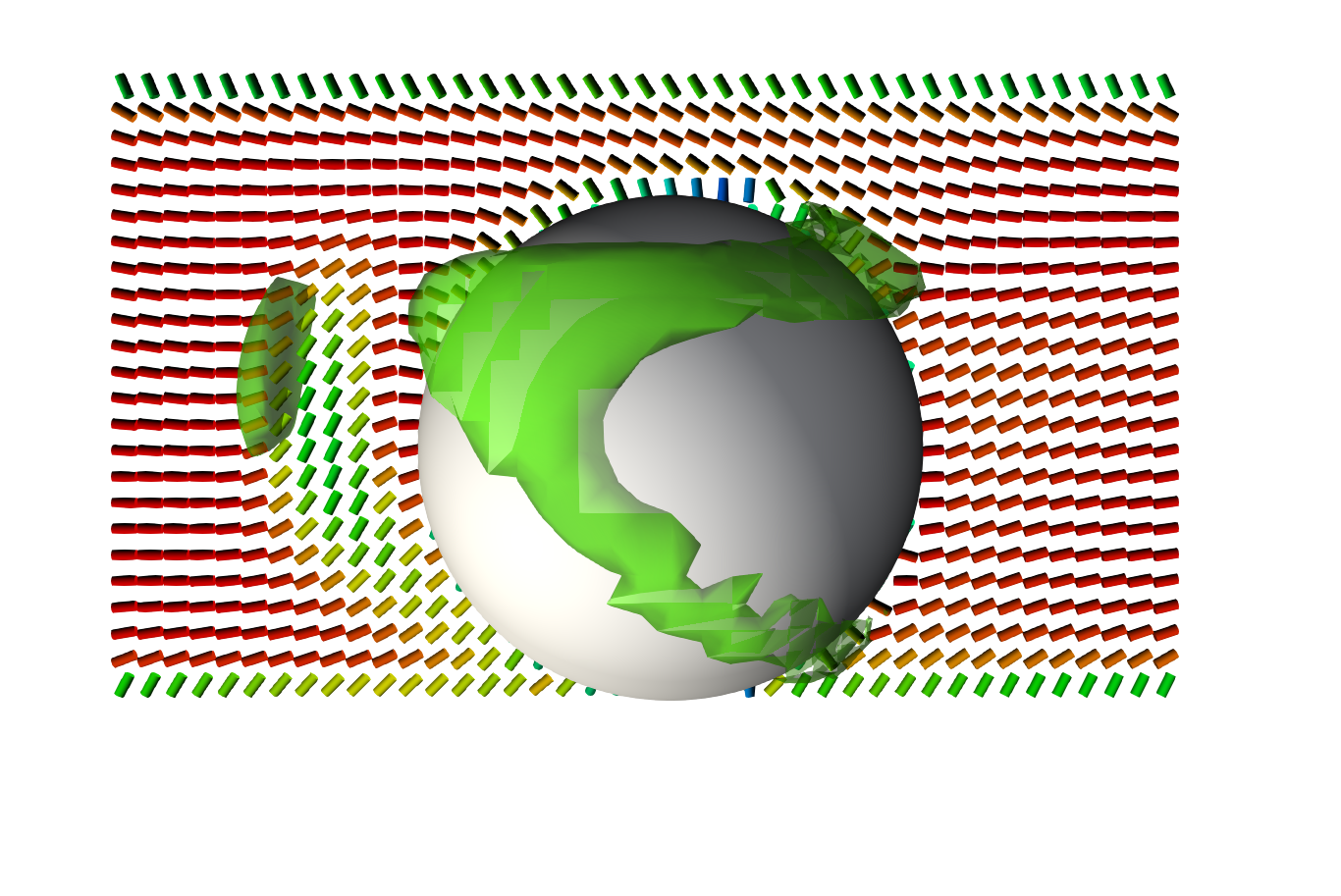}} & \includegraphics[width=0.055\linewidth, ]{colorbar_dir.png} \\
\multicolumn{2}{l}{\includegraphics[width=0.29\linewidth]{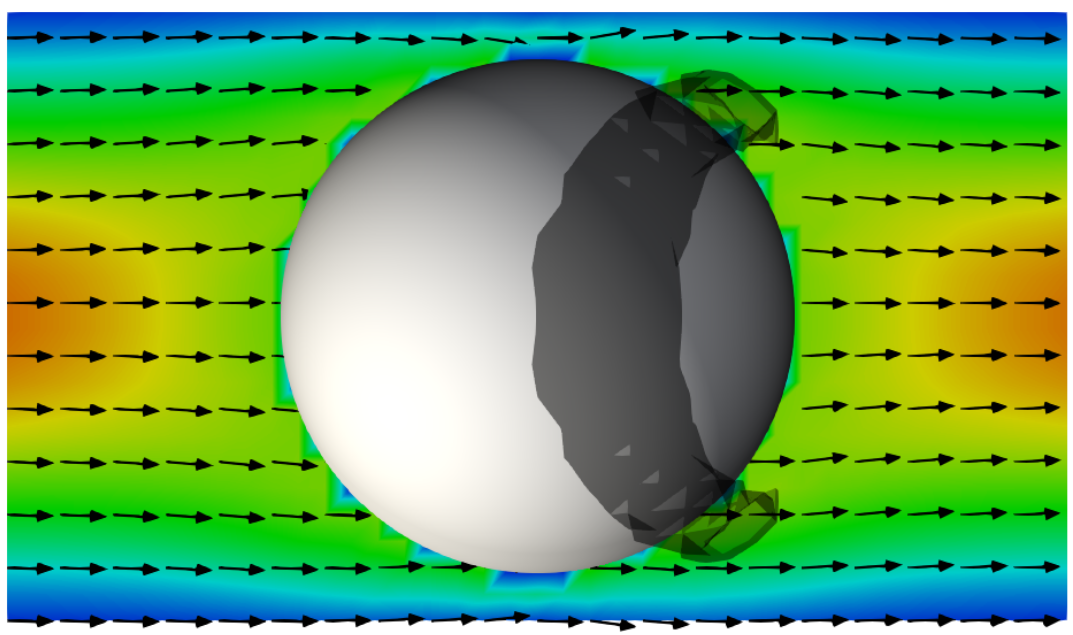}} &
\multicolumn{2}{l}{\includegraphics[width=0.29\linewidth]{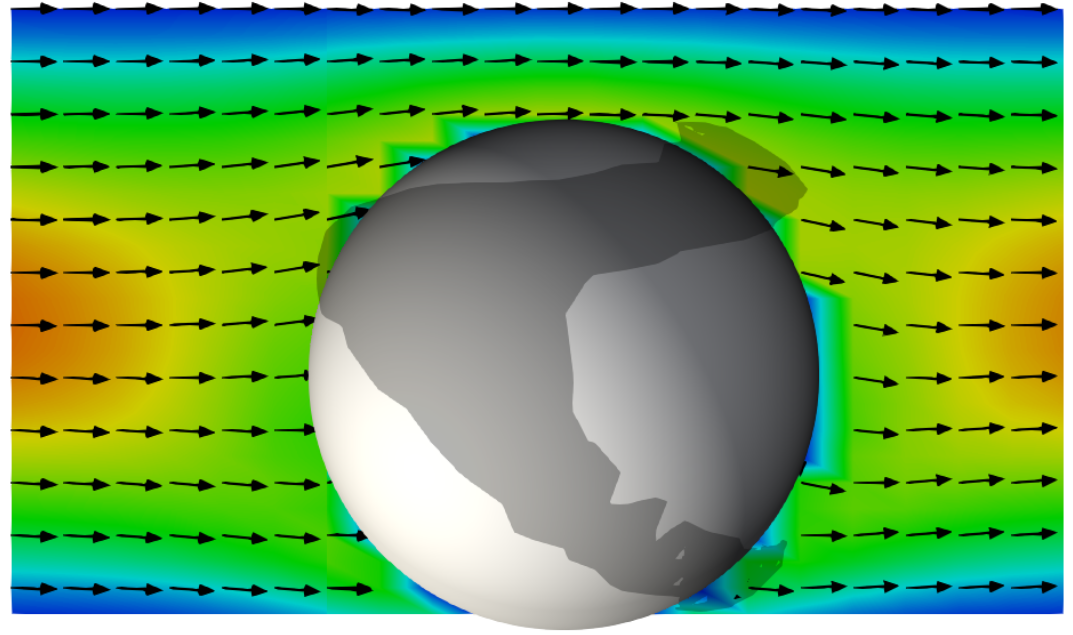}} &
\multicolumn{2}{l}{\includegraphics[width=0.29\linewidth]{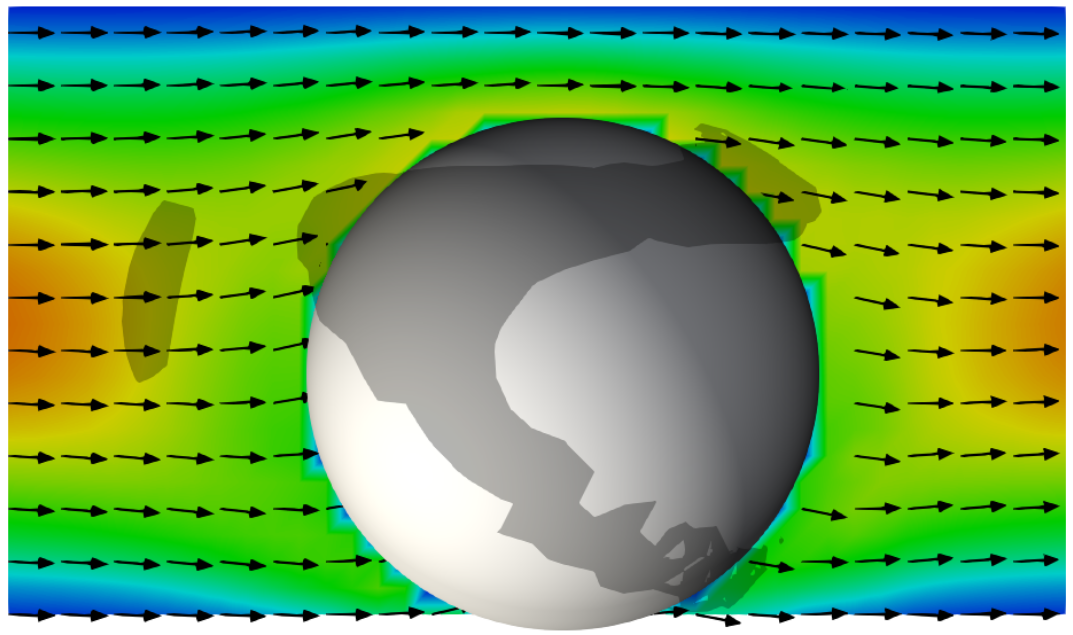}} &
\hspace*{0.5mm}\raisebox{1.5mm}{\includegraphics[width=0.055\linewidth]{colorbar_xz.png}} \\
\rev{Er}$= 22.32$ & \multicolumn{1}{r}{\includegraphics[trim={3.5cm 3.5cm 37.5cm 25.5cm},clip,scale=0.2]{yz_axis.png}} & \rev{Er}$=64.85 $ & \multicolumn{1}{r}{\includegraphics[trim={3.5cm 3.5cm 37.5cm 25.5cm},clip,scale=0.2]{yz_axis.png}} & \rev{Er}$= 86.06$ & \multicolumn{1}{r}{\includegraphics[trim={3.5cm 3.5cm 37.5cm 25.5cm},clip,scale=0.2]{yz_axis.png}} & \\
\multicolumn{2}{l}{\includegraphics[trim={4cm 2.5cm 4cm 2.5cm }, clip, width=0.29\linewidth]{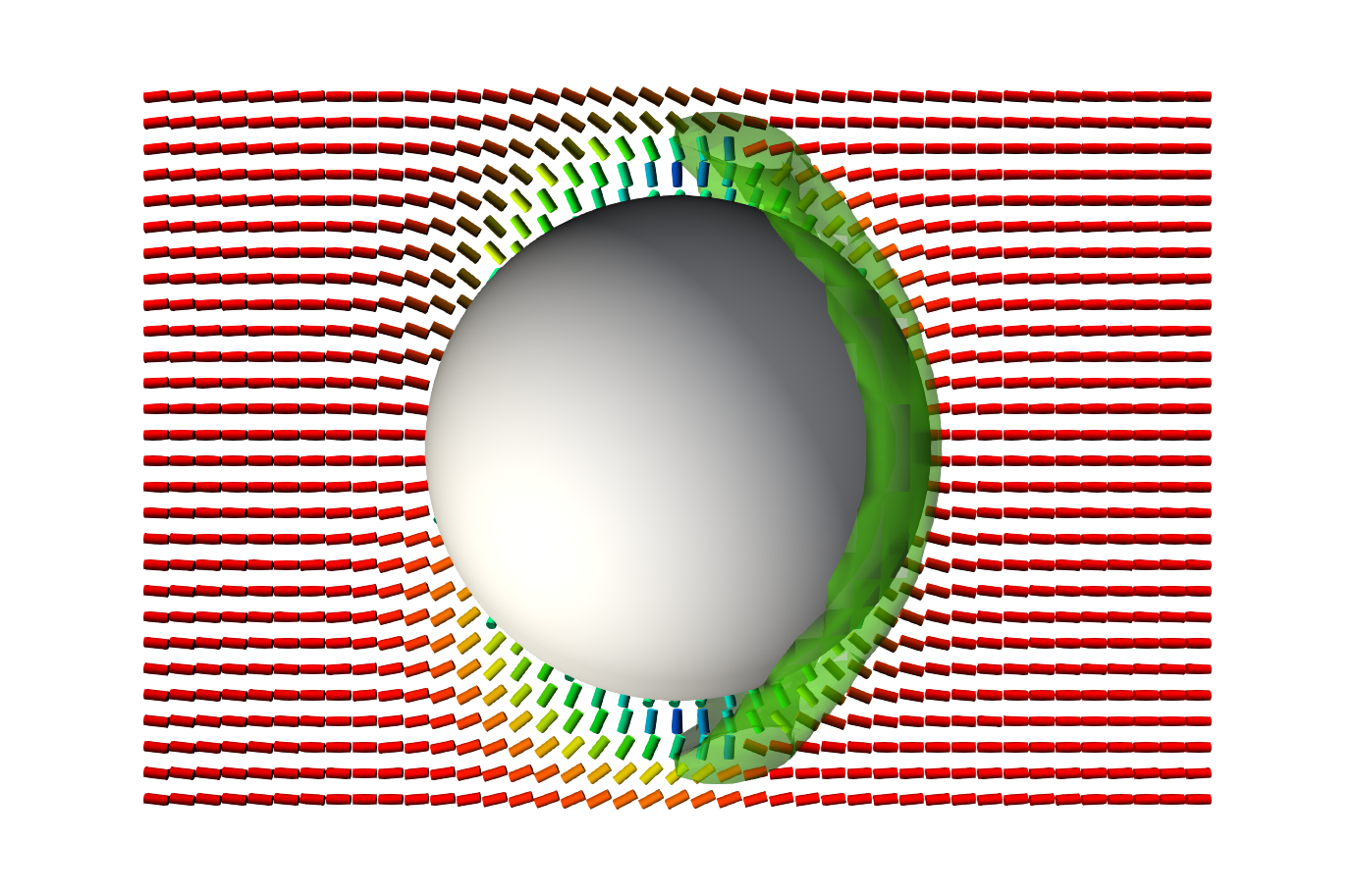}} & 
\multicolumn{2}{l}{\includegraphics[trim={2cm 2cm 4cm 2cm},clip,width=0.29\linewidth]{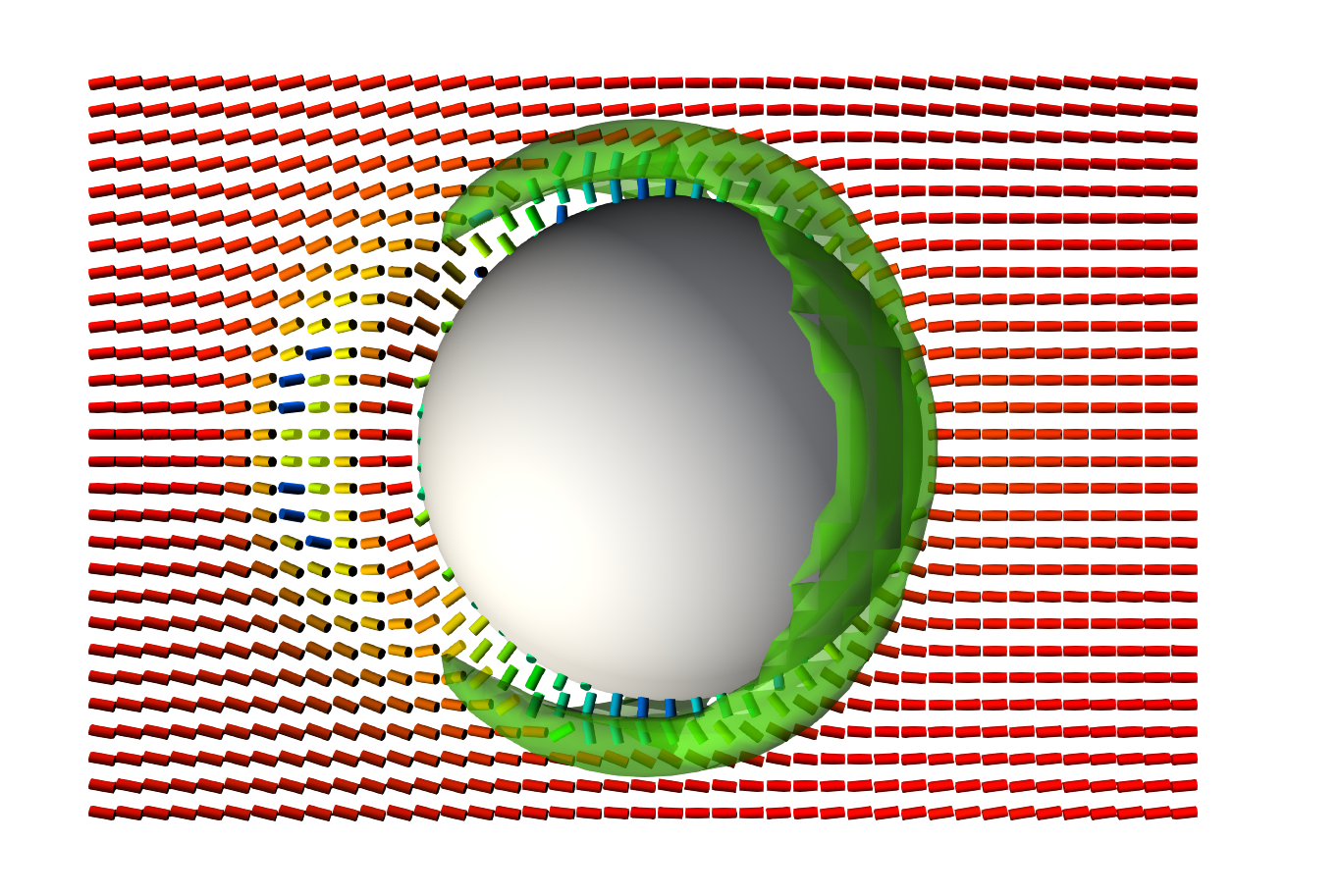}} & 
\multicolumn{2}{l}{\includegraphics[trim={4cm 3cm 6cm 3cm},clip,width=0.29\linewidth]{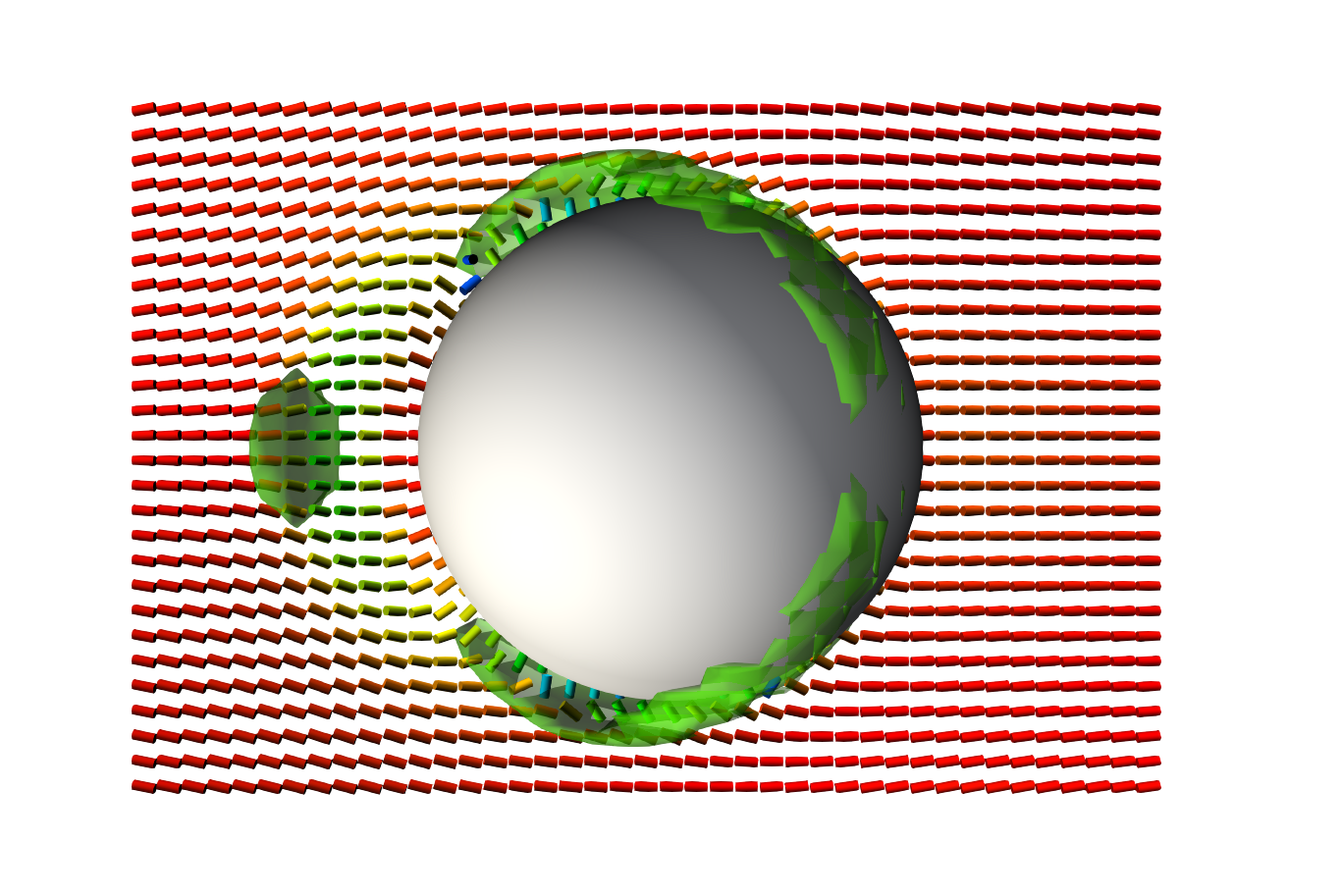} } & 
\raisebox{2.5mm}{\includegraphics[width=0.055\linewidth]{colorbar_dir.png}} \\
\multicolumn{2}{l}{\includegraphics[width=0.29\linewidth]{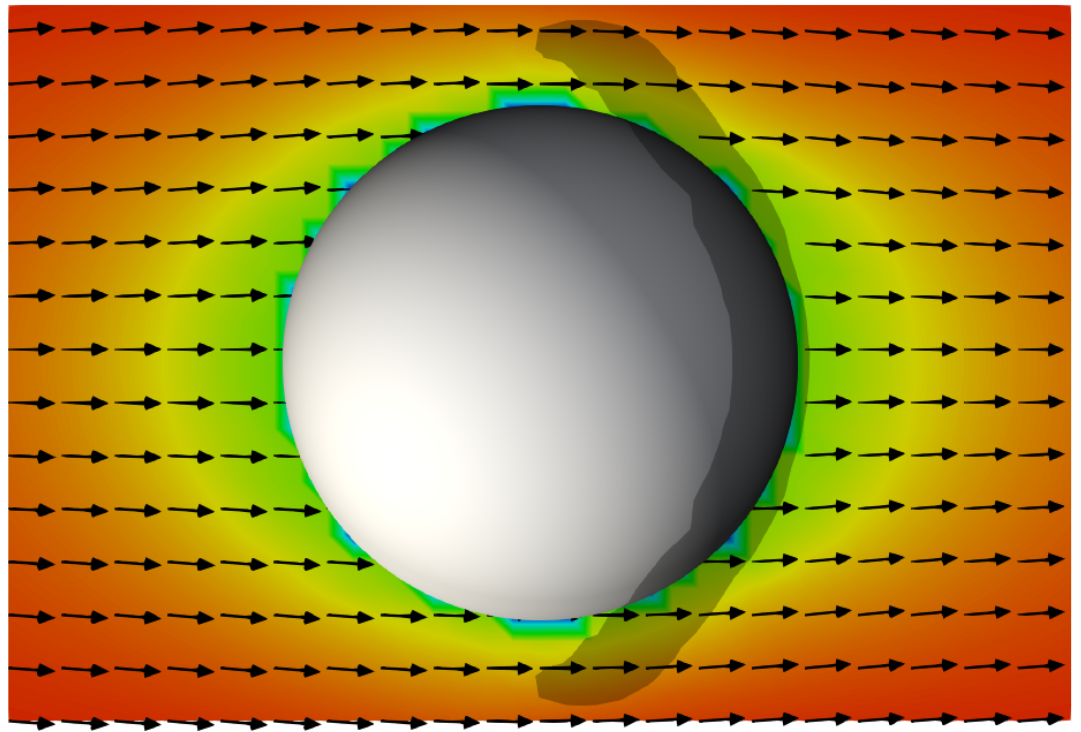}} &
\multicolumn{2}{l}{\includegraphics[width=0.29\linewidth]{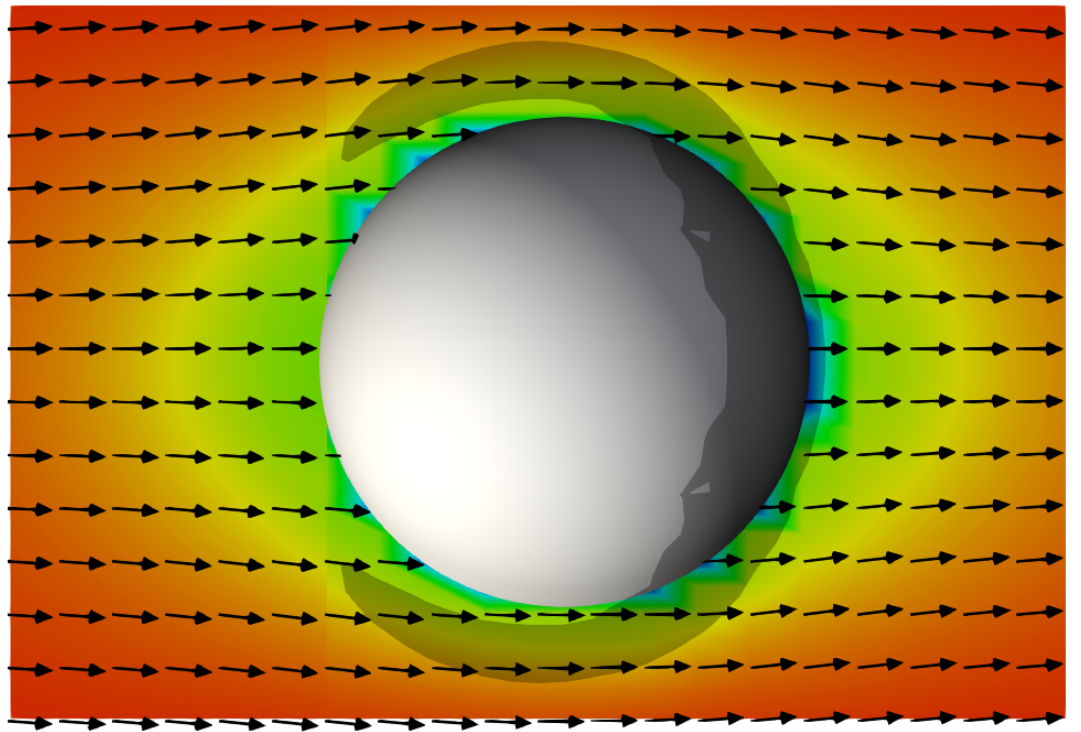}} &
\multicolumn{2}{l}{\includegraphics[width=0.29\linewidth]{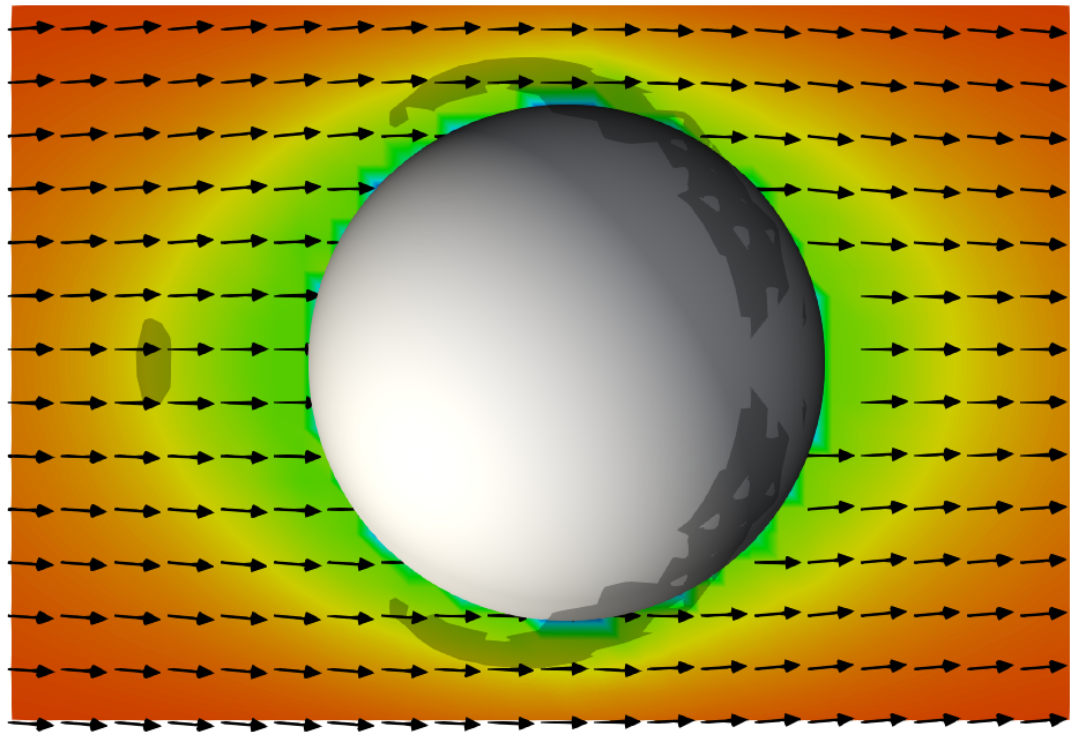}} &
\hspace*{0.5mm}\raisebox{2.5mm}{\includegraphics[width=0.055\linewidth]{colorbar_yz.png}} \\
\end{tabular}
\caption{Director field, defect structure and \rev{fluid velocity profiles} for confinement ratio $2R/L_x= 0.8$ and anchoring parameter $\omega=48$ after the bend-to-splay transition
for increasing Ericksen numbers \rev{Er}$=22.32, 64.85$ and $86.06$, respectively.
The first and third row show the director field $\bm{d}$ with the magnitude $d_z$ of its $z$-component indicated through the colour code. \rev{The second and fourth row show the magnitude of the fluid velocity $u(x,z)$ and $u(y,z)$ through the centre of the particle, normalised to the maximum velocity $u_c$ at the centre line of the duct, where arrows give a sense of the vectorial dependence of the fluid velocity field.} 
The images in the \rev{two top rows} represent slices through the middle of the channel in the $xz$-plane (narrowest duct dimension and flow direction) and have the view along the negative $y$-dimension. Those in the \rev{two bottom rows} show slices in the $yz$-plane (widest and narrowest duct dimension) and have the view in positive $x$-direction. The flow direction is from left to right in positive $z$-direction. The opacity of the defect rings (green isosurfaces) has been slightly reduced to enhance the visibility of the local director field.}
\label{fig7}
\end{figure*}

\begin{figure*}[htb]
\centering
\includegraphics[trim={0cm 0cm 0cm 0cm},clip,width=0.9\linewidth]{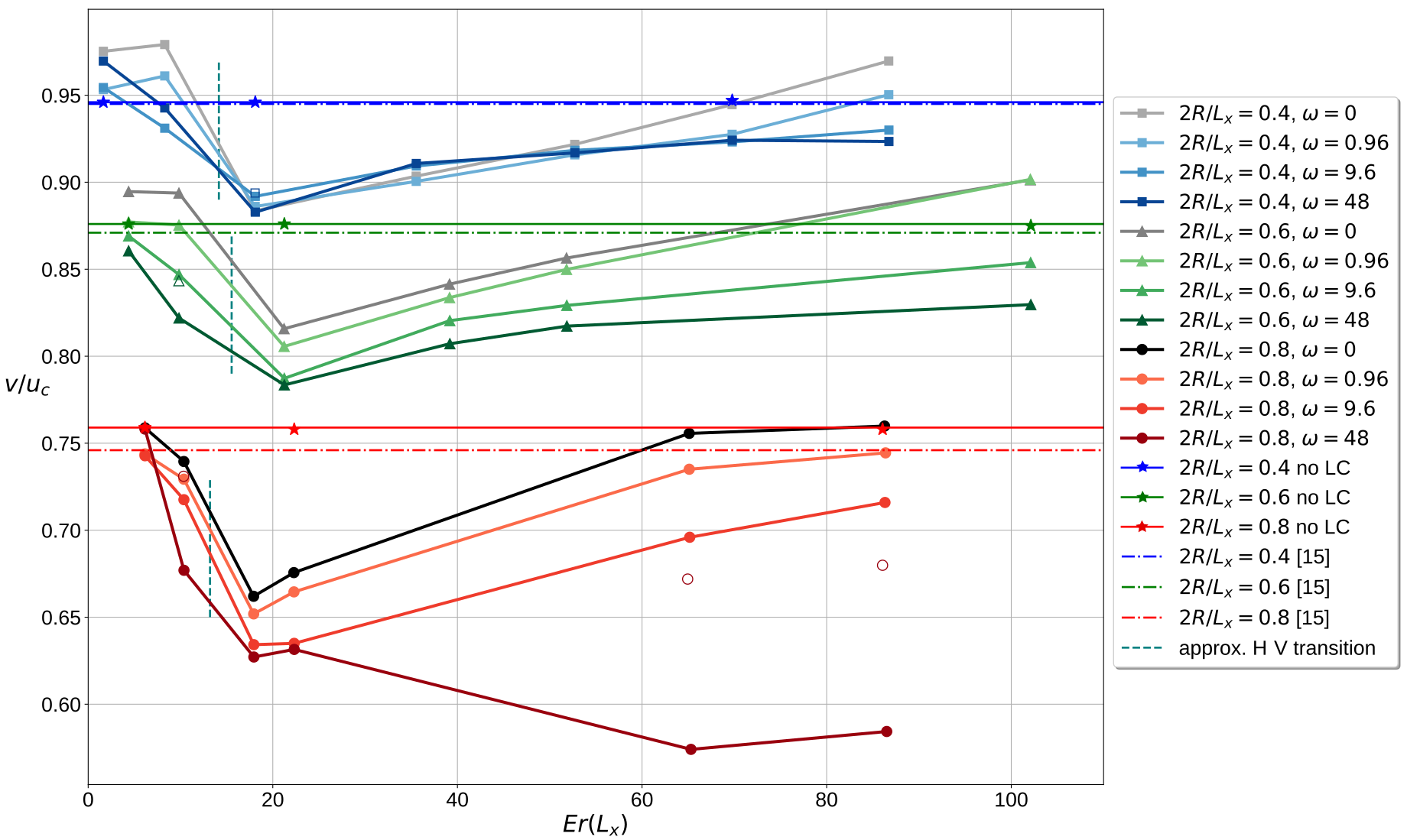}
\caption{Comparison of retardation ratios $v/u_c$ of particle velocity $v$ to fluid velocity $u_c$ at the centre of the rectangular duct for confinement ratios $2R/L_x=0.4$ (blue \rev{squares}), $0.6$ (green \rev{triangles}) and $0.8$ (red \rev{circles}) and different particle anchoring strengths. Horizontal lines show results in a Newtonian fluid from Staben et al. \cite{Staben2003} (dashed-dotted lines) and our approach \rev{in the isotropic phase (stars)}. Open symbols indicate cases where the colloidal particle has been fixed in $x$-direction for comparison as it would normally migrate away from the centre of the duct to either an off-centre position or to the walls. The vertical lines indicate the approximate position of the bend-to-splay transition.}
\label{fig8}
\end{figure*}

Before focusing on the director structure at high Ericksen numbers and large confinement in more detail (see Fig.~\ref{fig7}), we present briefly a synopsis of the defect rings at different confinement ratios and Ericksen numbers. Fig.~\ref{fig6} shows superimposed, vertically oriented defect rings as they occur after the bend-to-splay transition has taken place.
At the lowest confinement ratio $2R/L_x=0.4$, shown in Fig.~\ref{fig6} a), the defect ring remains relatively undistorted across a range of medium to high Ericksen numbers. However, comparing the images at the top with the view along the $y$-direction across the narrowest gap to those at the bottom with the view along the $x$-direction across the widest gap gives evidence that the shape of the Saturn ring defects is sensitive to confinement. When confined, the defect rings are located slightly downstream from the particle's equator, whereas they remain situated along the equator in the dimension of no or very small confinement ($2R/L_y=0.075$).
This feature becomes more pronounced as the confinement increases, discernible through the green defect rings at ratios $2R/L_x=0.6$ in Fig.~\ref{fig6} b), and more so at $2R/L_x=0.8$ in Fig.~\ref{fig6} c) where it results in the compressed appearance (Fig.~\ref{fig6} c) top row). This applies to lower (green isosurfaces) and medium (orange isosurfaces) Ericksen numbers. Increasing both confinement and Ericksen numbers leads to the aforementioned different appearance of the defect rings (purple and magenta isosurfaces).   

It is worth mentioning that the confinement ratios we studied are larger than those in similar studies \cite{Stieger2014, Stieger2015} ($2R/L_x= 0.25$ and $2R/L_x= 0.19$, respectively), where the lower confinement has been chosen to eliminate possible effects on the results. However, our case of $2R/L_x= 0.4$ is obviously already low enough to feature defect rings that appear undeformed and occur at unaltered relative positions to the particle. 

The director field, defect structure \rev{and magnitude of the fluid velocity} at high Ericksen numbers and the largest confinement ratio $2R/L_x=0.8$ are shown in Fig.~\ref{fig7}. At this confinement ratio the walls at the $x$-boundaries are close to the colloidal particle.
The \rev{two top rows} contain slices in the $xz$-plane (narrowest duct dimension and flow direction) at $y=L_y/2$ with walls at the $x$-boundaries at the top and bottom, whereas the \rev{two bottom rows} show slices in the $yz$-plane (widest duct dimension and flow direction) at $x=L_x/2$ cropped to the vicinity of the colloidal particle.
The director field in the first and third row is colour-coded with red indicating an orientation parallel to the flow or $z$-direction and blue indicating an orientation perpendicular to the flow direction or in $xy$-plane.
The left column shows the situation at moderately high Ericksen numbers \rev{Er}$=22.32$. 

The defect ring is vertically oriented, noticeably displaced \rev{downstream close to the walls at the boundary in $x$-direction} (see Fig.~\ref{fig7} first row first column), and situated at the equatorial region of the particle in the non-confined $y$-dimension (see Fig.~\ref{fig7} third row first column). 
The director field structure in $yz$-plane shows that flow alignment occurs in a short distance from the particle and entails a defect in the equatorial region. 
Focusing again on the director field in $xz$-plane reveals that the situation is different in the confined $x$-dimension. Here, the homeotropic anchoring conditions at the wall and particle surfaces prevent any kind of flow alignment in the narrow gap between the particle and the walls.
\rev{Considering the left-hand upstream side of the particle it becomes evident that both the normal anchoring conditions on the surface and the flow-alignment close to the surface work in the same sense and promote the same director orientation. This is different on the right-hand downstream side. While downstream directly right from the particle's centre flow-alignment and anchoring are also working in the same sense, this is not the case downstream above right and below right from the centre where flow-alignment invokes a northwest-southeast orientation of the director field, while surface anchoring promotes a northeast-southwest orientation.} This leads to the slight downstream displacement of the defect ring.

At higher Ericksen numbers \rev{Er}$=64.85$ and \rev{Er}$=86.06$ the particle migrates readily to one of the walls \cite{Lesniewska2022} and the shape of the defect changes markedly (see Fig.~\ref{fig7} first and third row, second and third column). The asymmetric positioning of the particle in the duct is only partly responsible for this. In fact, we observe large differential velocity between the particle and the fluid, which means the particle acts now increasingly as obstacle. \rev{Therefore, it is instructive to look again at fluid velocity profiles.}

\rev{The second and fourth row in Fig.~\ref{fig7} show the magnitude of the fluid velocity $u(x,z)=|\bm{u}(x,z)|$ in the $xz$-plane and $u(y,z)=|\bm{u}(y,z)|$ in the $yz$-plane normalised to the maximum velocity $u_c$ at the centre line of the duct.
The profiles in $xz$-plane (second row) show that compared to Fig.~\ref{fig3} where the confinement ratio is $2R/L_x=0.6$, the now larger confinement ratio of $2R/L_x=0.8$ leads to much lower relative fluid velocities upstream and downstream on the left and right of the particle. With increasing Ericksen number a region with enhanced flow velocities emerges immediately above the particle where the fluid is forced upwards (see Fig.~\ref{fig7} second row third column). 
The fluid velocity profiles in $yz$-plane (see Fig.~\ref{fig7} fourth row) demonstrate even further how the relative fluid velocity drops around the particle with increasing Ericksen number. However, what the colour code and normalisation to the peak flow velocity $u_c$ hide is that the velocity gradients in absolute terms are even larger for larger pressure gradient, a direct consequence of higher absolute values of the peak velocity $u_c$. In view of the director field and defect structure, it becomes evident that the regions with large fluid velocity gradients are also the regions where the director structure becomes noticeably distorted. This effect in combination with local flow-alignment and surface anchoring leads to local regions of low order, for instance the satellite region of very low order slightly upstream on the left of the particle (see Fig.~\ref{fig7} second and third column), and causes the defect ring to become extended further upstream, albeit never completely engulfing the particle.   
}

We conclude our study with an analysis of the \rev{advection} behaviour of the colloidal particle at different Ericksen numbers and confinement ratios and compare it to that in a simple Newtonian fluid. For this purpose we draw on the theoretical results obtained by Staben et al. \cite{Staben2003}, which have been reproduced in a number of studies.
\rev{While our Reynolds numbers are typically between ${\cal O}(10^{-2})$ and ${\cal O}(10^{-1})$ and therefore larger than those in Ref.~\cite{Staben2003}, it is worth emphasising that the latter results form still a suitable reference as both regimes can be classed as low-Reynolds number.} 

A suitable measure to characterise the \rev{advection} behaviour is the retardation ratio $v/u_c$ of particle velocity $v$ to fluid velocity $u_c$ at the centre \rev{line} of the duct. In an isotropic Newtonian fluid under Poiseuille flow this ratio is constant and depends only on the distance of the particle from the walls of the duct and the confinement ratio. In particular, $v/u_c$ is independent of the Reynolds number. Without confinement the retardation ratio $v/u_c$ is unity as the particle acts as a tracer and is simply advected with the fluid. At finite confinement ratios below $2R/L_x=1$ the movement of the particle is slowed down in the parabolic Poiseuille flow due to the no-slip boundary conditions on the walls of the duct. 

Fig.~\ref{fig8} shows the retardation ratio $v/u_c$ for different confinement ratios $2R/L_x$, particle anchoring strengths and Ericksen numbers \rev{Er}. 
Using the Ericksen number as abscissa has the advantage that the bend-to-splay transition occurs at similar values aiding the comparison across different confinement ratios. 
The straight, horizontal lines represent the results for a particle in a simple Newtonian fluid. Dashed-dotted lines give the results from \rev{Staben et al. \cite{Staben2003}} for particles at the centre of the duct. 
We measure retardation ratios of $v/u_c=0.946, 0.876$ and $0.759$ for confinement ratios $2R/L_x=0.4, 0.6$ and $0.8$, respectively, shown in Fig.~\ref{fig8} with solid lines. These results compare well with those of Staben et al., which are $v/u_c=0.945, 0.871$ and $0.746$ for the same confinement ratios and particles positioned at the centre of the duct.
It is worth emphasising that in our setup the largest confinement ratio $2R/L_x=0.8$ has less than 3 lattice sites between the particle surface and the walls on either side. Nevertheless, the relative deviation between ours and Staben's results for Newtonian host phases is less than $1.8 \%$ in the worst case, which means our method is remarkably accurate given the relatively sparse discretisation. However, it should be borne in mind that when modelling a liquid crystalline host phases the sparse discretisation affects also the tensor order parameter $\bm{Q}$ in addition to the fluid-solid interaction in a Newtonian host phase. While these limitations affect the results in Fig.~\ref{fig8} to a certain extent, there are nevertheless clear and robust trends that we will now discuss. 

At low Ericksen numbers we observe retardation ratios $v/u_c$ that are close or identical to their corresponding values in Newtonian fluids. Interestingly, and primarily for no or low particle anchoring strength and low confinement ratios $2R/L_x=0.4$ and $0.6$, the retardation ratio can be slightly larger in the nematic host phase than in the Newtonian host phase (see light and medium grey and blue data points in Fig.~\ref{fig8}).
\rev{We postulate this occurs because for a particular pressure gradient the peak flow velocity is lower in the flowing nematic than in the Newtonian fluid. But as the same pressure gradients acts across the particle, the latter does not slow down to the same degree in the flowing nematic, leading to comparably higher retardation ratios. For larger anchoring strengths or in higher confinement both additional elastic forces are exerted on the particle and the effective viscosity in the vicinity of the particle increases, both to the effect of slowing down the particle, resulting in smaller retardation ratios.}

As the Ericksen number increases, the nematic host phase undergoes a transition from the bend to the splay phase. This occurs at Ericksen numbers $8.30 < $\rev{Er}$ < 18.10\; (2R/Lx = 0.4), 9.84 < $\rev{Er}$ < 21.25\; (2R/Lx = 0.6)$ and $10.37 < $\rev{Er}$ < 17.95\; (2R/Lx = 0.8)$, respectively and is indicated by the vertical green dashed lines in Fig.~\ref{fig8}. The transition is accompanied by a noticeable drop in the retardation ratio, which reaches a minimum around Ericksen numbers \rev{Er}$\simeq20$, so just beyond the bend-to-splay transition. The minimum is smaller the larger the particle anchoring strength is, but only for medium and large confinement (see medium grey and green curves as well as black and red curves in Fig.~\ref{fig8}) and not so for small confinement (see light grey and blue curves in Fig.~\ref{fig8}). 

Beyond Ericksen numbers in the range of \rev{Er}$\simeq20$ the retardation ratio $v/u_c$ begins to increase again, giving rise to an overall non-monotonous dependence on the Ericksen number. This is the case across all confinement ratios, and the retardation ratios begin to flatten out towards higher Ericksen numbers, approaching or reaching the values of Newtonian fluids again. This non-monotonous behaviour is therefore a consequence of the decreasing importance of liquid crystalline elasticity and consistent with the idea that with higher the Ericksen numbers the liquid crystal behaves rheologically more like a simple fluid.

Regarding how the retardation ratio $v/u_c$ depends on the particle anchoring strength the same tends as for the minima prevail. Higher anchoring strengths entail smaller retardation ratios unless the confinement is small. For our largest confinement ratio $2R/L_x=0.8$ and strongest particle anchoring strength $w_{part}=0.05$ we observe a very strong decrease. This, however, originates also from the migration of the particles to the walls. The two empty circles in Fig.~\ref{fig8} (and similar empty symbols at the two other confinement ratios) permit us to estimate how the trend would continue if the particles had been prevented from leaving the region of maximum flow velocity at the centre of the duct.

In order to explain these findings, we have to look at several separate mechanisms: First of all, there is the transition from the bend to the splay state, which all particles regardless of their anchoring conditions are subject to. The data points for vanishing particle anchoring strength $w_{part}=0$ (light, medium and dark grey in Fig.~\ref{fig8}) are indicative of this. The transition causes the general reduction of the retardation ratios from their initially approximately Newtonian values at low Ericksen numbers to their minima around \rev{Er}$\simeq20$. The reason for this decrease is the drop in apparent viscosity and increase in flow velocity $u_c$ around the centre of the duct, whereas the regions of the particle closer to the walls act as anchor and do not allow the particle to pick up velocity $v$ at the same proportion.

The second mechanism at work is the reorientation of the defect ring at the bend-to-splay transition, provided the particle anchoring strength is large enough for a defect to emerge. The vertical orientation of the defect ring with its ring plane perpendicular to the flow direction and walls increases the effective particle radius in the narrowest duct dimension and therefore the effective confinement ratio. This leads to lower retardation ratio $v/u_c$ the larger the anchoring strength is. However, our results suggest this is only the case provided the confinement is not too large.
For instance, at $2R/L_x=0.4$ there is very little difference between vanishing and very strong particle anchoring up to Ericksen numbers \rev{Er}$\simeq60$, while at $2R/L_x=0.6$ and $0.8$ differences are clearly visible at all Ericksen numbers.
This subtlety can be understood by realising that at the different confinement ratios and flow velocities both velocity and order parameter gradients differ across the particle diameter. At a given flow velocity the gradients are largest in large confinement and vice versa. At a given confinement ratio the velocity gradient is largest at large flow velocities and Ericksen numbers. It is precisely this nonlinear order-flow coupling and the interactions between flow and order structure in the vicinity of the particle that cause the observed minor variations in the retardation ratio. 

Finally, there is also the possibility of a direct interaction with the wall anchoring when Ericksen numbers and confinement ratios are large. \rev{In these situations the colloidal particle shows a tendency to leave the centre of the duct and migrate to the wall regions. There, the \rev{advection} velocity and therefore the retardation ratio are reduced as a result of the no-slip boundary conditions at the walls.} 

\section{Conclusions}\label{sec5}

In microfluidic setups of particle suspensions, confinement is often necessary as it allows a certain degree of lateral control over the particle positions, for instance when techniques like confocal or polarised microscopy are used. Our present study has primarily the goal to address some knowledge gaps as to how defects influence and alter the \rev{advection} behaviour of colloidal particles in moderate and large confinement.

In homeotropic anchoring conditions at the walls and surface of the particle the director field is in the H- or bend state at low Ericksen numbers and has a Saturn ring defect which is oriented parallel to the walls. Increasing the confinement changes the appearance of the defect ring downstream. It can either thicken the defect ring, or invoke a migration to off-centre off-wall positions which we identify with the weak attractor region in our previous work \cite{Lesniewska2022}. The latter entails a slight distortion and tilt away from the centre plane.    

At moderately high Ericksen numbers around \rev{Er}$\simeq20$ we observe the transition from the bend or H-state to the splay or V-state. This leads generally to a reorientation of the defect ring with a ring plane perpendicular to the flow direction and walls. The defect ring is slightly peeled off downstream in the confined dimension, but sits at the particle's equatorial region in the unconfined dimension, giving it the appearance of an open mouth when viewed from the flow direction. These features are retained at higher Ericksen numbers and in lower confinement. Highly confined particles show a strong tendency to migrate to the walls, a behaviour we observed also in our previous work \cite{Lesniewska2022}. This leads to a highly asymmetric defect and induces a satellite region of low order upstream of the particle, which acts partly as an obstacle and forces the flow to \rev{slow down and} divert around it. \rev{Compared to our previous work we do not observe migration to the walls for all but the highest Ericksen numbers and confinement ratios that we tested. Therefore, increased confinement entails stabilisation of trajectories at the centre of the duct.}

The interaction between nematic order and flow on one hand, and the fluid-solid interaction on the other hand results in a non-monotonous dependence of the retardation ratio, the ratio of particle \rev{advection} velocity to the maximum velocity at the centre of the duct, on the Ericksen number.  
When the Ericksen number is low, the retardation ratio is close to values observed in a Newtonian host phase in all confinement ratios and particle anchoring conditions. This is also the case for vanishing or low anchoring strength and at high Ericksen numbers, where the nematic liquid crystal behaves increasingly like a simple Newtonian fluid as the relative importance of elastic effects decreases.
Intermediate Ericksen numbers, however, are characterised by a pronounced minimum in the retardation ratio. We attribute this to a combination of two effects: Firstly there is the bend-to-splay transition, to which particles in all anchoring conditions are subject. Secondly, the defect ring undergoes a reorientation from horizontal alignment with the ring plane parallel to the walls to a vertical orientation, which has the ring plane perpendicular to the flow direction and the walls. This increases the effective particle radius and therefore the confinement. The second effect is only present when the defect ring is properly formed, i.e. for stronger particle anchoring strengths, and when the confinement is lower. This is because the increased retardation that the particle experiences is a consequence of the interaction of the defect with the gradients of the flow velocity and liquid crystalline order. 

The present study leaves some questions untouched, for example how planar degenerate or hybrid anchoring conditions affects the defect morphology and \rev{advection} behaviour of the particles. Planar degenerate wall anchoring is fully compatible with the flow alignment that takes place at higher Ericksen numbers. Hence, there is no bend-to-splay transition, rather a more gradual transition to a state where the director field is flow-aligned at the Leslie angle. Furthermore, planar degenerate anchoring conditions on the particle surface invoke topologically different boojum defects that occur at low Ericksen numbers symmetrically upstream and downstream of the particle on an axis that goes through the centre of the particle. Similarly, particles with homeotropic anchoring conditions as used in the present study, but larger diameters, have also topologically different defects. Instead of the half-integer bulk defect loops with topological charge $-1/2$ they have dipolar full integer satellite defects with topological charge $-1$ that sit in a distance from the particle surface. It is not clear how these topological differences would affect the advection and migration behaviour. 

\section*{Conflicts of Interest}
There are no conflicts to declare.

\section*{Acknowledgements}

We would like to thank Timm Kr\"uger and Uro\v{s} Tkalec for very helpful discussions. We acknowledge support from EPSRC under Grant No. EP/R513349/1. M.L. acknowledges funding from the Mac Robertson Postgraduate Travel Scholarship.
For the purpose of complying with UKRI's open access policy, the authors have applied a Creative Commons Attribution (CC BY) license to any Author Accepted Manuscript version arising from this submission.
This work used the ARCHIE-WeSt High Performance Computer (\href{https://www.archie-west.ac.uk}{www.archie-west.ac.uk}) based at the University of Strathclyde.

\section*{Data Availability}

\rev{Data underpinning this publication are openly available from the University of Strathclyde KnowledgeBase research information portal.}



\balance


\bibliography{references.bib} 

\providecommand*{\mcitethebibliography}{\thebibliography}
\csname @ifundefined\endcsname{endmcitethebibliography}
{\let\endmcitethebibliography\endthebibliography}{}
\begin{mcitethebibliography}{47}
\providecommand*{\natexlab}[1]{#1}
\providecommand*{\mciteSetBstSublistMode}[1]{}
\providecommand*{\mciteSetBstMaxWidthForm}[2]{}
\providecommand*{\mciteBstWouldAddEndPuncttrue}
  {\def\EndOfBibitem{\unskip.}}
\providecommand*{\mciteBstWouldAddEndPunctfalse}
  {\let\EndOfBibitem\relax}
\providecommand*{\mciteSetBstMidEndSepPunct}[3]{}
\providecommand*{\mciteSetBstSublistLabelBeginEnd}[3]{}
\providecommand*{\EndOfBibitem}{}
\mciteSetBstSublistMode{f}
\mciteSetBstMaxWidthForm{subitem}
{(\emph{\alph{mcitesubitemcount}})}
\mciteSetBstSublistLabelBeginEnd{\mcitemaxwidthsubitemform\space}
{\relax}{\relax}

\bibitem[Chin \emph{et~al.}(2011)Chin, Laksanasopin, Cheung, Steinmiller,
  Linder, Parsa, Wang, Moore, Rouse, Umviligihozo, Karita, Mwambarangwe,
  Braunstein, van~de Wijgert, Sahabo, Justman, El-Sadr, and Sia]{Chin2011}
C.~D. Chin, T.~Laksanasopin, Y.~K. Cheung, D.~Steinmiller, V.~Linder, H.~Parsa,
  J.~Wang, H.~Moore, R.~Rouse, G.~Umviligihozo, E.~Karita, L.~Mwambarangwe,
  S.~L. Braunstein, J.~van~de Wijgert, R.~Sahabo, J.~E. Justman, W.~El-Sadr and
  S.~K. Sia, \emph{Nat. Med.}, 2011, \textbf{17}, 1015--U138\relax
\mciteBstWouldAddEndPuncttrue
\mciteSetBstMidEndSepPunct{\mcitedefaultmidpunct}
{\mcitedefaultendpunct}{\mcitedefaultseppunct}\relax
\EndOfBibitem
\bibitem[Hu \emph{et~al.}(2014)Hu, Wang, Wang, Li, Pingguan-Murphy, Lu, and
  Xu]{Hu2014}
J.~Hu, S.~Wang, L.~Wang, F.~Li, B.~Pingguan-Murphy, T.~J. Lu and F.~Xu,
  \emph{Biosens. Bioelectron.}, 2014, \textbf{54}, 585--597\relax
\mciteBstWouldAddEndPuncttrue
\mciteSetBstMidEndSepPunct{\mcitedefaultmidpunct}
{\mcitedefaultendpunct}{\mcitedefaultseppunct}\relax
\EndOfBibitem
\bibitem[Zhang \emph{et~al.}(2013)Zhang, Chan, and Leong]{Zhang2013}
Y.~Zhang, H.~F. Chan and K.~W. Leong, \emph{Adv. Drug Deliv. Rev.}, 2013,
  \textbf{65}, 104--120\relax
\mciteBstWouldAddEndPuncttrue
\mciteSetBstMidEndSepPunct{\mcitedefaultmidpunct}
{\mcitedefaultendpunct}{\mcitedefaultseppunct}\relax
\EndOfBibitem
\bibitem[Golombek \emph{et~al.}(2018)Golombek, May, Theek, Appold, Drude,
  Kiessling, and Lammers]{Golombek2018}
S.~K. Golombek, J.-N. May, B.~Theek, L.~Appold, N.~Drude, F.~Kiessling and
  T.~Lammers, \emph{Adv. Drug Deliv. Rev.}, 2018, \textbf{130}, 17--38\relax
\mciteBstWouldAddEndPuncttrue
\mciteSetBstMidEndSepPunct{\mcitedefaultmidpunct}
{\mcitedefaultendpunct}{\mcitedefaultseppunct}\relax
\EndOfBibitem
\bibitem[Seemann \emph{et~al.}(2012)Seemann, Brinkmann, Pfohl, and
  Herminghaus]{Seemann2012}
R.~Seemann, M.~Brinkmann, T.~Pfohl and S.~Herminghaus, \emph{Rep. Prog. Phys.},
  2012, \textbf{75}, 016601\relax
\mciteBstWouldAddEndPuncttrue
\mciteSetBstMidEndSepPunct{\mcitedefaultmidpunct}
{\mcitedefaultendpunct}{\mcitedefaultseppunct}\relax
\EndOfBibitem
\bibitem[Mark \emph{et~al.}(2010)Mark, Haeberle, Roth, von Stetten, and
  Zengerle]{Mark2010}
D.~Mark, S.~Haeberle, G.~Roth, F.~von Stetten and R.~Zengerle, \emph{Chem. Soc.
  Rev.}, 2010, \textbf{39}, 1153--1182\relax
\mciteBstWouldAddEndPuncttrue
\mciteSetBstMidEndSepPunct{\mcitedefaultmidpunct}
{\mcitedefaultendpunct}{\mcitedefaultseppunct}\relax
\EndOfBibitem
\bibitem[Qin \emph{et~al.}(2010)Qin, Xia, and Whitesides]{Qin2010}
D.~Qin, Y.~Xia and G.~M. Whitesides, \emph{Nat. Protoc.}, 2010, \textbf{5},
  491--502\relax
\mciteBstWouldAddEndPuncttrue
\mciteSetBstMidEndSepPunct{\mcitedefaultmidpunct}
{\mcitedefaultendpunct}{\mcitedefaultseppunct}\relax
\EndOfBibitem
\bibitem[Sackmann \emph{et~al.}(2014)Sackmann, Fulton, and Beebe]{Sackmann2014}
E.~K. Sackmann, A.~L. Fulton and D.~J. Beebe, \emph{Nature}, 2014,
  \textbf{507}, 181--189\relax
\mciteBstWouldAddEndPuncttrue
\mciteSetBstMidEndSepPunct{\mcitedefaultmidpunct}
{\mcitedefaultendpunct}{\mcitedefaultseppunct}\relax
\EndOfBibitem
\bibitem[Squires and Quake(2005)]{Squires2005}
T.~M. Squires and S.~R. Quake, \emph{Rev. Mod. Phys.}, 2005, \textbf{77},
  977--1026\relax
\mciteBstWouldAddEndPuncttrue
\mciteSetBstMidEndSepPunct{\mcitedefaultmidpunct}
{\mcitedefaultendpunct}{\mcitedefaultseppunct}\relax
\EndOfBibitem
\bibitem[{Di Carlo} \emph{et~al.}(2007){Di Carlo}, Irimia, Tompkins, and
  Toner]{DiCarlo2007}
D.~{Di Carlo}, D.~Irimia, R.~G. Tompkins and M.~Toner, \emph{Proc. Natl. Acad.
  Sci. U. S. A.}, 2007, \textbf{104}, 18892--18897\relax
\mciteBstWouldAddEndPuncttrue
\mciteSetBstMidEndSepPunct{\mcitedefaultmidpunct}
{\mcitedefaultendpunct}{\mcitedefaultseppunct}\relax
\EndOfBibitem
\bibitem[Amini \emph{et~al.}(2014)Amini, Lee, and Di~Carlo]{Amini2014}
H.~Amini, W.~Lee and D.~Di~Carlo, \emph{Lab Chip}, 2014, \textbf{14},
  2739\relax
\mciteBstWouldAddEndPuncttrue
\mciteSetBstMidEndSepPunct{\mcitedefaultmidpunct}
{\mcitedefaultendpunct}{\mcitedefaultseppunct}\relax
\EndOfBibitem
\bibitem[Zhang \emph{et~al.}(2016)Zhang, Yan, Yuan, Alici, Nguyen,
  Ebrahimi~Warkiani, and Li]{Zhang2016}
J.~Zhang, S.~Yan, D.~Yuan, G.~Alici, N.-T. Nguyen, M.~Ebrahimi~Warkiani and
  W.~Li, \emph{Lab Chip}, 2016, \textbf{16}, 10--34\relax
\mciteBstWouldAddEndPuncttrue
\mciteSetBstMidEndSepPunct{\mcitedefaultmidpunct}
{\mcitedefaultendpunct}{\mcitedefaultseppunct}\relax
\EndOfBibitem
\bibitem[Ho and Leal(1974)]{Ho1974}
B.~Ho and L.~Leal, \emph{J. Fluid. Mech.}, 1974, \textbf{65}, 365--400\relax
\mciteBstWouldAddEndPuncttrue
\mciteSetBstMidEndSepPunct{\mcitedefaultmidpunct}
{\mcitedefaultendpunct}{\mcitedefaultseppunct}\relax
\EndOfBibitem
\bibitem[Ganatos \emph{et~al.}(1980)Ganatos, Pfeffer, and
  Weinbaum]{Ganatos1980}
P.~Ganatos, R.~Pfeffer and S.~Weinbaum, \emph{J. Fluid. Mech.}, 1980,
  \textbf{99}, 755--783\relax
\mciteBstWouldAddEndPuncttrue
\mciteSetBstMidEndSepPunct{\mcitedefaultmidpunct}
{\mcitedefaultendpunct}{\mcitedefaultseppunct}\relax
\EndOfBibitem
\bibitem[Staben \emph{et~al.}(2003)Staben, Zinchenko, and Davis]{Staben2003}
M.~E. Staben, A.~Z. Zinchenko and R.~H. Davis, \emph{Phys. Fluids}, 2003,
  \textbf{15}, 1711–1733\relax
\mciteBstWouldAddEndPuncttrue
\mciteSetBstMidEndSepPunct{\mcitedefaultmidpunct}
{\mcitedefaultendpunct}{\mcitedefaultseppunct}\relax
\EndOfBibitem
\bibitem[Staben and Davis(2005)]{Staben2005}
M.~E. Staben and R.~H. Davis, \emph{Int. J. Multiph. Flow}, 2005, \textbf{31},
  529--547\relax
\mciteBstWouldAddEndPuncttrue
\mciteSetBstMidEndSepPunct{\mcitedefaultmidpunct}
{\mcitedefaultendpunct}{\mcitedefaultseppunct}\relax
\EndOfBibitem
\bibitem[Nikoubashman \emph{et~al.}(2013)Nikoubashman, Likos, and
  Kahl]{Nikoubashman2013}
A.~Nikoubashman, C.~N. Likos and G.~Kahl, \emph{Soft Matter}, 2013, \textbf{9},
  2603--2613\relax
\mciteBstWouldAddEndPuncttrue
\mciteSetBstMidEndSepPunct{\mcitedefaultmidpunct}
{\mcitedefaultendpunct}{\mcitedefaultseppunct}\relax
\EndOfBibitem
\bibitem[Stark and Ventzki(2002)]{Stark2002}
H.~Stark and D.~Ventzki, \emph{Europhys. Lett.}, 2002, \textbf{57},
  60--66\relax
\mciteBstWouldAddEndPuncttrue
\mciteSetBstMidEndSepPunct{\mcitedefaultmidpunct}
{\mcitedefaultendpunct}{\mcitedefaultseppunct}\relax
\EndOfBibitem
\bibitem[Grollau \emph{et~al.}(2003)Grollau, Abbott, and de~Pablo]{Grollau2003}
S.~Grollau, N.~L. Abbott and J.~J. de~Pablo, \emph{Phys. Rev. E}, 2003,
  \textbf{67}, 011702\relax
\mciteBstWouldAddEndPuncttrue
\mciteSetBstMidEndSepPunct{\mcitedefaultmidpunct}
{\mcitedefaultendpunct}{\mcitedefaultseppunct}\relax
\EndOfBibitem
\bibitem[G\^{a}rlea and Mulder(2015)]{Garlea2015}
I.~C. G\^{a}rlea and B.~M. Mulder, \emph{Soft Matter}, 2015, \textbf{11},
  608--614\relax
\mciteBstWouldAddEndPuncttrue
\mciteSetBstMidEndSepPunct{\mcitedefaultmidpunct}
{\mcitedefaultendpunct}{\mcitedefaultseppunct}\relax
\EndOfBibitem
\bibitem[G\^{a}rlea \emph{et~al.}(2016)G\^{a}rlea, Mulder, Alvarado, Dammone,
  Aarts, Lettinga, Koenderink, and Mulder]{Garlea2016}
I.~C. G\^{a}rlea, P.~Mulder, J.~Alvarado, O.~Dammone, D.~G. Aarts, M.~P.
  Lettinga, G.~H. Koenderink and B.~M. Mulder, \emph{Nat. Commun.}, 2016,
  \textbf{7}, 12112\relax
\mciteBstWouldAddEndPuncttrue
\mciteSetBstMidEndSepPunct{\mcitedefaultmidpunct}
{\mcitedefaultendpunct}{\mcitedefaultseppunct}\relax
\EndOfBibitem
\bibitem[Tkalec and Mu{\v{s}}evi{\v{c}}(2013)]{Tkalec2013}
U.~Tkalec and I.~Mu{\v{s}}evi{\v{c}}, \emph{Soft Matter}, 2013, \textbf{9},
  8140--8150\relax
\mciteBstWouldAddEndPuncttrue
\mciteSetBstMidEndSepPunct{\mcitedefaultmidpunct}
{\mcitedefaultendpunct}{\mcitedefaultseppunct}\relax
\EndOfBibitem
\bibitem[{\v{C}}opar \emph{et~al.}(2021){\v{C}}opar, Ravnik, and
  {\v{Z}}umer]{Copar2021}
S.~{\v{C}}opar, M.~Ravnik and S.~{\v{Z}}umer, \emph{Crystals}, 2021,
  \textbf{11}, 956\relax
\mciteBstWouldAddEndPuncttrue
\mciteSetBstMidEndSepPunct{\mcitedefaultmidpunct}
{\mcitedefaultendpunct}{\mcitedefaultseppunct}\relax
\EndOfBibitem
\bibitem[Sulaiman \emph{et~al.}(2006)Sulaiman, Marenduzzo, and
  Yeomans]{Sulaiman2006}
N.~Sulaiman, D.~Marenduzzo and J.~M. Yeomans, \emph{Phys. Rev. E}, 2006,
  \textbf{74}, 041708\relax
\mciteBstWouldAddEndPuncttrue
\mciteSetBstMidEndSepPunct{\mcitedefaultmidpunct}
{\mcitedefaultendpunct}{\mcitedefaultseppunct}\relax
\EndOfBibitem
\bibitem[Urbanski \emph{et~al.}(2017)Urbanski, Reyes, Noh, Sharma, Geng,
  Jampani, and Lagerwall]{Urbanski2017}
M.~Urbanski, C.~G. Reyes, J.~Noh, A.~Sharma, Y.~Geng, V.~S.~R. Jampani and
  J.~P.~F. Lagerwall, \emph{J. Phys. Condens. Matt.}, 2017, \textbf{29},
  133003\relax
\mciteBstWouldAddEndPuncttrue
\mciteSetBstMidEndSepPunct{\mcitedefaultmidpunct}
{\mcitedefaultendpunct}{\mcitedefaultseppunct}\relax
\EndOfBibitem
\bibitem[Wiese \emph{et~al.}(2016)Wiese, Marenduzzo, and Henrich]{Wiese2016}
O.~Wiese, D.~Marenduzzo and O.~Henrich, \emph{Soft Matter}, 2016, \textbf{12},
  9223--9237\relax
\mciteBstWouldAddEndPuncttrue
\mciteSetBstMidEndSepPunct{\mcitedefaultmidpunct}
{\mcitedefaultendpunct}{\mcitedefaultseppunct}\relax
\EndOfBibitem
\bibitem[Mondal \emph{et~al.}(2018)Mondal, Griffiths, Charlet, and
  Majumdar]{Mondal2018a}
S.~Mondal, I.~M. Griffiths, F.~Charlet and A.~Majumdar, \emph{Fluids}, 2018,
  \textbf{3}, 39\relax
\mciteBstWouldAddEndPuncttrue
\mciteSetBstMidEndSepPunct{\mcitedefaultmidpunct}
{\mcitedefaultendpunct}{\mcitedefaultseppunct}\relax
\EndOfBibitem
\bibitem[Kos and Ravnik(2020)]{Kos2020}
{\v{Z}}.~Kos and M.~Ravnik, \emph{Sci. Rep.}, 2020, \textbf{10}, 1446\relax
\mciteBstWouldAddEndPuncttrue
\mciteSetBstMidEndSepPunct{\mcitedefaultmidpunct}
{\mcitedefaultendpunct}{\mcitedefaultseppunct}\relax
\EndOfBibitem
\bibitem[Fedorowicz \emph{et~al.}(2023)Fedorowicz, Prosser, and
  Sengupta]{Fedorowicz2023}
K.~Fedorowicz, R.~Prosser and A.~Sengupta, \emph{Soft Matter}, 2023,
  \textbf{19}, 7084--7092\relax
\mciteBstWouldAddEndPuncttrue
\mciteSetBstMidEndSepPunct{\mcitedefaultmidpunct}
{\mcitedefaultendpunct}{\mcitedefaultseppunct}\relax
\EndOfBibitem
\bibitem[Mondal \emph{et~al.}(2018)Mondal, Majumdar, and
  Griffiths]{Mondal2018b}
S.~Mondal, A.~Majumdar and I.~M. Griffiths, \emph{J. Colloid Interface Sci.},
  2018, \textbf{528}, 431--442\relax
\mciteBstWouldAddEndPuncttrue
\mciteSetBstMidEndSepPunct{\mcitedefaultmidpunct}
{\mcitedefaultendpunct}{\mcitedefaultseppunct}\relax
\EndOfBibitem
\bibitem[Stieger \emph{et~al.}(2014)Stieger, Schoen, and Mazza]{Stieger2014}
T.~Stieger, M.~Schoen and M.~G. Mazza, \emph{J. Chem. Phys.}, 2014,
  \textbf{140}, 054905\relax
\mciteBstWouldAddEndPuncttrue
\mciteSetBstMidEndSepPunct{\mcitedefaultmidpunct}
{\mcitedefaultendpunct}{\mcitedefaultseppunct}\relax
\EndOfBibitem
\bibitem[Stieger \emph{et~al.}(2016)Stieger, Püschel-Schlotthauer, Schoen, and
  Mazza]{Stieger2015}
T.~Stieger, S.~Püschel-Schlotthauer, M.~Schoen and M.~G. Mazza,
  \emph{Molecular Physics}, 2016, \textbf{114}, 259--275\relax
\mciteBstWouldAddEndPuncttrue
\mciteSetBstMidEndSepPunct{\mcitedefaultmidpunct}
{\mcitedefaultendpunct}{\mcitedefaultseppunct}\relax
\EndOfBibitem
\bibitem[Lesniewska \emph{et~al.}(2022)Lesniewska, Mottram, and
  Henrich]{Lesniewska2022}
M.~Lesniewska, N.~Mottram and O.~Henrich, \emph{Soft Matter}, 2022,
  \textbf{18}, 6942--6953\relax
\mciteBstWouldAddEndPuncttrue
\mciteSetBstMidEndSepPunct{\mcitedefaultmidpunct}
{\mcitedefaultendpunct}{\mcitedefaultseppunct}\relax
\EndOfBibitem
\bibitem[H\'{\i}jar(2020)]{Hijar2020}
H.~H\'{\i}jar, \emph{Phys. Rev. E}, 2020, \textbf{102}, 062705\relax
\mciteBstWouldAddEndPuncttrue
\mciteSetBstMidEndSepPunct{\mcitedefaultmidpunct}
{\mcitedefaultendpunct}{\mcitedefaultseppunct}\relax
\EndOfBibitem
\bibitem[de~Gennes and Prost(1993)]{deGennes}
P.-G. de~Gennes and J.~Prost, \emph{The Physics of Liquid Crystals}, Oxford
  University Press, New York, 1993\relax
\mciteBstWouldAddEndPuncttrue
\mciteSetBstMidEndSepPunct{\mcitedefaultmidpunct}
{\mcitedefaultendpunct}{\mcitedefaultseppunct}\relax
\EndOfBibitem
\bibitem[Wright and Mermin(1989)]{WrightMermin:1989}
D.~C. Wright and N.~D. Mermin, \emph{Rev. Mod. Phys.}, 1989, \textbf{61},
  385--432\relax
\mciteBstWouldAddEndPuncttrue
\mciteSetBstMidEndSepPunct{\mcitedefaultmidpunct}
{\mcitedefaultendpunct}{\mcitedefaultseppunct}\relax
\EndOfBibitem
\bibitem[Beris and Edwards(1994)]{BerisEdwards}
A.~N. Beris and B.~J. Edwards, \emph{Thermodynamics of flowing systems}, Oxford
  Univ. Press, 1994\relax
\mciteBstWouldAddEndPuncttrue
\mciteSetBstMidEndSepPunct{\mcitedefaultmidpunct}
{\mcitedefaultendpunct}{\mcitedefaultseppunct}\relax
\EndOfBibitem
\bibitem[{\v{S}}karabot \emph{et~al.}(2007){\v{S}}karabot, Ravnik, {\v{Z}}umer,
  Tkalec, Poberaj, Babi{\v{c}}, Osterman, and Musevi{\v{c}}]{Skarabot2007}
M.~{\v{S}}karabot, M.~Ravnik, S.~{\v{Z}}umer, U.~Tkalec, I.~Poberaj,
  D.~Babi{\v{c}}, N.~Osterman and I.~Musevi{\v{c}}, \emph{Phys. Rev. E}, 2007,
  \textbf{76}, 1--8\relax
\mciteBstWouldAddEndPuncttrue
\mciteSetBstMidEndSepPunct{\mcitedefaultmidpunct}
{\mcitedefaultendpunct}{\mcitedefaultseppunct}\relax
\EndOfBibitem
\bibitem[Marenduzzo \emph{et~al.}(2007)Marenduzzo, Orlandini, Cates, and
  Yeomans]{Marenduzzo2007}
D.~Marenduzzo, E.~Orlandini, M.~E. Cates and J.~M. Yeomans, \emph{Phys. Rev.
  E}, 2007, \textbf{76}, 1--18\relax
\mciteBstWouldAddEndPuncttrue
\mciteSetBstMidEndSepPunct{\mcitedefaultmidpunct}
{\mcitedefaultendpunct}{\mcitedefaultseppunct}\relax
\EndOfBibitem
\bibitem[Nguyen and Ladd(2002)]{Nguyen2002}
N.~Q. Nguyen and A.~J. Ladd, \emph{Phys. Rev. E}, 2002, \textbf{66},
  046708\relax
\mciteBstWouldAddEndPuncttrue
\mciteSetBstMidEndSepPunct{\mcitedefaultmidpunct}
{\mcitedefaultendpunct}{\mcitedefaultseppunct}\relax
\EndOfBibitem
\bibitem[Ladd(1994)]{Ladd:1994a}
A.~Ladd, \emph{J. Fluid. Mech.}, 1994, \textbf{271}, 285\relax
\mciteBstWouldAddEndPuncttrue
\mciteSetBstMidEndSepPunct{\mcitedefaultmidpunct}
{\mcitedefaultendpunct}{\mcitedefaultseppunct}\relax
\EndOfBibitem
\bibitem[Ladd(1994)]{Ladd:1994b}
A.~Ladd, \emph{J. Fluid. Mech.}, 1994, \textbf{271}, 311\relax
\mciteBstWouldAddEndPuncttrue
\mciteSetBstMidEndSepPunct{\mcitedefaultmidpunct}
{\mcitedefaultendpunct}{\mcitedefaultseppunct}\relax
\EndOfBibitem
\bibitem[Kr\"uger \emph{et~al.}(2017)Kr\"uger, Kusumaatmaja, Kuzmin, Shardt,
  Silva, and Viggen]{LBBook2017}
T.~Kr\"uger, H.~Kusumaatmaja, A.~Kuzmin, O.~Shardt, G.~Silva and E.~M. Viggen,
  \emph{The Lattice Boltzmann Method: Principles and Practice}, Springer
  International Publishing, Switzerland, 2017\relax
\mciteBstWouldAddEndPuncttrue
\mciteSetBstMidEndSepPunct{\mcitedefaultmidpunct}
{\mcitedefaultendpunct}{\mcitedefaultseppunct}\relax
\EndOfBibitem
\bibitem[Parodi(1970)]{Parodi1970}
O.~Parodi, \emph{J. Phys.}, 1970, \textbf{31}, 581--584\relax
\mciteBstWouldAddEndPuncttrue
\mciteSetBstMidEndSepPunct{\mcitedefaultmidpunct}
{\mcitedefaultendpunct}{\mcitedefaultseppunct}\relax
\EndOfBibitem
\bibitem[{\relax Ludwig GitHub repository}()]{Ludwig}
{\relax Ludwig GitHub repository},
  \url{https://github.com/ludwig-cf/ludwig}\relax
\mciteBstWouldAddEndPuncttrue
\mciteSetBstMidEndSepPunct{\mcitedefaultmidpunct}
{\mcitedefaultendpunct}{\mcitedefaultseppunct}\relax
\EndOfBibitem
\bibitem[Desplat \emph{et~al.}(2001)Desplat, Pagonabarraga, and
  Bladon]{Desplat2001}
J.~C. Desplat, I.~Pagonabarraga and P.~Bladon, \emph{Comput. Phys. Commun.},
  2001, \textbf{134}, 273--290\relax
\mciteBstWouldAddEndPuncttrue
\mciteSetBstMidEndSepPunct{\mcitedefaultmidpunct}
{\mcitedefaultendpunct}{\mcitedefaultseppunct}\relax
\EndOfBibitem
\bibitem[Adhikari \emph{et~al.}(2005)Adhikari, Stratford, Cates, and
  Wagner]{Adhikari2005}
R.~Adhikari, K.~Stratford, M.~E. Cates and A.~J. Wagner, \emph{Europhys.
  Lett.}, 2005, \textbf{71}, 473--479\relax
\mciteBstWouldAddEndPuncttrue
\mciteSetBstMidEndSepPunct{\mcitedefaultmidpunct}
{\mcitedefaultendpunct}{\mcitedefaultseppunct}\relax
\EndOfBibitem
\end{mcitethebibliography}
\bibliographystyle{rsc.bst} 

\end{document}